\documentclass[twocolumn,twocolappendix,numberedappendix]{openjournal}

\usepackage{xcolor}
\usepackage{textgreek}
\usepackage[utf8]{inputenc}
\usepackage[english]{babel}

\usepackage{hyperref}
\hypersetup{
    unicode, 
    colorlinks=true,
    linkcolor=linkcolor,
    citecolor=linkcolor,
    filecolor=linkcolor,
    urlcolor=linkcolor,
}
\usepackage{color,colortbl}
\definecolor{linkcolor}{rgb}{0.0,0.3,0.5}
\usepackage{tensind}
\tensordelimiter{?}
\DeclareGraphicsExtensions{.bmp,.png,.jpg,.pdf}
\usepackage{verbatim}
\usepackage[normalem]{ulem}
\usepackage{orcidlink}
\usepackage{soul}
\usepackage{algorithm}
\usepackage{algpseudocode}
\usepackage{amsmath}
\usepackage{booktabs}
\graphicspath{ {./figs/} }

\begin{document}
\title{Joint Fit of Galaxy Cluster Scaling Relations from Individual Weak Lensing Data: Comparison with Stacked Analysis and a Real Data Test}

\author{André Z. Vitorelli\orcidlink{0000-0002-9740-4591}}
\email{andre.vitorelli@usp.br}
\affiliation{Departamento de Astronomia, Instituto de Astronomia, Geofísica e Ciências Atmosféricas da Universidade de São Paulo, Cidade
Universitária, 05508-900, São Paulo, SP, Brazil.}

\author{Eduardo S. Cypriano\orcidlink{0000-0001-9196-7038}}
\email{eduardo.cypriano@iag.usp.br}
\affiliation{Departamento de Astronomia, Instituto de Astronomia, Geofísica e Ciências Atmosféricas da Universidade de São Paulo, Cidade
Universitária, 05508-900, São Paulo, SP, Brazil}

\begin{abstract}
Stacking is the standard route to signal-to-noise in cluster weak lensing, but it destroys the cluster-by-cluster information that scaling-relation studies need. We instead fit every cluster's individual $\Delta\Sigma$ profile simultaneously, tying its mass and concentration to its own richness, redshift and magnitude gap, and compare with two stacked analyses of the same data: a forward-modelled stack sharing the same model and priors, and the standard two-step analysis. The magnitude gap (the BCG-to-brightest-satellite magnitude difference) enters the concentration. In the fossil-group picture, large gaps trace earlier-assembled, more relaxed systems. On mock populations with triaxial haloes and miscentering, at DES Y3-like and LSST-like density, all three analyses recover the injected gap slope $\gamma_{\mathrm{cm}}$, the joint and forward-modelled fits with uncertainties that match the scatter between realisations. They are not equally precise: the joint fit scatters 1.2--1.3 times less than the forward-modelled stack, and 2 (DES-like) to 6.5 (LSST-like) times less than the two-step analysis, which is also biased in several other scaling-relation parameters. Applied to WaZP clusters with DES Y3 shears, the joint and forward-modelled fits agree on the gap coupling once the mass at fixed richness is allowed to depend on the gap, $\gamma_{\mathrm{cm}} = 0.065 \pm 0.033$ (forward-modelled) and $0.007^{+0.032}_{-0.033}$~mag$^{-1}$ (joint); the joint value is consistent with no coupling and the forward-modelled one is $2\sigma$ from zero, so neither is a detection. The mass-concentration slope is consistent with zero in both, as expected from the upturn of the \citet{diemer2019} relation over this mass range. A gap coupling obtained in a model where the gap enters only the concentration need not survive a modest extension of that model.
\end{abstract}


\maketitle

\section{Introduction}
\label{sec:intro}

Better understanding galaxy clusters is a key element to use them as probes of physical cosmology. Similarly, studying their mass accretion histories can inform us about the astrophysical processes behind the evolution of galaxies and self-gravitating systems in general. From the start, galaxy clusters provided insights to various discoveries in astrophysics -- from the discovery of dark matter by \citet{zwicky1933} to the hints of the accelerating universe in \citet{bahcallcen1992} and even earlier by \citet{hoesselgunnthuan}. More recently, clusters have also provided tests for new physics, for example, in tests of the collisionless assumption of the dark matter model \citep{clowe2006,randall2008,Rogerio}.

In recent years, ever larger wide field surveys of galaxies have been conducted \citep{york2000,descollaboration2016,dejong2013}, and more are under way or planned \citep{euclid2011,lsst2019,roman} for the benefit of cosmology. From this wealth of data we can also build large samples of galaxy clusters and measure statistical characteristics of their population and its evolution. The masses of galaxy clusters are of particular interest, since the distribution of cluster masses as a function of redshift can be derived from cosmological models \citep{pressschechter1974,tinker2008} and are particularly sensitive to cosmological parameters \citep{allenevrardmantz2011}. This, in turn, stems from clusters being the most massive structures to have formed, being close to the exponential end of the mass function.

 Cluster masses are not, however, directly measurable. Therefore, a number of indirect techniques have been used to assess them \citep{pratt2019}. In dynamical methods, such as in \citet{zwicky1933,biviano2006}, the velocity dispersion of satellite galaxies is connected to the mass through the virial theorem or the Jeans equation \citep{lokasmamon2003}. The caustic technique -- using the escape velocity envelope in redshift space -- extends these dynamical estimates beyond the virial radius without assuming equilibrium \citep{diaferiogeller1997}. X-ray observations yield masses by applying an hydrostatic equilibrium assumption to the temperature and density profiles of the intracluster medium \citep{sarazin1988,vikhlinin2006}. The thermal Sunyaev–Zel'dovich effect, measuring the integrated gas pressure through distortions of the CMB spectrum, provides a nearly redshift-independent mass proxy. Both strong \citep{kneibnatarajan2011} and weak lensing have used the distortion of background galaxies to constrain the foreground mass. Finally, the lensing effect can also be detected on the CMB \citep{lewischallinor2006,dodelson2004}.

Weak gravitational lensing (WL) has, in particular, been frequently used to calibrate mass measurements. Since WL is sensitive to the whole mass of the cluster, it can be used to calibrate measurements that are easier to perform but are subject to modelling biases. It has been also shown that weak lensing masses for relaxed clusters are biased by at most $5$--$10\%$ \citep{beckerkravtsov2011,oguirihamana2011}.

The use of cluster WL signals is not without difficulty. The shear signal expected around a cluster is of the order of $1$--$10\%$, below the intrinsic shape scatter of galaxies $\sigma_e \sim 0.25$--$0.3$. Because of this, individual masses measured by WL have an intrinsic $20$--$30\%$ scatter \citep{hoekstra2003,beckerkravtsov2011}. Projection effects further complicate the problem, as foreground halos will cause additional shear \citep{hoekstra2001,wu2022}. Finally, even the modelling of the mass distribution of individual halos is not unambiguous. The triaxial shape of the mass distribution, formed by sequences of mergers of smaller halos \citep{jingsuto2002}, and the existence of substructures are two examples that complicate this modelling.

To address the low SNR of WL by galaxy clusters, a strategy frequently employed is stacking \citep[e.g.][]{johnston2007,sheldon2009,simet2017,murata2018,mcclintock2019}. In stacking, galaxies at the background from different clusters are lumped together and used as if they were the background to the same cluster. This is then fitted to a parametric profile, from which an average mass is calculated. To be consistent, stacks are usually divided by redshift and richness. The ensemble profile mass, however, does not correspond to a strict average of the masses of the clusters in the stack, as the relationship between the cluster mass and the WL profile is non-linear.

In this work, we propose to replace the stacking of clusters with a joint fit, constrained by known relations such as that between mass and richness, and mass and concentration. This is not without precedent: \citet{mantz2010} develop multivariate scaling relations that have then been applied to the Weighing the Giants project \citep{vonderlinden2014,applegate2014}. \citet{bocquet2019} jointly fits 343 SZ-selected clusters using weak lensing and X-ray data for different subsets. And in \citet{bocquet2024,bocquet2024b}, they develop a per-cluster hierarchical population likelihood combining abundance and individual-cluster lensing, without richness-binned stacking, and use it for a cosmological analysis. \citet{chiu2022} calibrate the X-ray observable-to-mass relations of eFEDS clusters and groups with a similar population model fit to the weak-lensing profiles of individual systems. Finally, in the CoMaLit project \citep{sereno2015}, they develop a Bayesian hierarchical scaling-relation model to assess individual masses. To test and validate our work, we performed a simulated analysis of joint fits and applied it to a DES-like and an LSST-like scenario, comparing the performance with a stacked analysis. We then test the method on real data, using the shape catalogue from DES Y3 and the WaZP cluster catalogue from DES Y6 \citep{benoist2025}, both of which are publicly available.

To further motivate this work, we will explore how the scatter of the mass-concentration relation can be explained by additional observables. This has been suggested as early as \citet{wechsler2002} from simulations of the hierarchical mass accretion. We have first tried this in \citet{vitorelli2018}, by dividing stacks into percentiles of the magnitude gap -- the difference in magnitude between the brightest cluster galaxy (BCG) and the brightest satellite in the inner region. The intuition here comes from fossil groups: it is proposed that systems with extreme ($>2$) magnitude gaps correspond to relaxed systems, that did not have any significant mergers in a long time \citep{jones2003}. This is because as time passes, the most massive satellites will suffer gravitational drag and eventually fall and merge into the central BCG. The link is statistical rather than one-to-one: large gaps are transient, and many systems that have a large gap at one time lose it later \citep{Dariush15,deason2013}.

Conditioning the sample on magnitude gaps has a second, related motivation. In the standard picture, a halo's galaxy content depends only on its mass and redshift; \emph{assembly bias} is the name for departures from this expectation, where formation history leaves an additional imprint at fixed mass \citep{wechsler2002}. Correlations between the richness and secondary cluster characteristics that are connected to its assembly history could then be evidence of assembly bias (Section~\ref{sec:dmr}). In particular, \citet{hearin2013} found in mock catalogues and SDSS groups that, at fixed richness, systems with a larger magnitude gap are more massive, so that the gap carries information on mass beyond the richness. 

Section~\ref{sec:Model} describes the model, and Section~\ref{sec:Sims} the simulations used to test it. Section~\ref{sec:SimResults} validates the joint fit on those simulations and compares it against stacking on the same data, and Section~\ref{sec:DataTest} applies both methods to WaZP clusters with DES~Y3 shears. Section~\ref{sec:discussion} discusses the results and Section~\ref{sec:conclusion} concludes. The appendices detail the triaxial halo simulation (Appendix~\ref{ap:triaxial}) and the radial binning choices (Appendix~\ref{app:binning}), test the magnitude-gap dependence of the boost factor (Appendix~\ref{app:boostdm}), describe a paired test of per-cluster covariances (Appendix~\ref{app:recommend}), justify the choice of covariance for the real data (Appendix~\ref{app:analytic}), compare the joint and forward-modelled fits with the two-step analysis of the same real and simulated stacks (Appendix~\ref{app:twostep}), and compare marginal and maximum a posteriori summaries of the real-data fits (Appendix~\ref{app:map}). Throughout, we use the \citet{planck2018} cosmology as implemented in \textsc{astropy} ($h=0.6766$, $\Omega_{\rm m}=0.310$), with $\sigma_8=0.802$ for the two-halo term, and masses in $M_\odot$ without factors of $h$.

\section{Model}\label{sec:Model}

At the core of this work is the development of a Bayesian framework that takes us from data to a constrained model of the population of galaxy clusters. One of the many advantages of this approach is that a model can be fully specified, and then simulated, to test if we can constrain the relations we want. So, in this section we will describe our model in detail, and in the next, test it. 

The main target is to investigate the use of joint fits of parametric profiles to cluster WL data, conditioned on scaling relations that have been previously studied both in simulations and in real data. From these relations, and using direct observables for each galaxy cluster, one can propose individual parametric curves to fit each cluster, and probe the space of parameters of cluster scaling relations \citep{sereno2015,umetsu2020}.

Basically, the method works as follows: the sampler proposes a set of values for the parameters of the scaling relations. Using individual cluster observables that relate to the mass profile, we then have a mass profile for each cluster. The sum of the individual log-likelihoods is then the total likelihood for the parameters. In the following subsections, we look more closely at each of the relations being used, and the joint fitting method. 

\subsection{Halo Mass profile}
\label{sec:massprofile}
Following the spherical collapse model development \citep{gunngott}, the theory of formation of clusters was studied through simulations. With dark-matter-only simulations, \citet{nfw1996,nfw,moore1999} have shown early on that halos form with similar distributions. In \citet{nfw}, specifically, they parametrise this empirical distribution as

\begin{equation}
  \rho(r) = \frac{\rho_0}{(r/r_s)\left(1+r/r_s\right)^2}
\end{equation}

with density falling roughly as $r^{-1}$ in the inner ($\sim 100 ~{\rm kpc}$) region and faster, as $r^{-3}$, in the outer regions ($>1~{\rm Mpc}$). The scale radius $r_s$ defines where the profile behaves as $\rho \sim r^{-2}$, and defines a \emph{concentration} for halos, $c_\Delta := r_\Delta/r_s$. Here $r_\Delta$ is a cut-off radius to define the mass of the halo, as models for the mass distribution frequently have divergent mass. One frequent choice for this radius is the one that encloses a region in which the density is $200$ times larger than the background, $r_{200}$, which we will use, giving $c_{200}=r_{200}/r_s$.  Throughout this work the overdensity is defined with respect to the \emph{mean} matter density $\bar\rho_m(z)$ rather than the critical density, so that $M_{200}\equiv M_{200\rm m}$.

\subsection{Large-Scale Structure Contribution}
\label{sec:lss}

Outer radii depart from the single halo shape due to the clustering of clusters. To account for that, we can use a two-halo mass contribution described as in \citet{johnston2007}, given as

\begin{equation}
 \rho_{2h} = b(\nu) \overbrace{\Omega_{\rm m} \rho_{c,0} (1+z)^3}^{\bar \rho_m (z)} \xi^L(r,z)\,,
\end{equation}where $b(\nu)$ is the linear halo bias calculated at the density peak height of the halo $\nu = \delta_c/\sigma(M)$, where $\sigma(M)$ is the linear matter variance in the Lagrangian scale of the halo, that is, with $R=\left[3M/(4\pi \bar{\rho}_m)\right]^{1/3}$. We use \citet{tinker2010} as the parametric $b(\nu)$. Finally, the linear matter correlation function at redshift $z$ is

\begin{equation}
  \xi^L(r,z) = D(z)^2 \ \sigma_8^2 \ \xi^L_n\left[(1+z)r\right]\,,
\end{equation} where $\xi^L_n(r)$ is the linear correlation function at redshift zero, taken from a {\sc camb}-calculated \citep{CAMB} linear power spectrum, normalised to $\sigma_8=1$, and $D(z)$ is the growth function. 

The projected two-halo term is then
\begin{equation}
\begin{split}
  \Sigma_{2h}(R) &= B(z,M)\,\Sigma_l(R)\,,\\
  B(z,M) &:= b(z,M)\,\Omega_{\rm m}\,\sigma_8^2\,D(z)^2\,,
\end{split}
\end{equation}
with
\begin{equation}
\begin{split}
  \Sigma_l(R) &= (1+z)^2\,\rho_{c,0}\,W\!\left[(1+z)R\right]\,,\\
  W(R) &:= \int_{-\infty}^{\infty} {\rm d}y\; \xi^L_n\!\left(\sqrt{y^2 + R^2}\right)\,,
\end{split}
\end{equation}
where $R$ is the physical projected radius and $(1+z)R$ the comoving one. The sum of a one-halo NFW profile and a linear-bias two-halo term is an approximation in the transition region around the virial radius \citep{hayashi2008}.

\subsection{Effect of Miscentering}
\label{sec:misc}

In optical identification and selection of galaxy clusters, we usually identify a bright red galaxy with the cluster centre (the brightest cluster galaxy, BCG). However, astrophysical processes and algorithmic difficulties may result in an offset between the galaxy labelled as BCG and the true centre of the gravitational potential. In the DES redMaPPer cluster catalogue, \citet{zhang2019} found that around a quarter have significant offsets between the selected central galaxy and the centre of the potential well, most of them ($\sim 60 \%$) caused by the cluster finder and the rest by astrophysics. With a significant departure in the measured lensing profile, we need to account for that when using cluster WL data.

Miscentering cannot, however, be individually constrained, so we model it as an ensemble parameter. For a given offset radius $r_{\rm off}$ between the true centre and the assumed one, the azimuthal average surface mass at radius $r$ is 

\begin{equation}
\begin{split}
\Sigma_{\mathrm{off}} (r | r_{\mathrm{off}}) &= \frac{1}{2 \pi} \int_0^{2 \pi} \mathrm{d} \theta \\
&\quad\times \Sigma \left(\sqrt{r^2 + r_{\mathrm{off}}^2 - 2 r r_{\mathrm{off}} \cos\theta}\right)\,.
\end{split}
\end{equation} 

As we do not know the individual $r_{\rm off}$ for each cluster, we marginalise over its distribution, and estimate the parametrisation of this distribution from the data. We use a Rayleigh distribution as the model for the scattering in offset radii, given by

\begin{equation}
 P\left(r_{\mathrm{off}}\right) = \frac{r_{\mathrm{off}}}{\sigma_{\mathrm{off}}^2} \exp\left [-\frac{1}{2}\left(\frac{r_{\mathrm{off}}}{\sigma_{\mathrm{off}}} \right)^2 \right]\,,
\end{equation}

where $\sigma_{\rm off}$ is the scale of the miscentering offset. \citet{zhang2019} model the offsets with a Gamma distribution in units of the cluster radius; the Rayleigh distribution in physical units is our choice.

In addition, we use the fraction of miscentered clusters, $f_{\rm mis}$, as an ensemble parameter to be determined from the data. The model of each cluster is then the miscentering-averaged profile of Eq.~\ref{eq1} (Section~\ref{sec:fullmodel}), rather than a mixture of likelihoods over the cluster's unknown offset. For a single cluster this is an approximation, adequate when its profile is dominated by shape noise. It affects only the joint fit, since the stacked model averages the members' profiles in any case.

\subsection{Weak Lensing}
\label{sec:wl}

The lensing effect on background galaxies is related to the foreground surface mass density contrast $\Delta \Sigma (R) = \bar{\Sigma}_{r<R} - \Sigma(R)$. For a given pair of lens $i$ and background galaxy $j$, the tangential shear is

\begin{equation}
    \gamma_t = \frac{\Delta \Sigma}{\Sigma_{\rm crit}}\,,
\end{equation}

where

\begin{equation}
    \Sigma_{\rm crit} = \frac{c^2}{4\pi G} \frac{D_S}{D_L D_{LS}}\,
\end{equation}

and $D_S$, $D_L$, and $D_{LS}$  are the angular diameter distances to the source, to the lens, and between lens and source, respectively. The model for the surface mass density $\Sigma(R)$ for an NFW-like mass distribution can be found in \citet{wrightbrainerd2000}. The quantity measured is the reduced shear $g_t=\gamma_t/(1-\kappa)$, with $\kappa=\Sigma/\Sigma_{\rm crit}$. Our model and simulations use $\gamma_t$, an approximation whose effect on the real data we quantify in Section~\ref{sec:robust}. Section~\ref{sec:datavector} describes how the observed data vector $\widehat{\Delta\Sigma}(r)$ and its covariance are constructed from a background galaxy catalogue.

\subsection{Mass-Richness relation}

One of the most immediate mass-observable relations in clusters is the mass-richness relation. As the galaxy density is a tracer of mass density, we expect more massive clusters to have more galaxies. Indeed, the mass-richness relation of clusters has been investigated in several works, both from theoretical and observational perspectives \citep{berlindweinberg2002,kravtsov2004}.

In principle, a halo occupation distribution (HOD) can be used as a model for the connection between halo mass and its population of galaxies \citep{cooraysheth2002}. However, the true number of member galaxies is not observed. Instead, given a particular cluster finder algorithm, we have a richness estimator that attaches membership probabilities to some galaxies around each halo \citep{rykoff2014,bellagamba2018}. Thus, the mass-richness relation is also a property of the survey, not only of the physical properties of the system. Nevertheless, given a single survey, the richness can be calibrated as a mass proxy by WL \citep{mcclintock2019} consistently.


A general model of the mass-richness relation can be written as

\begin{equation}
\label{eq:mass-richness}
\begin{split}
    \log_{10} M_{200}(\lambda,z) ={}& \log_{10} M_0 + \alpha_{M\lambda}\log_{10}(\lambda/\lambda_{\rm piv})\\
    &+ \beta_{M\lambda}\log_{10}\bigl((1+z)/(1+z_{\rm piv})\bigr)\,,
\end{split}
\end{equation}

where we have dependencies with the richness $\lambda$, and redshift $z$, about pivots $\lambda_{\rm piv}=40$ and $z_{\rm piv}=0.35$.

\subsection{Mass-Concentration Relation}

In the hierarchical structure formation paradigm smaller halos form earlier, while larger ones form later, mainly through mergers \citep{lacey1993,wechsler2002}. After a halo forms, it keeps the mass profile of Section~\ref{sec:massprofile} \citep{bullock2001}.

In this way, a natural intuition for a mass-concentration relation appears as a clock for halo formation time: as the density of the universe decreases with time, the $r_{200}$ grows larger just by its definition. For a halo that stops accreting, $r_s$ stays nearly constant \citep{bullock2001,wechsler2002}, which in turn leads to concentration increasing with $r_\Delta$. This process is called \emph{pseudo-evolution} in \citet{diemer2013}. With that, halo concentrations have been linked to mass, redshift, and cosmology.

The universality of halo shape as a function of mass, when mass and concentration are related, has been shown to be more complicated, however. At fixed mass and redshift, data has shown that concentrations have significant scatter \citep{neto2007}, largely attributable to the diversity of mass accretion histories among halos of similar mass \citep{wechsler2002}. The original \citet{bullock2001}-type relation was also later shown to mispredict the evolution of concentration for massive, still-assembling halos. More recent calibrations correct for this trend while preserving the qualitative picture that concentration encodes formation time \citep{diemer2019}.

We will begin by considering a simple, power-law model for the concentration as a function of mass and redshift:

\begin{equation}
\begin{split}
\log_{10} c (M,z) ={}& \log_{10} c_{0}
+ \alpha_{\rm cm}\log_{10}\!\left(\frac{M}{M_{\rm piv}}\right)\\
&+ \beta_{\rm cm}\log_{10}\!\left(\frac{1+z}{1+z_{\rm piv}}\right)\,,
\end{split}
\end{equation}

where again we separate dependencies with mass $\alpha_{\rm cm}$ and redshift $\beta_{\rm cm}$.

To go further, as we want to investigate the relationship if the scatter in this relation can be explained by direct observables, we model the normalisation of the relation as a function of an observable as

\begin{equation}
\begin{split}
\log_{10} c (M,z,\Delta m) ={}& \log_{10} c_{0}
+ \alpha_{\rm cm}\log_{10}\!\left(\frac{M}{M_{\rm piv}}\right)\\
&+ \beta_{\rm cm}\log_{10}\!\left(\frac{1+z}{1+z_{\rm piv}}\right) + \gamma_{\rm cm}\,\Delta m\,,
\end{split}
\label{eq:c-m}
\end{equation}

where, following our previous tests in \citet{vitorelli2018}, the observable is the magnitude gap $\Delta m$ (in magnitudes) between the brightest cluster galaxy and the brightest satellite, so that $\gamma_{\rm cm}$ is in mag$^{-1}$ and $c_0$ is the concentration at $\Delta m=0$. The term is linear in $\Delta m$ because the magnitude gap is already a logarithmic quantity. The mass and redshift pivots are $M_{\rm piv}=10^{14}\,M_\odot$ and $z_{\rm piv}=0.35$.

\subsection{The full model of cluster mass profiles}
\label{sec:fullmodel}

Putting each contribution together, the full model for the mass distribution of galaxy clusters is

\begin{equation} \label{eq1}
\begin{split}
  \Delta \Sigma(r|M_{200},&c_{200},f_{\rm mis},\sigma_{\mathrm{off}}) = \\
  &(1-f_{\rm mis})\,\Delta \Sigma_{\mathrm{NFW}}(r|M_{200},c_{200}) \\
  &+ f_{\rm mis}\, \Delta\Sigma_{\mathrm{NFW}}^{\mathrm{off}}(r|M_{200},c_{200},\sigma_{\mathrm{off}}) \\
  &+ \Delta \Sigma_{2h}(r | M_{200})
 \end{split}
\end{equation}

where $f_{\rm mis}$ is the fraction of miscentered clusters. For each single cluster, the mass is estimated from the mass-richness relation and the concentration subsequently by the mass-concentration relation. In turn, $f_{\rm mis}$ and $\sigma_{\rm off}$ are nuisance parameters. Given the spread of the mass-richness relation, we also fit its normalisation $M_{0}$ (Eq.~\ref{eq:mass-richness}) rather than fixing it to a calibration. 

The model signal $\Delta\Sigma$ is evaluated at the logarithmic centre of each radial bin, whereas the measurement is an area-weighted average over each annulus. This approximation is much cheaper in the joint fit, which evaluates every radius of every cluster; Appendix~\ref{app:binning} quantifies it and shows that it affects mainly the mass normalisation.

\subsection{Data Vector}
\label{sec:datavector}

The joint and stacked fits use the same estimator of $\Delta\Sigma$, applied either to a single cluster or to a stack. Each cluster $c$ contributes the background galaxies $j$ selected behind it (Section~\ref{sec:background}), each with tangential ellipticity $e_{t,j}$, measurement weight $w_j$, tangential shear response $R_j$ (Section~\ref{sec:shearcal}), and critical surface density $\Sigma_{{\rm crit},j}$ of the lens-source pair. Radial bins are logarithmically spaced, with bin centres $R_k$ and edges at the geometric mean between centres. For each cluster and radial bin $k$ we compute
\begin{equation}
\begin{split}
  T_{c,k} &= \sum_{j\in c,\,k} \frac{e_{t,j}\,w_j}{\Sigma_{{\rm crit},j}}\,,
  \qquad
  K_{c,k} = \sum_{j\in c,\,k} \frac{R_j\,w_j}{\Sigma_{{\rm crit},j}^2}\,,\\
  n_{c,k} &= \#\{j\in c,\,k\}\,,
\end{split}
  \label{eq:stacksums}
\end{equation}
so that $K_{c,k}$ is the response-weighted lensing weight of the cluster in that bin. The data vector of a single cluster, used by the joint fit, and that of a stack $s$ of clusters, used by the stacked fit, are
\begin{equation}
\begin{split}
    \widehat{\Delta \Sigma}_{c,k} &= B_{c,k}\,\frac{T_{c,k}}{K_{c,k}}\,,\\
    \widehat{\Delta\Sigma}_{s,k} &= B_{s,k}\,\frac{\sum_{c\in s} T_{c,k}}{\sum_{c\in s} K_{c,k}}\,,
\end{split}
    \label{eq:deltasigma}
\end{equation}
where the response $R_j$ removes the multiplicative shear bias, and the boost factor $B$ corrects the dilution of the signal by cluster members and foreground galaxies that pass the background selection. For a stack, the boost factor is measured directly from the excess of source counts,
\begin{equation}
  B_{s,k} = \frac{\sum_{c\in s} n_{c,k}}{\bar n_s\,A_k}\,,
  \label{eq:boostfactor}
\end{equation}
which compares the galaxy count in bin $k$ with that expected at the stack's mean source density, $\bar n_s = \sum_{c,k} n_{c,k}/A_{\rm tot}$, over the bin's area $A_k$ (in practice evaluated with a random catalogue at that density). A single cluster has too few sources for this measurement, so $B_{c,k}$ is a smooth fit to the boost factors of the stack the cluster belongs to (Section~\ref{sec:boost}).

\subsection{Covariances}
\label{sec:covariance}

The stacked and joint fit data vectors differ in their covariance estimation. For a single cluster, part of the covariance comes from the shape noise of its own background galaxies, and we test two ways of estimating it. The first is an analytic shape-noise covariance of the estimator,
\begin{equation}
\mathrm{Var}(\widehat{\Delta\Sigma}_{c,k}) =
B_{c,k}^2\,\frac{\sigma_e^2\sum_{j\in c,\,k} \left(w_j/\Sigma_{{\rm crit},j}\right)^2}{K_{c,k}^2}\,,
\label{eq:covariance}
\end{equation}
where the per-component shape dispersion $\sigma_e$ is the weighted root-mean-square of the source ellipticities over the whole selected source sample, in the same (response-uncorrected) units as $T_{c,k}$. It is diagonal, since every galaxy contributes to a single radial bin and the shape noise of different galaxies is independent, and it uses no individual galaxy shape.

The second is a bootstrap over the cluster's own background galaxies: the sources are resampled with replacement, $\widehat{\Delta\Sigma}_{c,k}$ is recomputed for each resample, and the data vector and its covariance are the mean and covariance over the resamples. The boost factor, fixed in advance, scales the covariance by $B_{c,k}B_{c,k'}$. Within a radius $r_{\rm floor}$, where bins hold few sources and the bootstrap variance can be spuriously small, the variance is floored at the value of Eq.~\ref{eq:covariance}.

In these single-cluster covariances, bins with too few sources are masked. For the analytic covariance, bins with fewer than three sources are excluded from that cluster's likelihood (Appendix~\ref{app:binning}); for the bootstrap covariance, a bin is used when its estimate is finite in at least $90\%$ of the resamples and its variance is positive, which also keeps some bins with two sources.

For a stack, we bootstrap-resample its member clusters $N_{\rm boot}=500$ times: each resample recomputes $\widehat{\Delta\Sigma}_{s,k}$ -- numerator, denominator, and boost factor together -- from the resampled clusters' sums, so the covariance $\mathbf{C}_s$ includes the uncertainty in the boost factor and its correlation with the signal, as well as the cluster population variance within the stack. The data vector is the mean over resamples. Unlike the per-cluster bootstrap, this covariance is estimated from the cluster population and thousands of galaxies per bin, so it barely responds to the fluctuation of the stacked profile itself.

\subsection{Likelihoods}
\label{sec:jointfit}

The joint model assigns masses and concentrations to each cluster through the overall mass-richness and mass-concentration relations, given a choice of their parameters and a richness and observable value. For each cluster $i$ the residual is $\mathbf{r}_i(\theta) = \mathbf{d}_i - \mathbf{m}_i(\theta)$, where $\mathbf{d}_i$ is the $\Delta\Sigma$ data vector and $\mathbf{m}_i(\theta)$ is the corresponding model prediction at parameter vector $\theta = (\log_{10} M_0, \alpha_{M\lambda}, \beta_{M\lambda}, c_0, \alpha_{\rm cm}, \beta_{\rm cm}, \gamma_{\rm cm}, f_{\rm mis}, \sigma_{\rm off})$ -- the three mass-richness parameters of Eq.~\ref{eq:mass-richness}, the four mass-concentration parameters of Eq.~\ref{eq:c-m}, and the two ensemble miscentering nuisance parameters, nine in total. The concentration normalisation $c_0$ is sampled linearly, not in $\log_{10}c_0$. For a cluster with the analytic covariance $\mathbf{C}_i$, the log-likelihood is the Gaussian
\begin{equation}
  \ell_i(\theta) = -\frac{1}{2}\Bigl[
    k\ln(2\pi)
    + \ln|\mathbf{C}_i|
    + \mathbf{r}_i^\top \mathbf{C}_i^{-1} \mathbf{r}_i
  \Bigr],
  \label{eq:loglike}
\end{equation}
with $k$ radial bins per cluster.

When the covariance is estimated by the galaxy bootstrap, the Gaussian is no longer the appropriate likelihood. The bootstrap variance $s_{i,k}^2$ of a bin is itself estimated from the $n_{i,k}$ sources in it, which can be as few as two, so we replace the Gaussian of each bin by the likelihood of \citet{sellentin2016}, which marginalises over the true variance given the estimated one. Since every galaxy contributes to a single radial bin, the bootstrap covariance is also diagonal up to sampling noise (Section~\ref{sec:covariance}), and this is applied bin by bin,
\begin{equation}
  \ell_i(\theta) = -\sum_{k}\frac{n_{i,k}}{2}\,\ln\!\left[1+\frac{r_{i,k}^2}{(n_{i,k}-1)\,s_{i,k}^2}\right],
  \label{eq:shlike}
\end{equation}
a Student-$t$ that tends to the Gaussian for many sources and down-weights large residuals in sparse bins; bins whose variance was floored at the analytic value are not estimates and keep the Gaussian form.

The joint log-likelihood is the sum over the $N$ clusters of the sample,
\begin{equation}
  \ell(\theta) = \sum_{i=1}^{N} \ell_i(\theta)\,,
  \label{eq:jointlike}
\end{equation}
with $\ell_i$ given by Eq.~\ref{eq:loglike} or Eq.~\ref{eq:shlike} according to the covariance.

\subsection{Likelihood Bootstrapping}
\label{sec:likeboot}

The covariance matrix for single clusters as described above does not take into account all the covariance, because it is only calculated from the shape noise of the background for each cluster separately. Neither the analytic nor the bootstrap in a single cluster includes the shear of uncorrelated large-scale structure \citep{hoekstra2003} or the intrinsic variation of profiles at fixed mass \citep{gruen2015}, which a bootstrap over the member clusters of a stack does include.

To understand this better, we can write the total variance for $\hat\gamma_{\rm cm}$ as

\begin{equation}
  \mathrm{Var}[\hat\gamma_{\rm cm}]
  \;\approx\;
  \underbrace{\sigma_{\rm shape}^2}_{\text{shape noise}}
  \;+\;
  \underbrace{\sigma_{\rm sample}^2}_{\text{cluster sample variance}}\,.
  \label{eq:varboot}
\end{equation}

To estimate this total variance, we have adopted a simple strategy: to run the inference multiple times, on different random draws from the cluster sample.

Given the dataset $\mathcal{D} = \{(\mathbf{d}_i, \mathbf{C}_i, z_i, \lambda_i, \Delta m_i)\}_{i=1}^{N}$, we draw $N$ indices with replacement:

\begin{equation}
  \mathbf{I}^{(b)} = (I_1^{(b)}, \ldots, I_N^{(b)}),
  \qquad
  I_j^{(b)} \overset{\mathrm{iid}}{\sim} \mathrm{Uniform}\{1,\ldots,N\}\,.
\end{equation}

Each bootstrap log-likelihood for draw $b$ is
\begin{equation}
  \ell^{(b)}(\theta)
  = \sum_{j=1}^{N} \ell_{I_j^{(b)}}(\theta)
  = \sum_{i=1}^{N} m_i^{(b)}\,\ell_i(\theta)\,,
\end{equation}

where $m_i^{(b)} = \#\{j : I_j^{(b)} = i\}$ is the bootstrap multiplicity of cluster $i$ in draw $b$.  The vector $(m_1^{(b)},\ldots,m_N^{(b)})$ is drawn from a $\mathrm{Multinomial}(N;\,1/N,\ldots,1/N)$ distribution. The expected value is $\mathbb{E}[m_i^{(b)}]=1$ and the variance is $\mathrm{Var}[m_i^{(b)}] = 1 - 1/N \approx 1$, so on average each cluster enters once, with Poisson-like fluctuations. 

We fit each of $N_b=50$ resamples and summarise the joint fit by the median and the 16th and 84th percentiles of the $N_b$ per-bootstrap posterior medians. Because resampling the clusters also resamples their background galaxies, the scatter of these medians measures both terms of Eq.~\ref{eq:varboot} together. Pooling the per-bootstrap posteriors instead, as a mixture, would add the shape noise a second time, since each posterior is itself as wide as the shape-noise term; such a bagged posterior is wider than the sampling distribution when the model is correct \citep{huggins2019}.

\subsection{Forward-Modelled Stacking}
\label{sec:stackmodel}

Most stacked WL calibrations of cluster scaling relations follow a two-step procedure: a mean profile is fit to each stack, and the scaling relation is then fit to the per-stack parameter posteriors, with the expected mean mass of a stack modelled as a weighted average over its members \citep[e.g.][]{johnston2007,melchior2017,mcclintock2019,bellagamba2019,lesci2022}, possibly including intrinsic scatter \citep{sereno2020}. \citet{mcclintock2019}, in particular, calibrate the difference between the mean mass and the mass inferred from a stacked profile with simulated stacks built cluster by cluster. 

An alternative to this is fitting the scaling relations directly to the stacked profiles, by either taking into account the population distribution or a member-by-member model. In the population approach, the model profile of a stack is an integral of the halo profile over the mass function and the distribution of richness at fixed mass, usually together with the cluster abundance \citep{murata2018,murata2019}. \citet{payerne2025} compare this one-step approach with the two-step one on simulations. 

In the member-by-member approach of \citet{simet2017}, closest to ours, the model of each stack is a weighted sum of the model profiles of its members, each with a mass drawn from the mass-richness relation at its own richness and a concentration from the mass-concentration relation. Our model here has two differences: each member's model is weighted, in every radial bin, exactly as its galaxies are in the stacked estimator, rather than with one weight per cluster; and every member's secondary observable, here the magnitude gap, enters its own model, so that the stacked fit constrains the same parameters as the joint fit. Appendix~\ref{app:twostep} compares our forward-modelled fit and the joint fit with the two-step analysis of the same real and simulated stacks.

The sample is binned into stacks by redshift, richness, and magnitude gap, and each stack's profile and covariance are measured as in Section~\ref{sec:datavector}. Individual cluster profiles $\Delta\Sigma_c(R_k\,|\,\theta)$ are the same as in the joint fit: each mass from its own richness and redshift (Eq.~\ref{eq:mass-richness}), its concentration from that mass, its redshift, and its magnitude gap (Eq.~\ref{eq:c-m}), the NFW profile with the miscentering mixture of Eq.~\ref{eq1} (the same $f_{\rm mis}$ and $\sigma_{\rm off}$ for all clusters). The stacked model is the average of its members' profiles, weighted as the estimator weights their galaxies in each bin:
\begin{equation}
  \Delta\Sigma_{s,k}^{\rm model}(\theta) = \frac{\sum_{c\in s} K_{c,k}\,\Delta\Sigma_c(R_k\,|\,\theta)}{\sum_{c\in s} K_{c,k}}\,.
  \label{eq:stackmodel}
\end{equation}
This is the expectation of Eq.~\ref{eq:deltasigma} when each galaxy's tangential ellipticity estimates $R_j\,\Delta\Sigma_c/\Sigma_{{\rm crit},j}$ and the boost factor removes the dilution exactly. The weights $K_{c,k}$ include every galaxy that passes the background selection, contaminants included, which is exact when the contamination fraction is the same for all members of a stack. Every member's richness, redshift, and magnitude gap therefore enters the model, not only the stack's mean.

\paragraph{Likelihood.} With $\mathbf{r}_s(\theta)$ the residual between $\widehat{\Delta\Sigma}_s$ and Eq.~\ref{eq:stackmodel} over the stack's valid bins, the log-likelihood is
\begin{equation}
\begin{split}
  \ell(\theta) &= -\frac{1}{2}\sum_{s} \mathbf{r}_s^\top\,\widehat{\mathbf{C}_s^{-1}}\,\mathbf{r}_s\,,\\
  \widehat{\mathbf{C}_s^{-1}} &= \frac{N_s-p_s-2}{N_s-1}\left(\mathbf{C}_s + \epsilon\,\mathbf{I}\right)^{-1},
\end{split}
  \label{eq:stacklike}
\end{equation}
where $p_s$ is the number of valid bins and the \citet{hartlap2007} factor debiases the inverse of the estimated covariance, with $N_s$ the number of member clusters of the stack, the number of independent realisations behind the cluster bootstrap; $\epsilon=10^{-6}$ times the mean diagonal of $\mathbf{C}_s$ is a numerical regulariser. The likelihood of \citet{sellentin2016}, which marginalises over the estimated covariance, is an alternative with the same $N_s$; Section~\ref{sec:robust} tests it. The covariances do not depend on $\theta$, so the normalisation is constant and omitted. The parameters $\theta$ are exactly those of the joint fit, so that, under the same priors and sampler, the two methods differ only in whether the data are stacked before the likelihood is evaluated.

A bootstrap over $N$ member clusters gives a covariance of rank at most $N-1$, so a stack with fewer members than radial bins has a singular covariance, and a stack only somewhat larger has a poorly determined one whose inverse can dominate Eq.~\ref{eq:stacklike}. We therefore use only stacks with at least $50$ members, and recommend a threshold of at least a few times the number of radial bins in general.

\section{Simulations}
\label{sec:Sims}

We test the methods on simulated data: we generate plausible populations of galaxy clusters and background source catalogues, and apply the cluster shear to the sources. Our goal with the simulated test was not absolute realism and, as such, we do not run full physics simulations. What we test is if we can recover the parameters from the scaling relations, especially $\gamma_{\rm cm}$, given plausible data. 

\begin{figure*}
    \centering
    \includegraphics[width=\textwidth]{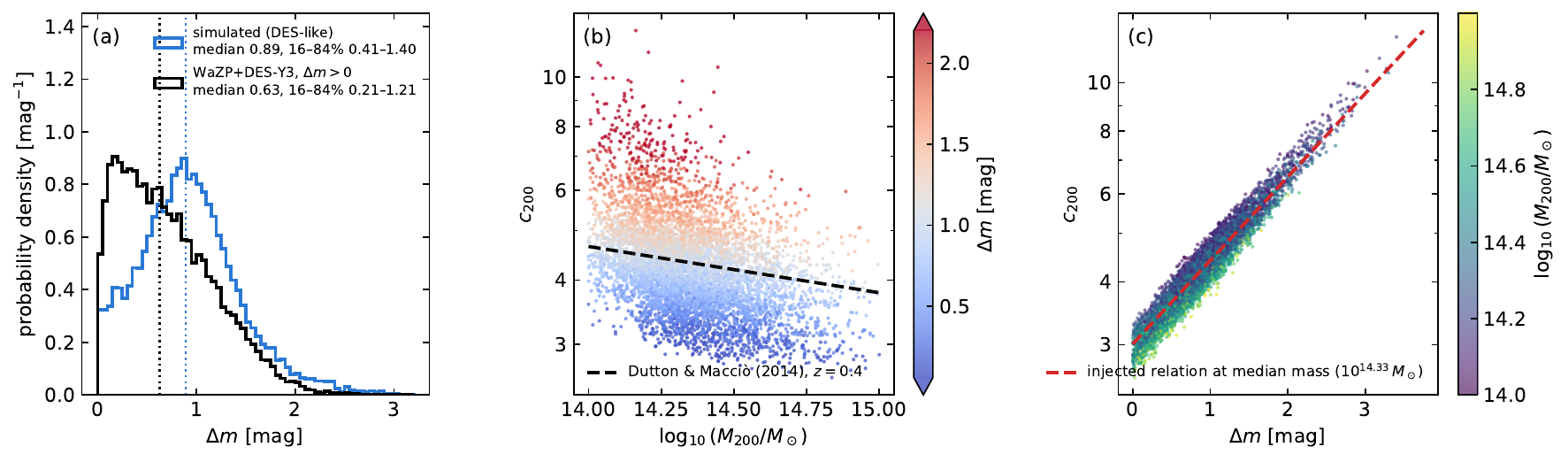}
    \caption{Example simulated DES-like population. (a) Magnitude-gap distribution of the simulated clusters (blue) against that of the real WaZP+DES-Y3 clusters with $\Delta m>0$ (black); dotted lines mark the medians. (b) $c_{200}$ against halo mass, coloured by magnitude gap (blue: small gaps, red: large gaps). (c) $c_{200}$ against $\Delta m$, coloured by halo mass. The injected truth is $c_0=3.23$, $\alpha_{\rm cm}=-0.092$, $\beta_{\rm cm}=-0.400$, $\gamma_{\rm cm}=0.167$~mag$^{-1}$. The dashed line in panel (b) is the \citet{duttonmaccio2014} baseline evaluated at $z=0.4$; the dashed line in panel (c) is the injected $c_{200}$--$\Delta m$ relation (Eq.~\ref{eq:c-m}) evaluated at the median simulated halo mass, illustrating the coupling the joint fit is designed to recover.}
    \label{fig:example_pop_sim}
\end{figure*}

The pool of $N_\mathrm{pool}$ candidate clusters is generated as follows:

\begin{enumerate}
  \item \textbf{Redshift.}  Draw $z \sim \mathrm{Uniform}[0.1,\,0.6]$. Combined with the mass draw below, which follows the shape of the mass function at each redshift but not its normalisation, this ignores both the comoving volume and the redshift evolution of the halo abundance. We do not include them because, in the absence of a selection function, which we leave for later work, they would lead to an excess of clusters at high redshift.

  \item \textbf{Halo mass.}  Draw $\log_{10}(M_{200}/M_\odot)$ from the \citet{tinker2008} halo mass function at the cluster's redshift, normalised within the range $\log_{10}(M_{200}/M_\odot) \in [14.0,\,15.0]$.

  \item \textbf{Central galaxy.} The luminosity of the central galaxy follows the central luminosity-halo mass relation of \citet{zheng2007}, with a lognormal scatter of $0.12$~dex.

  \item \textbf{Satellite galaxies.} The satellites form a Poisson process in absolute magnitude, whose mean cumulative number brighter than $M_r$ interpolates, linearly in $\ln N$, the \citet{zheng2007} halo occupation distribution (HOD) satellite numbers of the SDSS $M_r-5\log_{10}h<-18$, $-21$ and $-22$ samples at the cluster's halo mass, extrapolated brightward with the slope between the last two. The brightest and second-brightest satellites are placed at cumulative numbers $E_1$ and $E_1+E_2$, with $E_1$ and $E_2$ independent unit exponentials, as for the first two events of a Poisson process. Clusters without a satellite brighter than $M_r-5\log_{10}h=-18$ are rejected.

  \item \textbf{Magnitude gap.} As an observer would define it, the brightest galaxy of the cluster is taken as the BCG, whether it is the central or a satellite, and
        \begin{equation}
          \Delta m = m_2 - m_1 \quad[\mathrm{mag}]\,,
        \end{equation}
        where $m_1$ and $m_2$ are the magnitudes of the brightest and second-brightest galaxies, so that $\Delta m\geq0$. The resulting gaps have a median of $0.89$~mag and a $16$--$84\%$ range of $0.41$--$1.40$~mag, as wide as the real WaZP gaps ($0.21$--$1.21$~mag, Section~\ref{sec:deltam}) but about $0.25$~mag larger (Fig.~\ref{fig:example_pop_sim}a). We do not tune the model to remove this offset, which plausibly arises from what the simulation leaves out: projected interlopers, which can supply a spurious bright satellite, the difference between the $r$ band of the HOD and the $i$ band of the WaZP gaps, and the identification of the central galaxy by the cluster finder.

  \item \textbf{NFW concentration.} The concentration is modelled as
        \begin{equation}
          \log_{10} c = \log_{10} c_\mathrm{DM14}(z,M)
          + \gamma_{\rm cm}^{\rm true}\,\left(\Delta m - 0.9\,{\rm mag}\right)
          \,,
         \end{equation}
        centred near the median simulated gap, so that the mean concentration changes little with $\gamma_{\rm cm}^{\rm true}$; the fit model has no such pivot, and its $c_0$ absorbs the centring.
        The \citet{duttonmaccio2014} baseline concentration is
        \begin{equation}
          \log_{10} c_\mathrm{DM14} = a(z)
          + b(z)\,\log_{10}\!\left(\frac{M_{200}\,h}{10^{12}\,M_\odot}\right),
        \end{equation}
        with $a(z)=0.520+0.385\,\exp(-0.617\,z^{1.21})$ and $b(z)=-0.101+0.026\,z$. This form is calibrated for a critical overdensity. We evaluate it at $M_{200}$ only as a baseline for the injected truth, which does not affect the recovery test, since the injected $c_0$, $\alpha_{\rm cm}$ and $\beta_{\rm cm}$ the fit is compared against are measured from the simulated population itself.

  \item \textbf{Richness.}  Each cluster's richness is assigned by inverting the mass-richness relation of Eq.~\ref{eq:mass-richness} at its true mass and redshift, with the \citet{mcclintock2019} values $\log_{10}(M_0/M_\odot)=14.489$, $\alpha_{M\lambda}=1.356$ and $\beta_{M\lambda}=-0.30$. The injected relation therefore has no scatter and no magnitude-gap dependence. It is a relation for the mean of $\log_{10}M$, whereas \citet{mcclintock2019} calibrate the mean mass. The distinction is irrelevant for this closure test.

  \item \textbf{Selection.}  The pool is drawn in batches and filtered to the clusters that would enter the real analysis: richness $\lambda > 25$ (matching the WaZP cosmology-sample cut of Section~\ref{sec:DataTest}) and $\Delta m > 0$. Batches are generated until $N$ clusters pass, and the catalogue is truncated to exactly $N$: $15{,}967$ for the DES-like configuration, the size of the real sample before its $\Delta m>0$ cut (Section~\ref{sec:DataTest}), and $80{,}000$ for the LSST-like one (Section~\ref{sec:SimResults}). The selection is therefore matched to the real sample's cuts rather than to an assumed parametric $n(z)$.
\end{enumerate}

One important caveat of this work is that we do not simulate here the spatial distribution of clusters in the plane of the sky. The investigation of effects arising from clustering of clusters, alignments, and substructures will be relegated to future work.

\begin{figure*}
    \centering
    \includegraphics[width=0.9\textwidth]{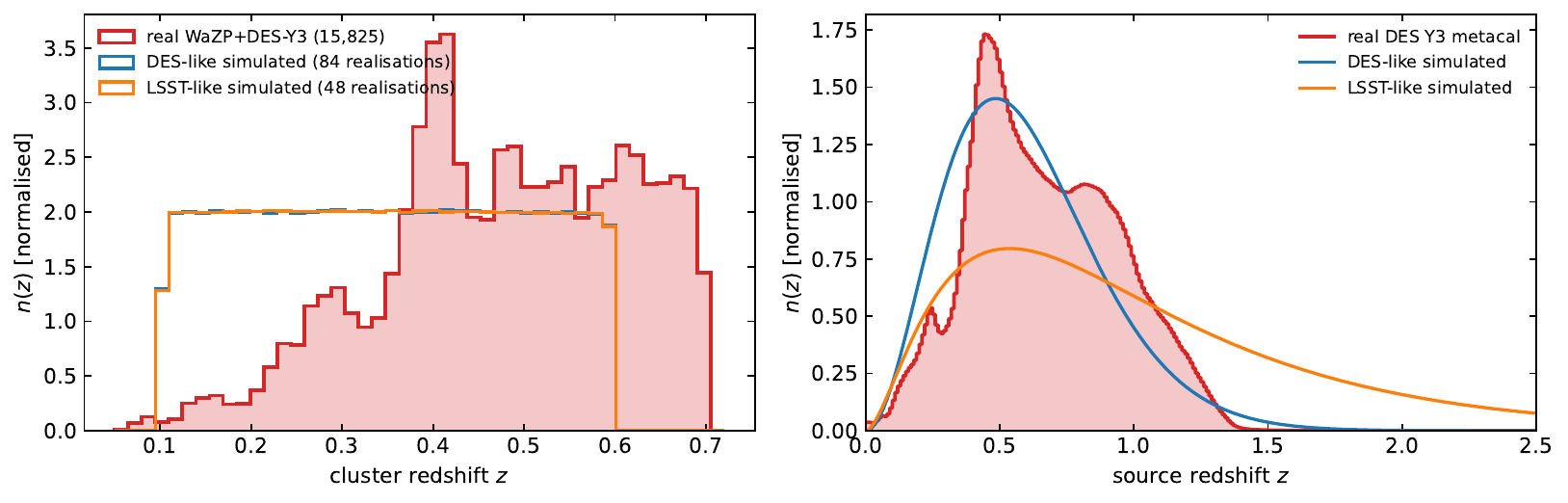}
    \caption{Redshift distributions of the simulated campaigns against the real data: clusters (left, pooling the $84$ DES-like and $48$ LSST-like realisations, against the $15{,}825$ real WaZP+DES-Y3 clusters with $\Delta m>0$; the simulated distributions are uniform in $0.1<z<0.6$ by construction and identical for the two configurations) and background sources (right, against the full DES~Y3 \textsc{metacalibration} catalogue).}
    \label{fig:des_nz_sim}
\end{figure*}

The background galaxies have intrinsic ellipticities drawn per component from a Gaussian of dispersion $0.27$, redrawn until $|e|<1$, and each receives a metacalibration-like response $R=1-(s/1.5)^2$ (clipped to $[0.05,0.99]$) from a lognormal size proxy $s$. The shear is applied with the exact composition $e_{\rm obs}=(e+Rg)/(1+Rg^*e)$, for which the mean over an isotropic distribution is exactly $Rg$ \citep{seitzschneider1997}. Sources are measured within a projected radius of $5.3$~Mpc of the (possibly miscentered) cluster centre, matching the outermost radial bin of either grid. The source counts are Poisson-distributed around a mean density of $6.7$ galaxies per square arcmin for DES-like and $30$ for LSST-like simulations. The DES-like galaxy density is measured directly from the DES~Y3 source catalogue used in Section~\ref{sec:DataTest}, over its occupied-footprint area. Source redshifts are drawn from a Smail-Schechter distribution, $P(z)\propto(z/z_0)^{\alpha}\exp[-(z/z_0)^{\beta}]$ truncated at $z=2.5$, with $(\alpha,\beta,z_0)=(2.0,1.5,0.40)$ for the DES-like configuration, chosen to approximate the real DES~Y3 source distribution, and $(2.0,0.68,0.11)$ for the deeper LSST-like one. The resulting distributions, and a comparison with real data for the DES case is shown in Fig.~\ref{fig:des_nz_sim}. The simulated profiles contain no two-halo term and the source catalogues no contaminating cluster members or foreground galaxies, so the fits to the simulations omit the two-halo term and use $B_{c,k}=1$.

\subsection{Simulated Triaxiality}
\label{sec:triaxial}

Galaxy clusters are expected to be triaxial due to the collisionless nature of Dark Matter \citep{jingsuto2002}. This is a particular problem for our modelling, as we do not include information to recover the particular orientation of each cluster. Because of this, we need to test if fitting a spherical NFW profile on the sample carries any biases. In the future this can be expanded into using the BCG as a cluster shape prior, as studied in \citet{herbonnet2022,zhang2023}.

\begin{figure*}
    \centering
    \includegraphics[width=0.75\textwidth]{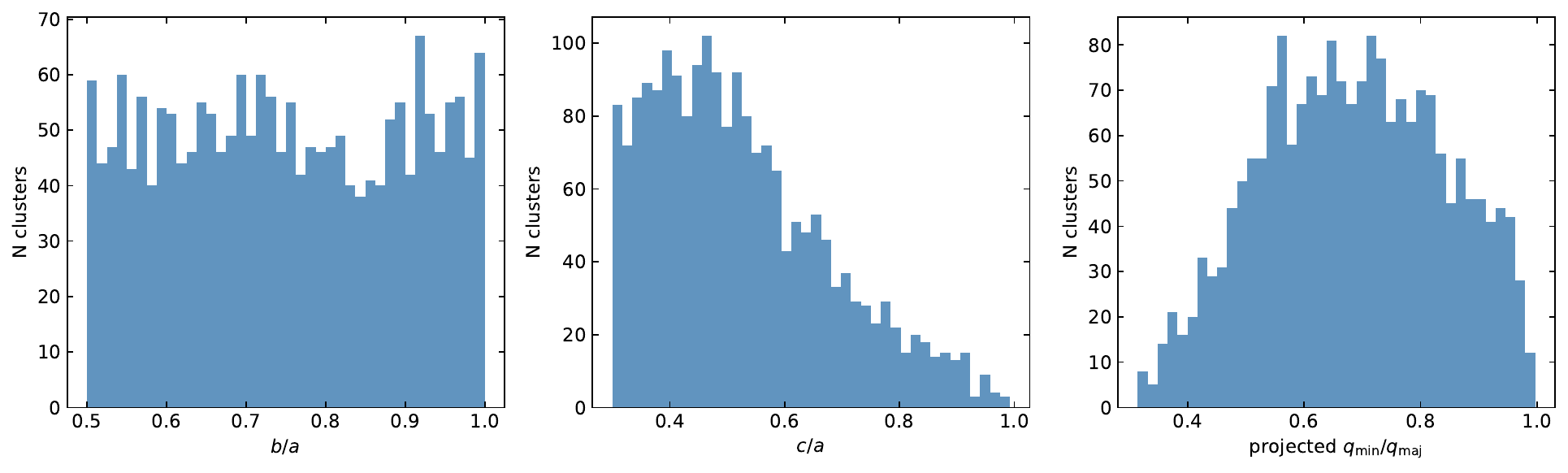}
    \caption{Distribution of the injected triaxial axis ratios $b/a$, $c/a$, and the resulting projected axis ratio $q_{\rm min}/q_{\rm maj}$ on the sky.}
    \label{fig:triaxiality_demo}
\end{figure*}

Each simulated cluster is a triaxial NFW halo, and each background galaxy receives the lensing signal of its projection: the projected surface density is elliptical, with an amplitude that depends on the orientation (larger when the halo is elongated along the line of sight), and the shear is that of this elliptical mass distribution, computed exactly \citep{schramm1990,keeton2001}. The halo mass and concentration keep their spherical meaning, and its total projected mass is the same for every orientation. The full method for projection, and its effect on the measured $\Delta\Sigma$, are given in Appendix~\ref{ap:triaxial}. Each cluster is assigned random triaxial axis ratios and a random line-of-sight (LOS) orientation, drawn independently (Fig.~\ref{fig:triaxiality_demo}):

\begin{itemize}
  \item Intermediate-to-major axis ratio: $b/a \sim \mathrm{Uniform}[0.5,\,1.0]$.
  \item Minor-to-major axis ratio: $c/a \sim \mathrm{Uniform}[0.3,\,b/a]$
        (ensures $c/a \leq b/a$).
  \item LOS polar angle: $\theta_\mathrm{los} = \arccos(U)$ with
        $U \sim \mathrm{Uniform}[0,\,1]$ (isotropic).
  \item LOS azimuthal angle: $\phi_\mathrm{los} \sim \mathrm{Uniform}[0,\,2\pi]$.
\end{itemize}

\subsection{Miscentering}
\label{sec:miscenter}

\begin{figure}
    \centering
    \includegraphics[width=0.8\linewidth]{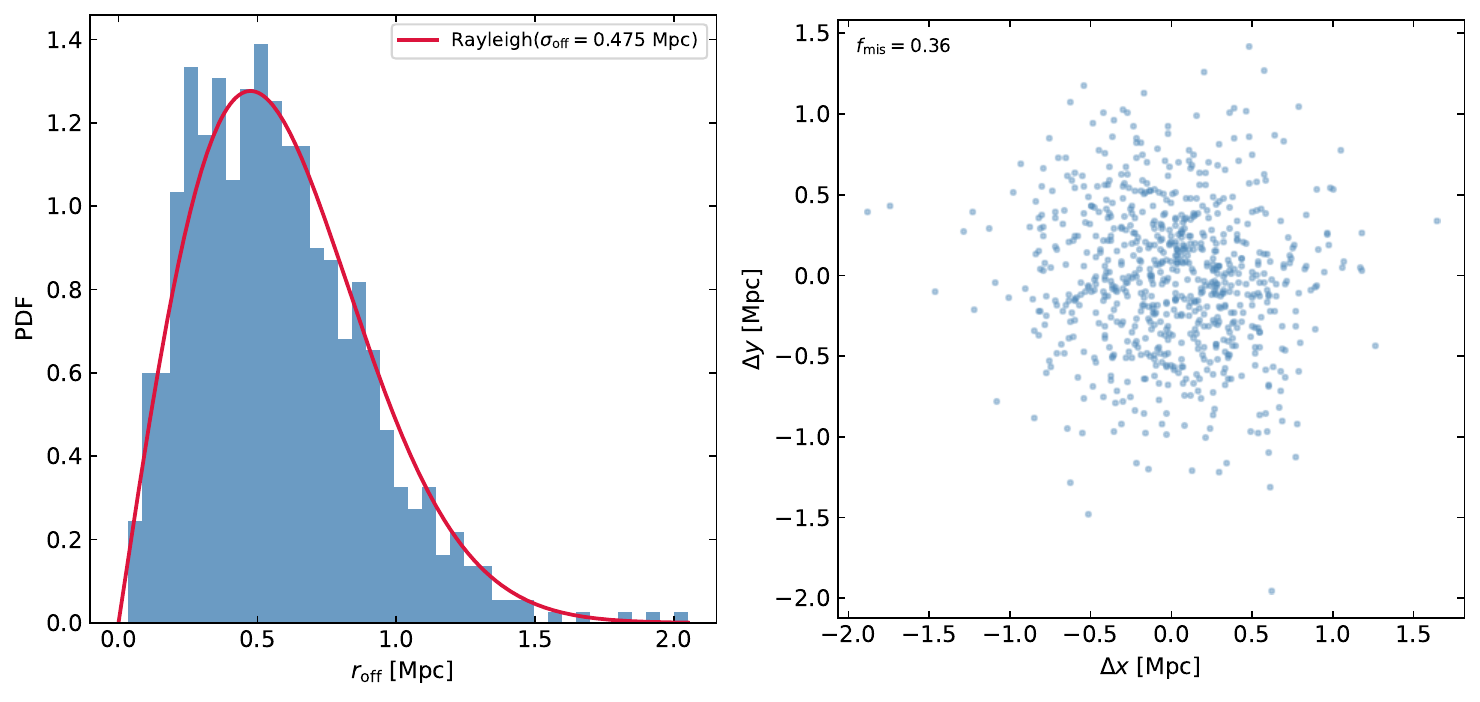}
    \caption{Simulated offset magnitude distribution (left, with the injected Rayleigh model overlaid) and offset positions on the sky (right) for the miscentered fraction of the sample.}
    \label{fig:miscentering_sim}
\end{figure}

We simulate miscentered clusters simply by shifting the centre by choosing a random offset (Fig.~\ref{fig:miscentering_sim}) with properties 

\begin{itemize}
  \item Centred fraction: $F_\mathrm{cen} = 0.64$; miscentered fraction:
        $F_\mathrm{mis} = 1 - F_\mathrm{cen} = 0.36$.
  \item Offset magnitude for miscentered clusters:
        $r_\mathrm{off} \sim \mathrm{Rayleigh}(\sigma_\mathrm{off})$ with
        $\sigma_\mathrm{off} = 0.32\,h^{-1}$~Mpc.
        Using $h = 0.6766$ gives
        $\sigma_\mathrm{off} \approx 0.473$~Mpc.
  \item Offset direction: $\phi_\mathrm{mis} \sim \mathrm{Uniform}[0,\,2\pi]$.
  \item Centred clusters: $r_\mathrm{off} = 0$.
\end{itemize}

These miscentering values and distributions need not correspond to those of the real data.

\section{Simulation Results}
\label{sec:SimResults}

We test both methods on independent realisations of each configuration, with new cluster populations, source catalogues, shapes, orientations and centring offsets in every realisation. We use seven injected values $\gamma_{\rm cm}^{\rm true} = n/36$~mag$^{-1}$, $n = 0, 1, \ldots, 6$, spanning a range that keeps the concentrations physically plausible: at $\gamma_{\rm cm}^{\rm true}=6/36$~mag$^{-1}$ a $10^{14}\,M_\odot$ halo at $z=0.3$ with the $99$th-percentile gap ($2.4$~mag) has $c_{200}\approx8.6$, against $4.8$ at the median gap. Only $\gamma_{\rm cm}$ varies between the simulated populations; the other injected scaling-relation parameters are the same in all realisations. For each value of $\gamma_{\rm cm}^{\rm true}$ we create 6 independent realisations for the LSST-like configuration (twelve at $\gamma_{\rm cm}^{\rm true}=1/36$~mag$^{-1}$) and twelve for the DES-like one, $48$ and $84$ in total. Each realisation is fit once by the joint fit and once by the forward-modelled stacked fit (Section~\ref{sec:stackmodel}), with the same priors and sampler, so the two methods see exactly the same data. We sample each posterior with \textsc{nautilus} \citep{lange2023} to an effective sample size of 3000, for which the Monte Carlo error of the posterior median and 68\% bounds is expected to be below $0.03\sigma$. The scatter between independent realisations measures the full uncertainty of each method and tests its reported posterior width.

All nine parameters are free, under weakly informative, physically bounded priors (Table~\ref{tab:joint_priors}) that place every physically plausible value within about $2\sigma$ of their centre, with a width at least several times that of the corresponding posterior.

\begin{table}[H]
\centering
\caption{Priors of every simulated fit and of the stacked fits (the real-data joint fit changes only the $c_0$ prior, Section~\ref{sec:jointresults}). $\mathcal{N}(\mu,\sigma)$ is a Gaussian, truncated at the bounds given; $\mathcal{U}(a,b)$ is uniform.}
\label{tab:joint_priors}
\begin{tabular}{ll}
\toprule
Parameter & Prior \\
\midrule
$c_0$ & $\mathcal{U}(1,15)$ \\
$\alpha_{\rm cm}$ & $\mathcal{N}(-0.1,0.5)$ \\
$\beta_{\rm cm}$ & $\mathcal{N}(-0.4,2)$ \\
$\gamma_{\rm cm}$ & $\mathcal{N}(0,1)$ \\
$\log_{10}(M_0/M_\odot)$ & $\mathcal{N}(14.489,0.5)$ \\
$\alpha_{M\lambda}$ & $\mathcal{N}(1.356,0.5)$ \\
$\beta_{M\lambda}$ & $\mathcal{N}(-0.30,2)$ \\
$f_{\rm mis}$ & $\mathcal{U}(0,1)$ \\
$\sigma_{\rm off}$ [Mpc] & $\mathcal{U}(0,1)$ \\
\bottomrule
\end{tabular}
\end{table}

The two configurations use different radial grids, each matched to what its source density can support. The LSST-like configuration measures $\widehat{\Delta\Sigma}$ in 8 logarithmically spaced bins spanning $0.16$--$3.98$~Mpc. At DES-like density that grid is not usable: its innermost bins frequently hold too few background sources for a usable measurement (Appendix~\ref{app:binning}), and which bins fail varies from cluster to cluster. In testing, this heterogeneous masking pattern destabilised the bound construction of the nested sampler. The DES-like configuration therefore matches the real-data measurement used later in Section~\ref{sec:DataTest}: 6 wider bins spanning $0.25$--$3.98$~Mpc. Both configurations apply the analytic shape-noise covariance of Eq.~\ref{eq:covariance} (Section~\ref{sec:covariance}), with $\sigma_e=0.269$ measured from the simulated sources.

The stacked fits use the same stack binning in the simulations and in the real data (Section~\ref{sec:stackedresults}). The stacks split the sample into 3 redshift bins ($0.1$--$0.3$--$0.5$--$0.7$), 4 richness bins ($25$--$35$--$55$--$85$--$\infty$), and, within each redshift-richness cell, 3 magnitude-gap terciles, so that each cell contributes three points to the magnitude-gap dependence. Each stack's profile is measured in 16 logarithmic radial bins spanning $0.16$--$3.98$~Mpc. Stacks with at least 50 members are used.

\begin{figure*}
    \centering
    \includegraphics[width=0.9\textwidth]{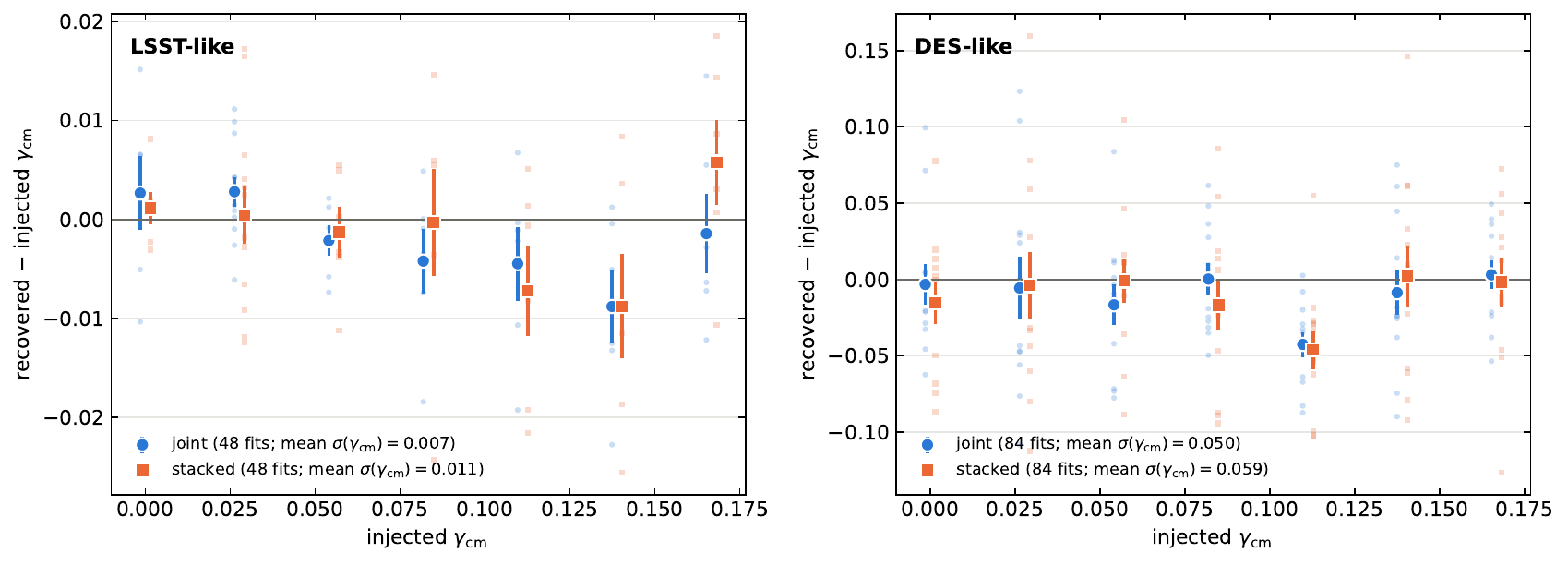}
    \caption{Recovered minus injected $\gamma_{\rm cm}$ for independent realisations of the LSST-like (left) and DES-like (right) simulations, for the joint fit (circles) and the forward-modelled stacked fit (squares). Small points are individual realisations; large markers are the mean and its standard error at each injected value (offset horizontally for visibility). The legends give the number of fits and the mean posterior width of each method.}
    \label{fig:gamma_scan_DES}
\end{figure*}

Both methods recover the injected coupling at both densities (Figure~\ref{fig:gamma_scan_DES} and Table~\ref{tab:gamma_recovery}). The table gives the slope and offset of a straight-line fit of recovered against injected values over all realisations, the injected value of each realisation being the slope measured from its own population (Section~\ref{sec:Sims}). Every offset is consistent with zero, and every slope but one with unity.

\begin{table*}
\centering
\caption{Recovery of $\gamma_{\rm cm}$ in the simulations: slope and offset of the straight-line fit of recovered against injected values over all realisations, standard deviation of $\gamma_{\rm cm}$ between realisations (the measured precision), its ratio to the mean posterior width, and the mean posterior 68\% half-width $\sigma(\gamma_{\rm cm})$ (the reported uncertainty).}
\label{tab:gamma_recovery}
\begin{tabular}{llccccc}
\toprule
Configuration & Fit & Slope & Offset [mag$^{-1}$] & Scatter [mag$^{-1}$] & Scatter/width & $\sigma(\gamma_{\rm cm})$ [mag$^{-1}$] \\
\midrule
LSST-like (48) & joint & $0.947\pm0.020$ & $+0.002\pm0.002$ & $0.0082$ & $1.11$ & $0.0074$ \\
 & stacked & $0.984\pm0.028$ & $+0.000\pm0.003$ & $0.0105$ & $0.98$ & $0.0107$ \\
DES-like (84) & joint & $0.983\pm0.093$ & $-0.009\pm0.009$ & $0.0468$ & $0.94$ & $0.0498$ \\
 & stacked & $1.011\pm0.114$ & $-0.013\pm0.011$ & $0.0578$ & $0.98$ & $0.0592$ \\
\bottomrule
\end{tabular}
\end{table*}

The exception is the LSST-like joint fit, whose slope is $2.6\sigma$ below unity. Further investigation revealed that miscentering causes most of this deficit. Realisations at $\gamma_{\rm cm}^{\rm true}\in\{2/36,\,3/36,\,5/36,\,6/36\}$~mag$^{-1}$ with the same random seeds but without miscentering show no deficit: the mean offsets are then between $-0.004$ and $+0.001$ at every injected value. Switching the miscentering back on lowers the recovered $\gamma_{\rm cm}$ by $(3.0\pm0.7)\%$ of the true value on average, with a spread between injected values somewhat larger than the errors ($\chi^2=8.3$ for $3$ degrees of freedom). We have also tested if triaxiality had an impact, but repeating this test with triaxiality off changes the recovered $\gamma_{\rm cm}$ by only $-0.0004\pm0.0003$.

At DES-like density the recovery is unbiased within its larger uncertainties. The mean offset is within two standard errors of zero at every injected value except for one value, where both methods are low. That point lies $3.2$ (joint) and $2.8$ (stacked) standard errors below zero. Since both methods fit the same twelve realisations there, it is a fluctuation of those data sets rather than of either method.

Both methods report uncertainties that match their scatter between realisations, in both the DES and LSST scenarios: the ratio of the scatter of $\gamma_{\rm cm}$ between realisations to the mean posterior width lies between $0.94$ and $1.11$ as can be seen in Table \ref{tab:gamma_recovery}.

\begin{figure}
    \centering
    \includegraphics[width=1.0\linewidth]{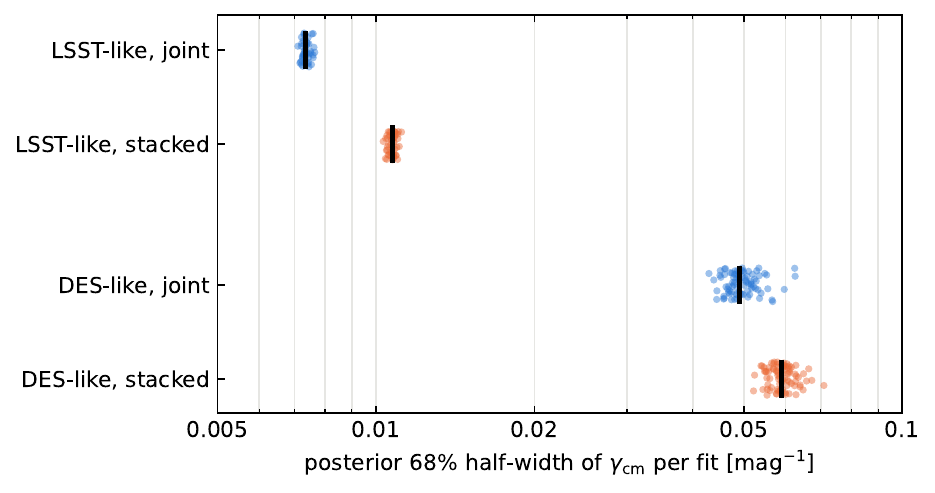}
    \caption{Posterior 68\% half-width of $\gamma_{\rm cm}$ of every simulated fit, the uncertainty each method reports (Table~\ref{tab:gamma_recovery} gives the measured scatter between realisations), for the joint (blue) and forward-modelled stacked (orange) fits at LSST-like and DES-like density; vertical bars mark the medians. The median widths are $6.7$ (joint) and $5.5$ (stacked) times smaller at LSST-like density.}
    \label{fig:gamma_width}
\end{figure}

For $\gamma_{\rm cm}$, the joint fit is the most precise method: its scatter between realisations is $1.28$ (LSST-like) and $1.23$ (DES-like) times smaller than that of the forward-modelled stacked fit, and $6.5$ and $2.0$ times smaller than that of the two-step analysis (Tables~\ref{tab:gamma_recovery} and~\ref{tab:width_ratios}; Figure~\ref{fig:gamma_width} shows the corresponding posterior widths).

A few additional caveats are: i) the richness seen by the fits is the noiseless inverse of the mass-richness relation, so the model's assumption of no scatter at fixed richness holds exactly (the selection at $\lambda>25$ uses a Poisson-scattered richness, but the fits receive the noiseless value); ii) the magnitude gap is known exactly, whereas real gaps carry measurement uncertainties (Section~\ref{sec:DataTest}), which attenuate a fitted slope \citep{kelly2007}; and iii) halo orientations are isotropic, without the orientation selection a cluster finder may introduce \citep{dietrich2014}. Each of these will be addressed in forthcoming work (Section~\ref{sec:conclusion}).

\subsection{Joint vs.\ Stacked Constraining Power}
\label{sec:jointvsstacked}

\begin{figure*}
    \centering
    \includegraphics[width=0.95\linewidth]{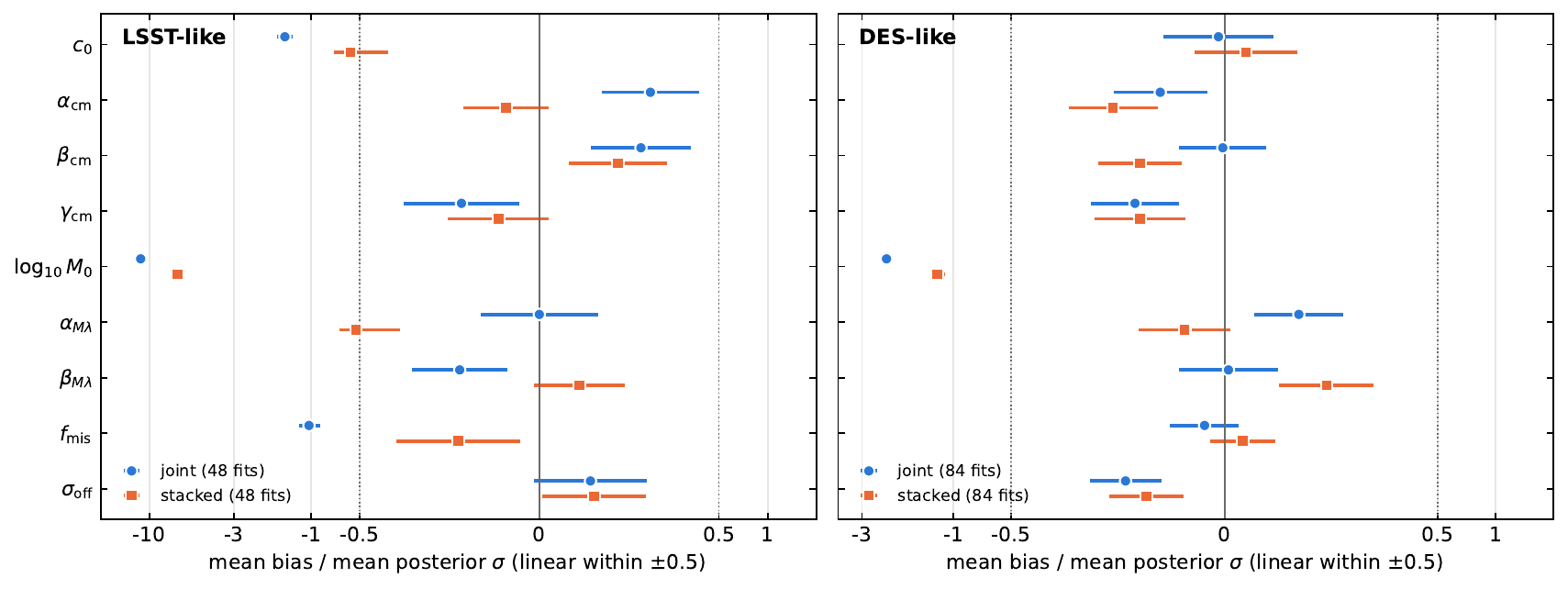}
    \caption{Mean bias of every parameter, in units of the mean posterior width (linear axis within $\pm0.5$, logarithmic beyond), over all independent realisations and injected $\gamma_{\rm cm}$ of the LSST-like (left) and DES-like (right) simulations, for the joint fit (circles) and the forward-modelled stacked fit (squares); error bars are standard errors of the mean.}
    \label{fig:sim_jvs_bias}
\end{figure*}

We now compare the two methods on all nine parameters, in both configurations. Table~\ref{tab:sim_bias} and Figure~\ref{fig:sim_jvs_bias} give the mean bias of every parameter over all realisations and injected values.

\begin{table*}
\centering
\caption{Mean bias (fit minus truth) $\pm$ its standard error over all realisations and injected values, and, in parentheses, the ratio of the scatter between realisations to the mean posterior width. LSST-like: 48 fits per method; DES-like: 84.}
\label{tab:sim_bias}
\begin{tabular}{lrrrr}
\toprule
 & \multicolumn{2}{c}{LSST-like} & \multicolumn{2}{c}{DES-like} \\
Parameter & Joint & Stacked & Joint & Stacked \\
\midrule
$\gamma_{\rm cm}$ [mag$^{-1}$] & $-0.002\pm0.001$ (1.11) & $-0.001\pm0.002$ (0.98) & $-0.011\pm0.005$ (0.94) & $-0.012\pm0.006$ (0.98) \\
$c_0$ & $-0.29\pm0.03$ (1.11) & $-0.16\pm0.04$ (1.04) & $-0.02\pm0.17$ (1.18) & $+0.08\pm0.19$ (1.10) \\
$\alpha_{\rm cm}$ & $+0.006\pm0.003$ (0.93) & $-0.003\pm0.004$ (0.83) & $-0.019\pm0.014$ (1.00) & $-0.043\pm0.017$ (0.96) \\
$\beta_{\rm cm}$ & $+0.026\pm0.013$ (0.96) & $+0.026\pm0.016$ (0.95) & $-0.00\pm0.07$ (0.94) & $-0.15\pm0.07$ (0.90) \\
$\log_{10}M_0$ & $-0.044\pm0.001$ (0.95) & $-0.029\pm0.001$ (1.02) & $-0.053\pm0.003$ (0.99) & $-0.033\pm0.003$ (0.93) \\
$\alpha_{M\lambda}$ & $-0.000\pm0.003$ (1.13) & $-0.013\pm0.004$ (0.98) & $+0.016\pm0.009$ (0.95) & $-0.013\pm0.014$ (0.99) \\
$\beta_{M\lambda}$ & $-0.011\pm0.007$ (0.92) & $+0.007\pm0.008$ (0.87) & $+0.00\pm0.04$ (1.06) & $+0.10\pm0.05$ (1.02) \\
$f_{\rm mis}$ & $-0.027\pm0.004$ (1.09) & $-0.007\pm0.005$ (1.20) & $-0.007\pm0.012$ (0.74) & $+0.007\pm0.013$ (0.70) \\
$\sigma_{\rm off}$ [Mpc] & $+0.002\pm0.003$ (1.09) & $+0.003\pm0.002$ (1.00) & $-0.036\pm0.013$ (0.77) & $-0.029\pm0.014$ (0.79) \\
\bottomrule
\end{tabular}
\end{table*}

Both methods recover the mass-richness slopes and the mass-concentration slopes, and their reported uncertainties track the realisation-to-realisation scatter on every parameter, with ratios between $0.7$ and $1.2$. The joint fit is the most precise on every scaling-relation slope: beyond $\gamma_{\rm cm}$, the scatter of the forward-modelled stacked estimates between realisations is $1.3$--$1.4$ times that of the joint ones at LSST-like density and $1.1$--$1.5$ times at DES-like density (Table~\ref{tab:width_ratios}). The standard two-step analysis of the same stacks (Appendix~\ref{app:twostep}) is far less precise, its scatter being $1.7$--$5.3$ times that of the joint fit at LSST-like density and $1.5$--$1.8$ times at DES-like density on these slopes, and it is biased in several of them.

The one large and consistent bias is in the mass normalisation, which both methods recover low (Table~\ref{tab:sim_bias}; Section~\ref{sec:simoffsets}).

At LSST-like density the joint fit also places $c_0$ slightly low ($1.4\sigma$ per fit, Table~\ref{tab:sim_bias}), with a correlated small deficit in $f_{\rm mis}$. The stacked fit shows the same direction at half the size. The main quantity we targeted, $\gamma_{\rm cm}$, remains unaffected.

\section{Data Tests}
\label{sec:DataTest}

We use the DES Y3 \citep{descollaboration2016} \textsc{metacalibration} shape catalogue, whose shape measurements and calibrations are detailed in \citet{Gatti2020}, to apply our methods to real data. In our selection, we keep sources with \texttt{flags}$=0$, $10<S/N<1000$, size ratio above $0.5$, $T>0$, and a positive DNF photometric redshift \citep{devicente2016}, which leaves $1.27\times10^{8}$ sources over an occupied footprint of $\approx5200$ square degrees.

\begin{figure}
    \centering
    \includegraphics[width=1.0\linewidth]{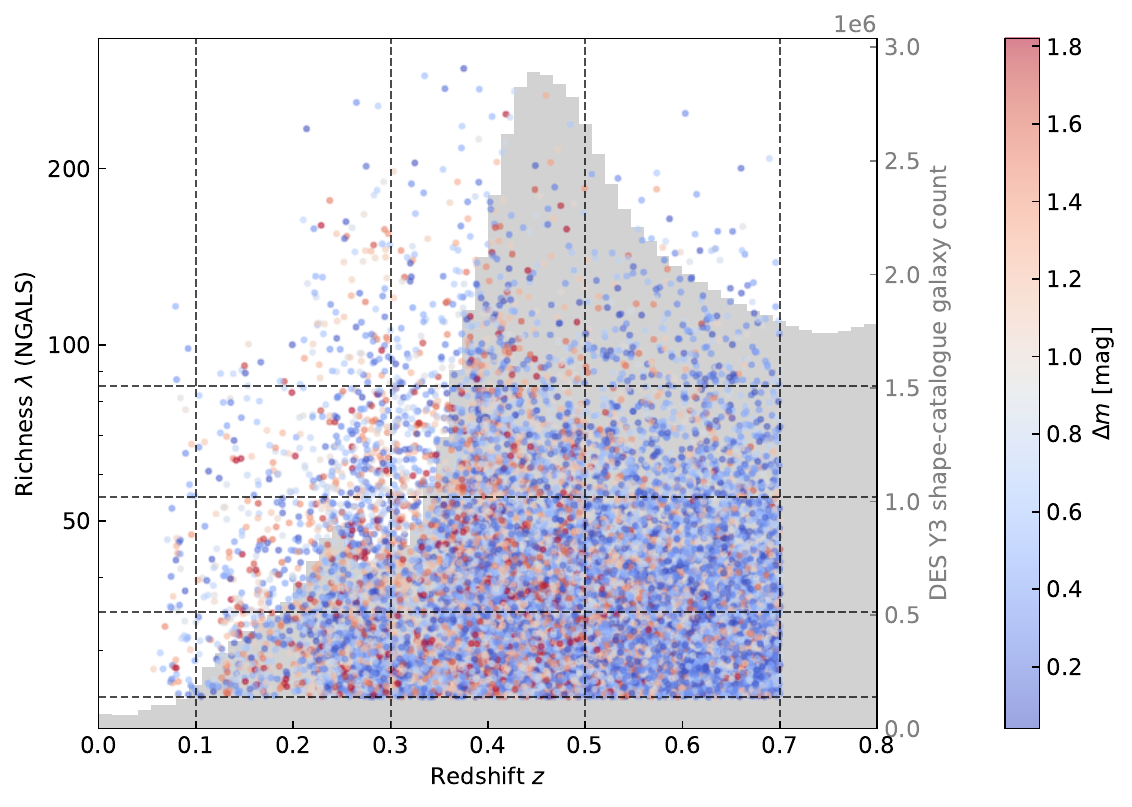}
    \caption{The WaZP cluster sample distribution across redshifts and richnesses ($N_{\rm gals}$, the WaZP richness; left Y axis). The colours blue to red indicate increasing magnitude gap. The grey histogram is the galaxy count in the DES Y3 shape catalogue as a function of redshift (right Y axis). The dashed black lines indicate the split in stacks for the stacked analysis.}
    \label{fig:population_zr}
\end{figure}

Our clusters are drawn from the DES~Y6 WaZP catalogue (version 5.0.12; \citealt{aguena2021,benoist2025}). From its more than $400{,}000$ cluster candidates we keep those with richness above $25$, the threshold of the WaZP cosmology sample of \citet{aguena2021}, and a central galaxy flagged by WaZP. No other catalogue flag is used. WaZP reports richness as $N_{\rm gals}$, the effective number of member galaxies; this is the quantity we use as $\lambda$ in the mass-richness relation of Eq.~\ref{eq:mass-richness} throughout, with $\lambda_{\rm piv}=40$. We also remove candidates at $z>0.7$, where spurious detections are more likely and the lensing signal is weak, so that they would add more noise than signal. This leaves $15{,}967$ clusters, of which $15{,}825$ have a positive magnitude gap (Figure~\ref{fig:population_dm}). The joint fit uses the $15{,}783$ of these with a usable lensing measurement, and the stacked fit the $15{,}628$ in stacks of at least 50 members. Figure~\ref{fig:population_zr} shows their distribution across redshift and richness, together with the galaxy count of the shape catalogue as a function of redshift.

\subsection{Magnitude Gaps}
\label{sec:deltam}

We have calculated the expected value for the magnitude gaps for each cluster following the same method in \citet{vitorelli2018}. The BCG is the member that WaZP flags as the central galaxy, and the satellites are the other members within a projected radius of $0.5$~Mpc of the WaZP centre. Ordering the satellite galaxies by increasing magnitude with membership probabilities $p_1, p_2, p_3, \ldots$, the probability of the $n$-th satellite being the brightest satellite galaxy is the probability that it is a member while every brighter satellite is not:

\begin{equation}
 P_n = p_n \prod_{i=1}^{n-1} (1-p_i)\,.
\end{equation}

\begin{figure}
    \centering
    \includegraphics[width=1.0\linewidth]{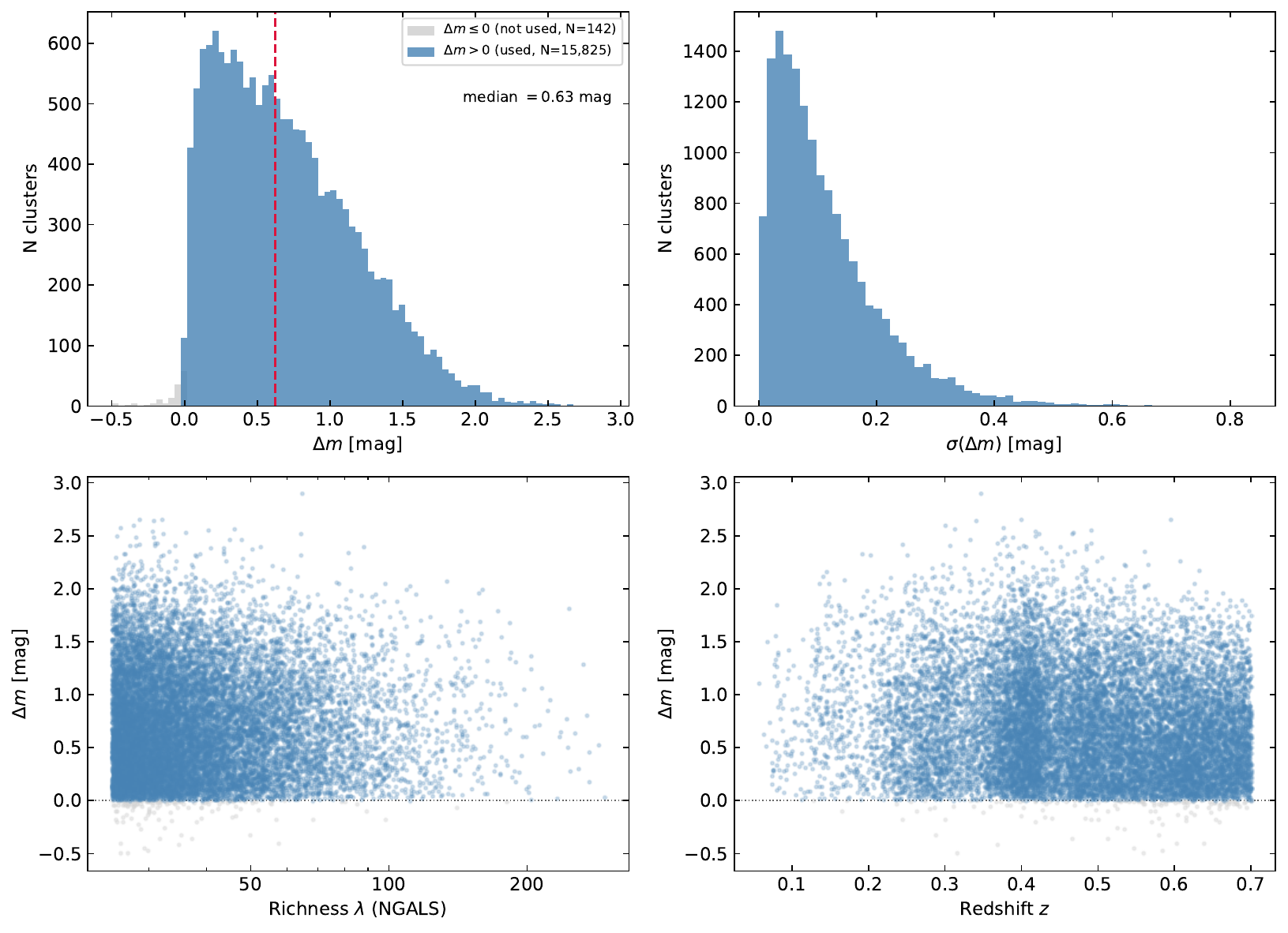}
    \caption{WaZP magnitude gap distribution (top left, median $+0.63$~mag over the $\Delta m>0$ sample used in the fit), per-cluster uncertainty (top right), and its dependence on richness and redshift (bottom row). The 142 clusters with $\Delta m\leq0$ (light grey; excluded from every fit in this work) are shown for reference only: their preferential scatter toward low richness and high redshift, visible in the bottom row, is the expected signature of noisier BCG/brightest-satellite identification in exactly the regime where it should be hardest.}
    \label{fig:population_dm}
\end{figure}

The expected magnitude gap and its uncertainty are then

\begin{equation}
\begin{split}
E\left[\Delta m \right] &= \sum_{n=1}^N P_n \left(m_n - m_{\rm BCG}\right)\,,\\
\sigma^2_{\Delta m} &= \sum_{n=1}^N P_n \left(m_n - m_{\rm BCG}\right)^2 - E\left[\Delta m\right]^2\,,
\end{split}
\end{equation}

where $m_n$ and $m_{\rm BCG}$ are the magnitudes of the $n$-th satellite and of the brightest cluster galaxy. The resulting distribution of magnitude gaps across the entire sample, and their per-cluster uncertainties, are shown in Fig.~\ref{fig:population_dm}.

\subsection{Shear Calibration}
\label{sec:shearcal}

Averaging the shapes of a galaxy sample does not directly yield the gravitational shear. Instead, for small shears, average shapes are related to the true shear linearly, i.e.:

\begin{equation}
\langle g_{\mathrm{obs}} \rangle = (1+m)\gamma + c \,,
\end{equation}

where $m$ and $c$ are respectively called multiplicative and additive biases. 

In \textsc{metacalibration}, biases are determined by testing how shear acts directly on real images \citep{metacal1}. From the \emph{shear response matrix} $R_{ij} = \partial e_i / \partial \gamma_j$ we can build an unbiased estimator for the true shear as

\begin{equation}
    \gamma = \langle R_{ij} \rangle^{-1} \langle e_{\mathrm{obs}} \rangle \,.
\end{equation}

To convert the observed shapes and responses into tangential ellipticities around halos, we apply rotations as

\begin{equation}
e_{\mathrm T} = -e_1 \cos(2\phi) - e_2 \sin(2\phi)\,,
\end{equation}

where $\phi$ is the azimuth\footnote{I.e.\ the angle between the great circle that connects the cluster centre and the background galaxy position in the sky, and the meridian that contains the cluster centre.}.

Correspondingly, we can transform the metacalibration response matrix to a multiplicative term (equivalent to $1+m$), as

\begin{equation}
\begin{split}
R_{\mathrm T} ={}& R_{11} \cos^2\left(2\phi\right) + R_{22}\sin^2\left(2\phi\right)\\
&+ \left( R_{12} + R_{21} \right) \sin\left(2\phi\right)\cos\left(2\phi\right)\,.
\end{split}
\end{equation}

The tangential ellipticity $e_{\mathrm T}$ and response $R_{\mathrm T}$ of each source are the $e_{t,j}$ and $R_j$ of Eq.~\ref{eq:stacksums}.

Two calibration terms that we do not apply act on every cluster alike and approximately cancel. We do not apply a \textsc{metacalibration} selection response for our source cuts. A selection response requires every selection to be repeated on the artificially sheared images, but our background selection uses DNF photometric redshifts from the unsheared photometry (Section~\ref{sec:background}), for which no sheared counterpart exists. For the DES~Y1 catalogue \citet{mcclintock2019} found $\langle R_{\rm sel}\rangle\approx0.013$--$0.015$ against a shear response $\langle R_\gamma\rangle\approx0.55$--$0.60$, so neglecting it raises $\Delta\Sigma$ by about $2.5\%$. The residual multiplicative bias of the Y3 catalogue, $m$ between $-0.006$ and $-0.037$ across its redshift bins and caused mainly by blending \citep{maccrann2022}, lowers it by a similar amount, in the joint and the stacked fit alike. With $\Delta\Sigma\propto M^{0.6}$ near the scale radius, each corresponds to about $0.02$~dex in the mass normalisation, comparable to its statistical uncertainty, but in opposite directions; neither changes the slopes or $\gamma_{\rm cm}$, since both are nearly independent of redshift and do not depend on the magnitude gap, so they do not affect our conclusions.

\subsection{Background selection}
\label{sec:background}

Shear profiles are measured around the WaZP cluster centres, not BCG positions, from the sources within a projected radius that encloses the outermost radial bin ($5.3$~Mpc for the per-cluster profiles). Background sources are selected behind each cluster by a redshift margin, keeping sources with $z_s > f\,z_l + \delta$. We use $\delta=0.2$ throughout. Source redshifts are the DNF point estimates, used both for this cut and for $\Sigma_{\rm crit}$. Cluster members and foreground galaxies that pass the cut dilute the signal and are corrected by the boost factor (Section~\ref{sec:boost}). The joint fit of Section~\ref{sec:jointresults}, and the boost-factor fits it uses, take $f=1.1$, which keeps enough sources in the bins of individual clusters. The stacked analysis of Section~\ref{sec:stackedresults} takes a stricter $f=1.2$, which the pooled sources of a stack can afford.

\subsection{Boost correction}
\label{sec:boost}

\begin{figure}
    \centering
    \includegraphics[width=1.0\linewidth]{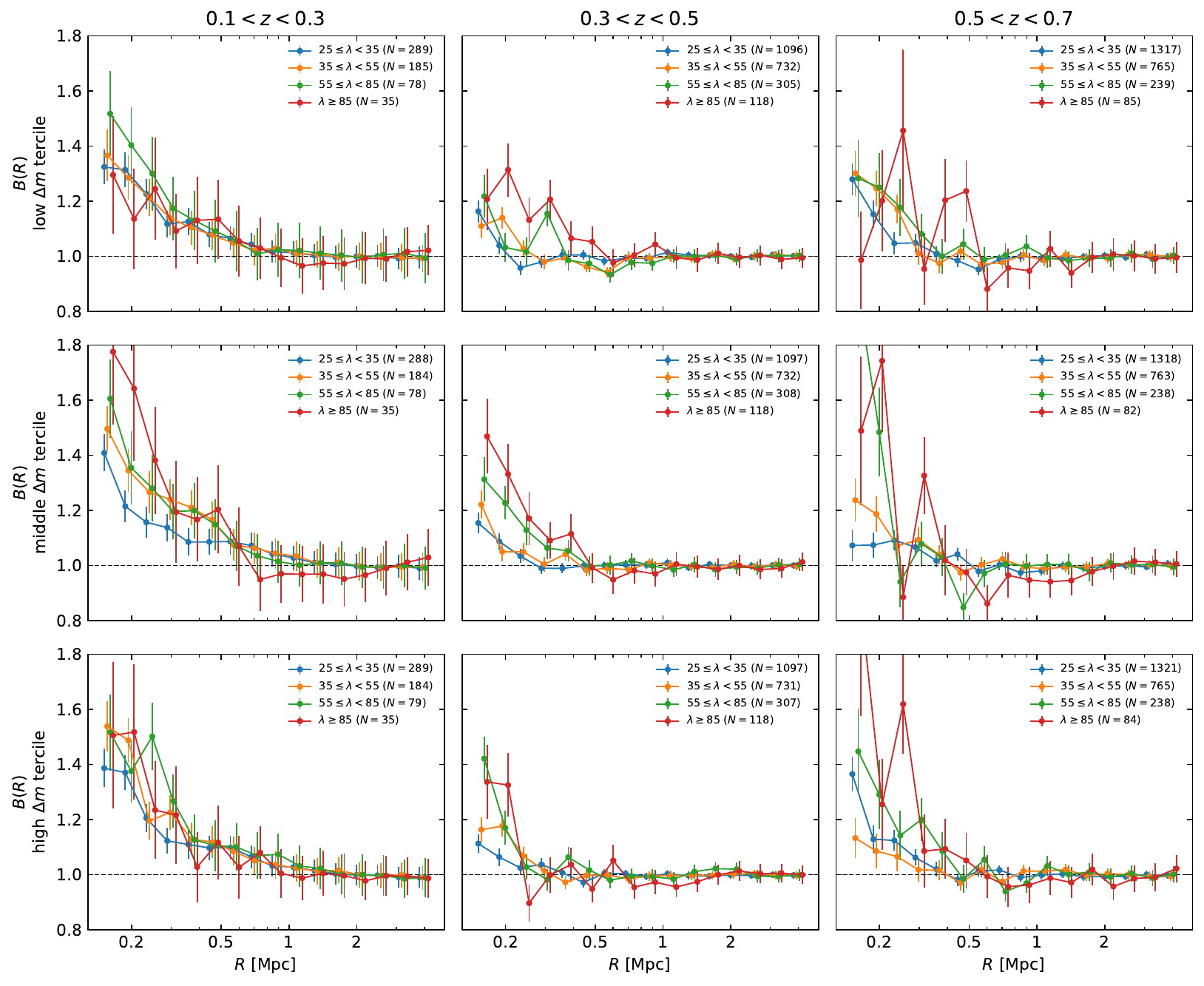}
    \caption{Boost factors $B(R)$ measured for each stack of the stacked analysis ($f=1.2$ background selection), by redshift (columns), magnitude-gap tercile (rows), and richness (colour; $N$ is the number of member clusters), with bootstrap error bars. All stacks approach $B(R)\to1$ at large $R$, as required. The three lowest-redshift, highest-richness stacks ($N=35$) are shown but excluded from the fit by the $N\geq50$ threshold.}
    \label{fig:boost_factors_f1}
\end{figure}

Figure~\ref{fig:boost_factors_f1} shows the boost factors (Eq.~\ref{eq:boostfactor}) measured for each stack of the stacked analysis. All approach unity beyond $\sim1$~Mpc, where contamination is negligible, with the largest corrections at small radius and low redshift, where member contamination is most severe.

\begin{figure}
    \centering
    \includegraphics[width=1.0\linewidth]{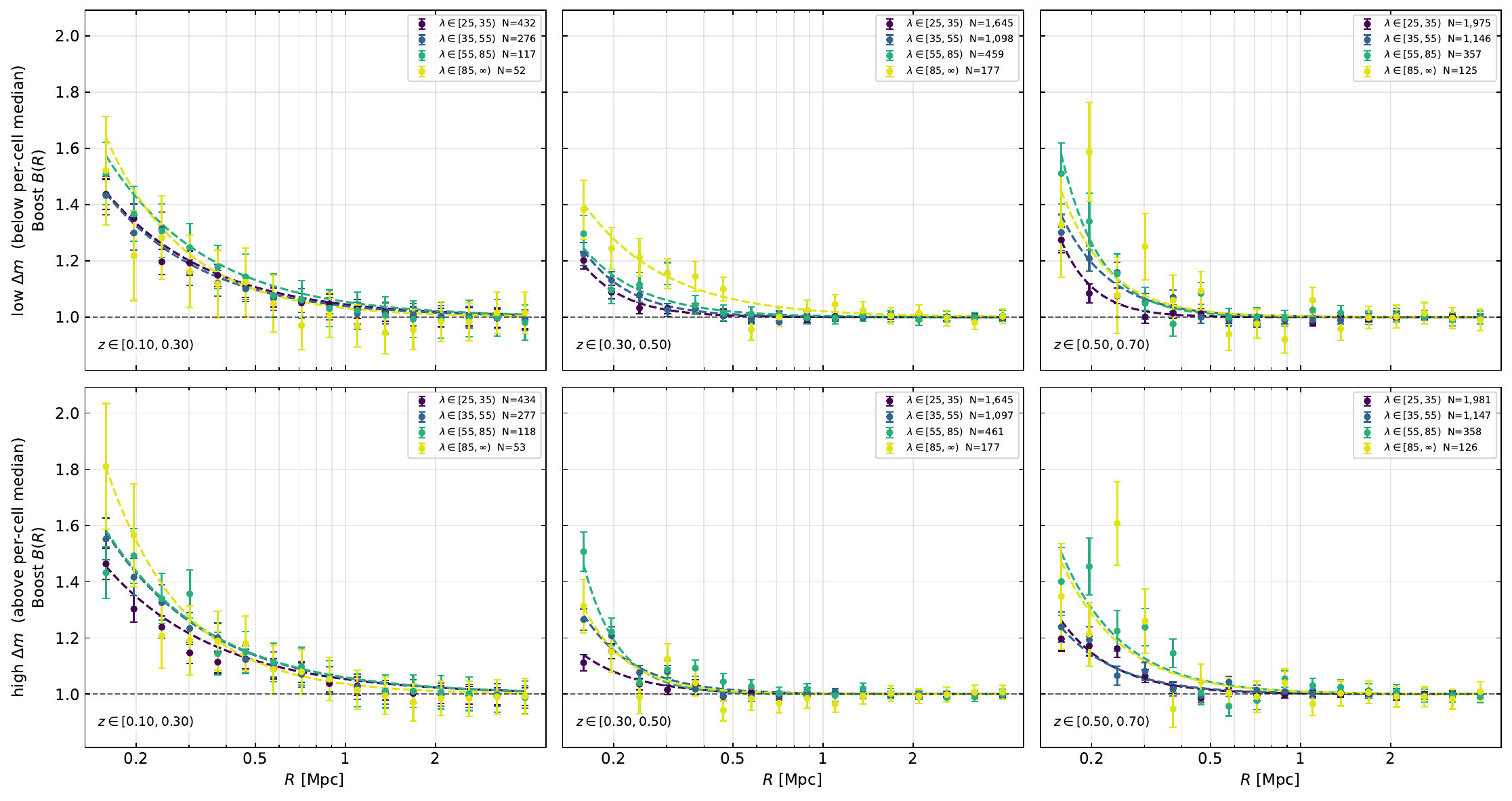}
    \caption{Boost-factor power-law fits (Eq.~\ref{eq:boost}, dashed curves) against the stacked $B(R)$ data used to fit them (points with bootstrap error bars), for all 24 stacks of the 2-way magnitude-gap binning, split by redshift, richness, and low/high magnitude gap. These are the fits that give the per-cluster boost factors $B_{c,k}$ of Eq.~\ref{eq:deltasigma}. Lower-richness (smaller, more numerous) stacks show a larger boost amplitude at small radius, where member contamination is most severe -- the same qualitative pattern as the independent measurement of Figure~\ref{fig:boost_factors_f1}.}
    \label{fig:boost_baseline}
\end{figure}

For the joint fit, the boost factor $B_{c,k}$ of each cluster (Eq.~\ref{eq:deltasigma}) comes from a power-law parametrisation of stacked boost factors,
\begin{equation}
B(R) = 1 + B_0\left(\frac{R}{R_s}\right)^{\alpha_B}\,,
\label{eq:boost}
\end{equation}
fit to each of 24 stacks of a coarser binning, with the redshift and richness bins of the stacked analysis and 2 magnitude-gap bins split at the median (Figure~\ref{fig:boost_baseline}). To each cluster the boost factor is assigned from that fit by its stack.

The boost correction is a point estimate, and the uncertainty of the boost fit is not propagated into $\mathbf{C}_i$. Refitting the joint model with boost factors redrawn from the bootstrap covariance of the stacked boosts changes $\gamma_{\rm cm}$ well within its uncertainty (Section~\ref{sec:robust}). This is the one respect in which the joint fit draws on a stacked quantity. However, $B(R)$ describes the contamination of the source sample, not the mass profile of individual clusters, so a shared boost estimate does not compromise the per-cluster nature of the fit.

Finally, we have tested the boost dependence on magnitude gaps. We found that the contamination depends at most weakly on the magnitude gap at fixed richness and redshift, as shown in Appendix~\ref{app:boostdm}.

\subsection{Stacked Fitting Results}
\label{sec:stackedresults}

We apply the forward-modelled stacked analysis of Section~\ref{sec:stackmodel} to the WaZP+DES-Y3 sample, with the binning of Section~\ref{sec:SimResults} (16 radial bins, $0.16$--$3.98$~Mpc) and the priors of Table~\ref{tab:joint_priors}. 33 of the 36 stacks have at least 50 members; the 3 excluded are the three magnitude-gap terciles of a single cell (lowest redshift, highest richness, 35 clusters each). The binning was fixed before any measurement, and we keep it rather than adjust the bins after seeing the results.

\begin{figure*}
    \centering
    \includegraphics[width=\linewidth]{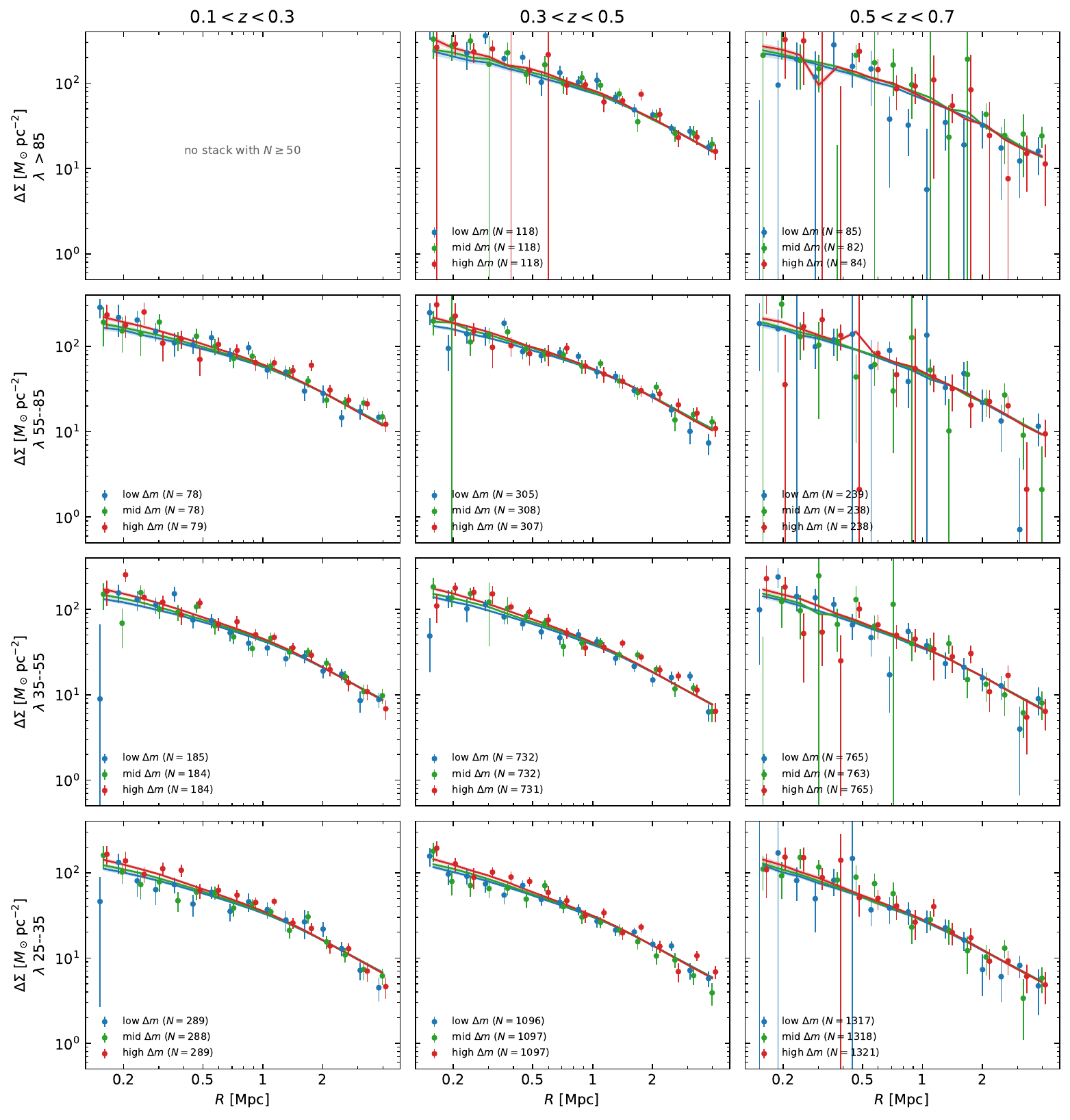}
    \caption{Boost-corrected stacked $\Delta\Sigma$ profiles of the real WaZP+DES-Y3 sample, by redshift (columns, increasing to the right) and richness (rows, increasing upward), so that each panel sits where its stack cell sits in Figure~\ref{fig:population_zr}, for the low, middle, and high magnitude-gap terciles of each cell (colours; points slightly offset in radius for visibility; $N$ is the number of member clusters). Curves and bands are the median and 68\% range of the forward-modelled stacked fit's prediction (Eq.~\ref{eq:stackmodel}) over its posterior, fit to all 33 stacks at once with the 9 parameters of the joint model.}
    \label{fig:stacked_forward_fit}
\end{figure*}

\begin{figure*}
    \centering
    \includegraphics[width=\textwidth]{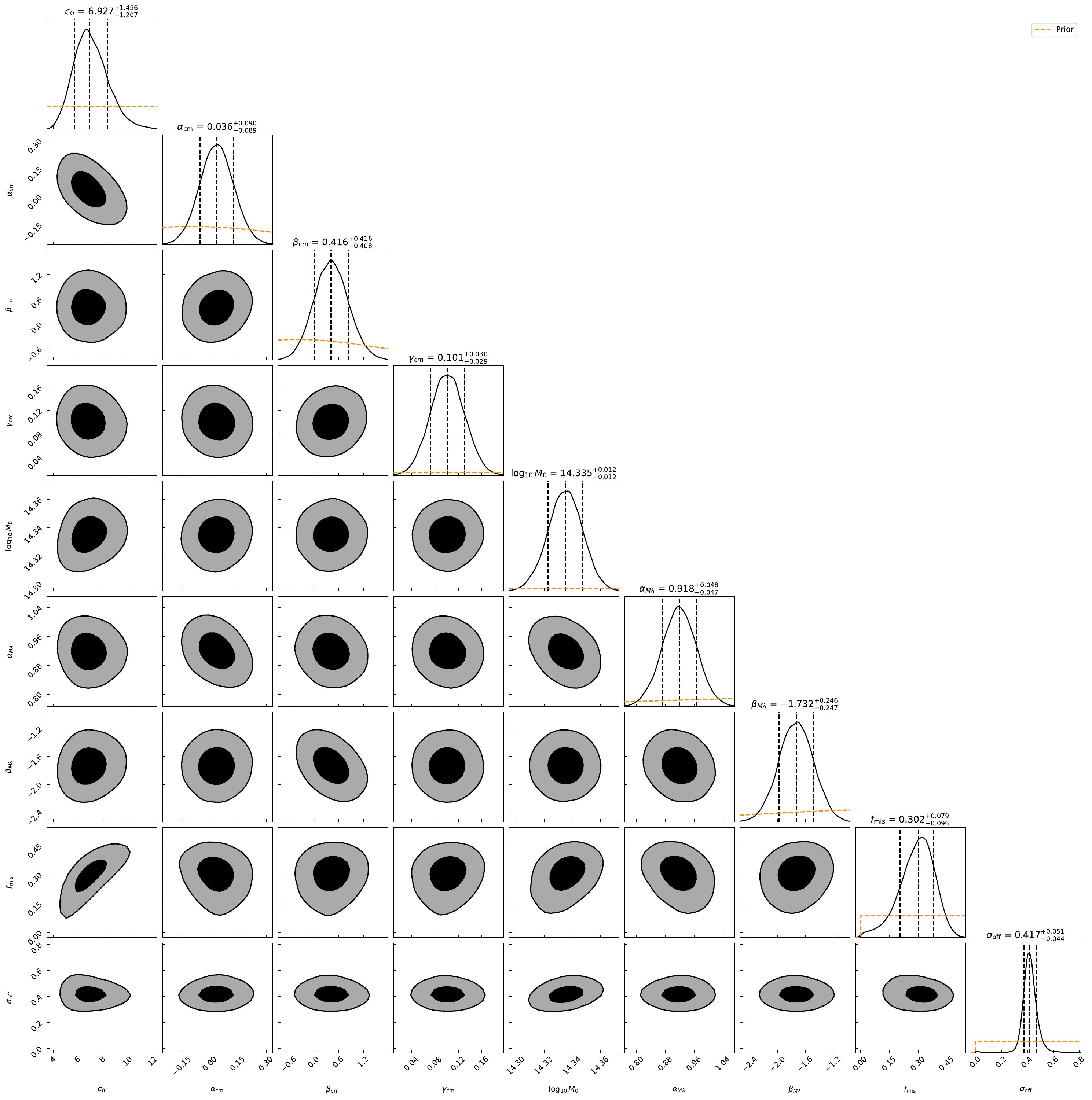}
    \caption{Posterior of the forward-modelled stacked fit to the real WaZP+DES-Y3 stacks, for the same 9 parameters as the joint fit; dashed orange curves are the priors.}
    \label{fig:stacked_full_param_corner}
\end{figure*}

We find that a single set of scaling-relation parameters describes all 33 stacks (15{,}628 clusters) well, with $\chi^2=509.9$ for $528$ data points at the posterior median. Figure~\ref{fig:stacked_forward_fit} shows the stacked profiles and the fitted model, and Figure~\ref{fig:stacked_full_param_corner} the posterior. Table~\ref{tab:real_results} gives the fitted parameters. In this analysis, the magnitude-gap coupling is positive, in the fossil-group direction -- at fixed mass and redshift, a larger magnitude gap is associated with higher concentration. The stacks also prefer the linear magnitude-gap term of Eq.~\ref{eq:c-m} to the power law in $\Delta m$ of \citet{vitorelli2018}, $\gamma_{\rm cm}\log_{10}(\Delta m/0.1\,{\rm mag})$: with the same number of parameters, the Bayesian evidence is higher by $\Delta\ln Z=1.7$ and $\chi^2$ at the posterior median lower by $4.1$. Section~\ref{sec:jointresults} compares these numbers with the joint fit.

\begin{figure*}
    \centering
    \includegraphics[width=0.95\linewidth]{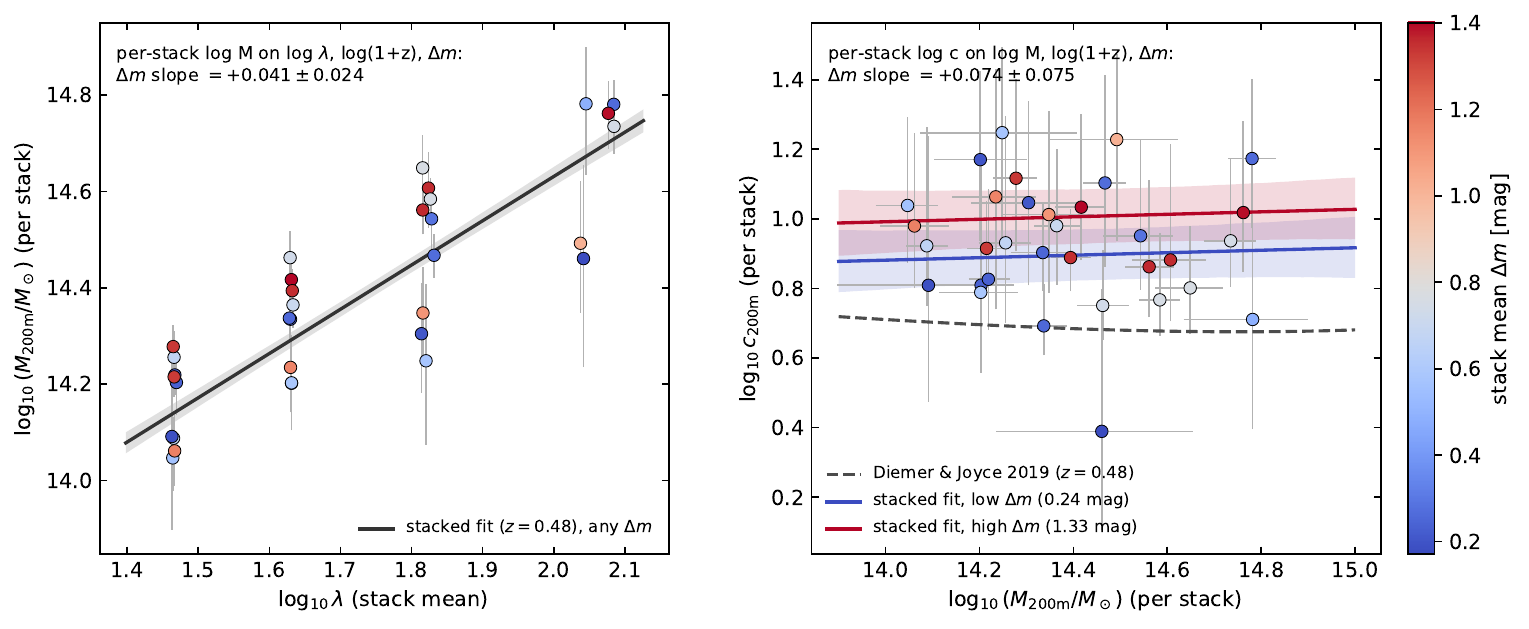}
    \caption{Mass against mean richness (left) and concentration against mass (right) of each of the 33 real WaZP+DES-Y3 stacks, from independent fits of each stack with free mass, concentration and miscentering parameters (two-halo term included, flat priors), with medians and 68\% intervals, coloured by the stack's mean magnitude gap. Left: the forward-modelled stacked mass-richness relation at the sample's mean redshift (solid line, with its 68\% posterior band; the model has no magnitude-gap term in this relation). The annotation in the panel gives the magnitude-gap slope of a weighted regression of the per-stack masses on richness, redshift and magnitude gap. Right: the forward-modelled stacked relation at the sample's mean redshift for a low and a high magnitude gap (the 17th and 83rd percentiles of the stacks' mean gaps; values in the legend), with their 68\% posterior bands, and the \citet{diemer2019} relation at the same redshift (dashed). The annotation in the panel gives the magnitude-gap slope of the analogous regression of the per-stack concentrations.}
    \label{fig:stacked_cm_dm}
\end{figure*}

Figure~\ref{fig:stacked_cm_dm} illustrates this result in the mass-richness and mass-concentration planes. Each stack is refit on its own, with its own mass, concentration and miscentering parameters, and coloured by its mean magnitude gap. These single-halo values are an illustration: the forward-modelled fit models every member cluster and never assigns a mass or concentration to a stack, so the lines are not fits to the points. With the miscentering free for each stack, the per-stack concentrations are too uncertain to show the magnitude-gap ordering by themselves (a weighted regression gives a magnitude-gap slope of $0.074\pm0.075$~mag$^{-1}$, Appendix~\ref{app:twostep}), while the forward-modelled fit, which shares the miscentering between stacks, measures it as the offset between the two lines.

A weighted regression of the per-stack masses on richness, redshift and magnitude gap gives a magnitude-gap slope of $1.7$ times its uncertainty (Figure~\ref{fig:stacked_cm_dm}, left), a weak indication that clusters with larger gaps are more massive at fixed richness (Section~\ref{sec:dmr}). The per-stack masses are much better determined than the concentrations, and their richness dependence follows the forward-modelled relation.

\subsection{Joint Fitting Results}
\label{sec:jointresults}

The data vector and covariance of each cluster are measured in the 6 radial bins of the DES-like configuration ($0.25$--$3.98$~Mpc, Section~\ref{sec:SimResults}) with the galaxy bootstrap of Section~\ref{sec:covariance} ($200$ resamples) and corrected with the boost factors of Section~\ref{sec:boost}. The variance is floored at the shape-noise value in bins at $R<0.5$~Mpc, where bins hold few sources (Figure~\ref{fig:bin_occupancy}). Bootstrap correlations between radial bins are consistent with sampling noise, so the covariances are effectively diagonal. We use the bootstrap rather than the analytic covariance for the real data because the analytic shape-noise covariance does not describe the noise of the real inner bins (Appendix~\ref{app:analytic}; the paired test of Appendix~\ref{app:recommend} gives further details).

The real-data joint fit uses the priors of Table~\ref{tab:joint_priors} with one change: the concentration-normalisation prior is a Gaussian centred on the \citet{duttonmaccio2014} value at the pivot, $\mathcal{N}(4.5,5)$ truncated at zero, instead of the flat prior on $[1,15]$. The miscentering parameters keep their flat priors. With the flat $c_0$ prior, or with other changes of the priors, the posterior does not shift beyond a fraction of its width.

\begin{figure*}
    \centering
    \includegraphics[width=\textwidth]{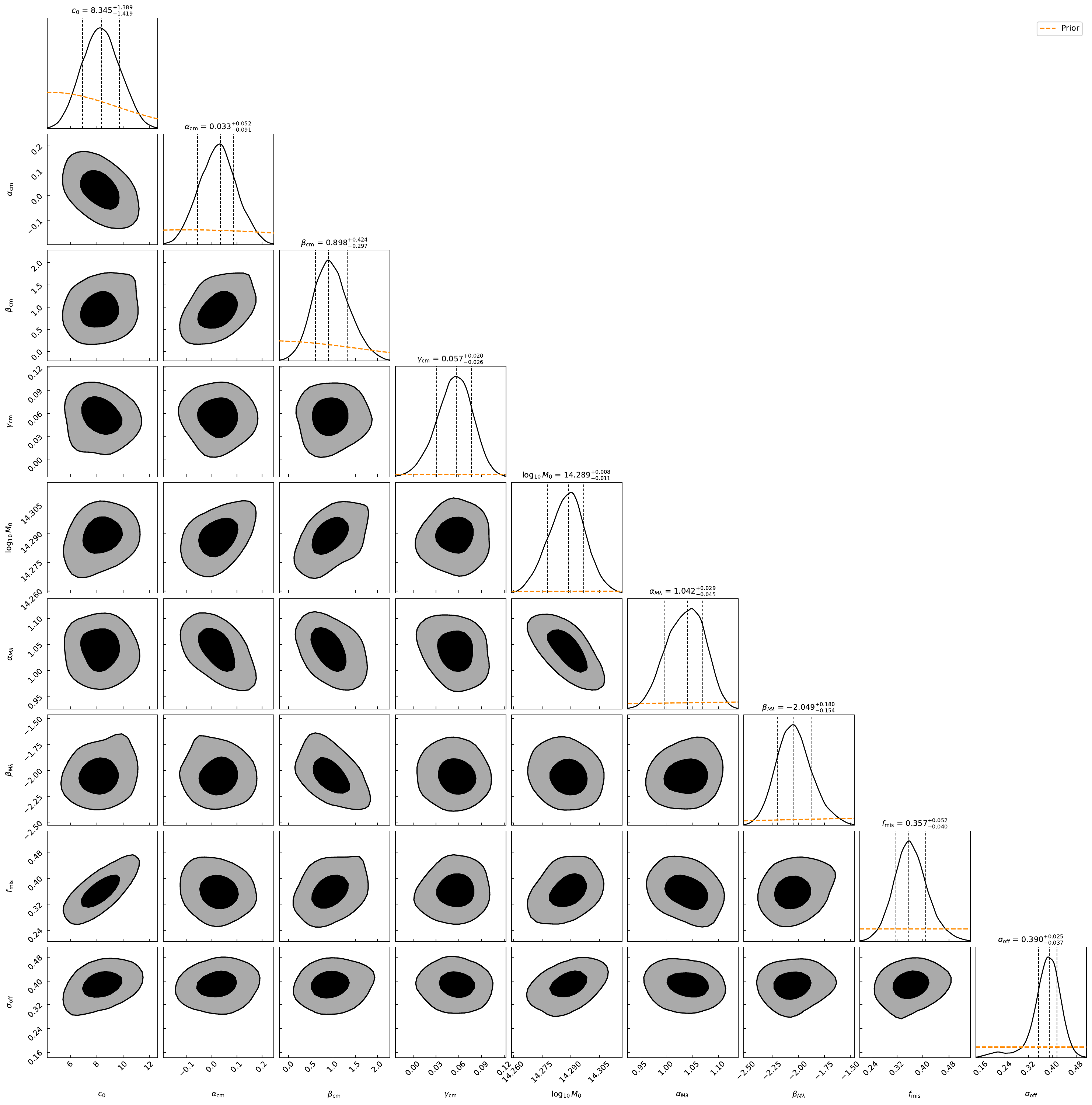}
    \caption{Joint fit, all 9 model parameters: distribution of the 50 per-bootstrap posterior medians (Section~\ref{sec:likeboot}), shown as a smoothed bootstrap that preserves their mean and covariance. Dashed orange curves are the priors; dashed black lines mark the 16th, 50th and 84th percentiles of the medians, as quoted above each panel and in Table \ref{tab:real_results}.}
    \label{fig:joint_fit_posterior}
\end{figure*}

\begin{figure}
    \centering
    \includegraphics[width=1.0\linewidth]{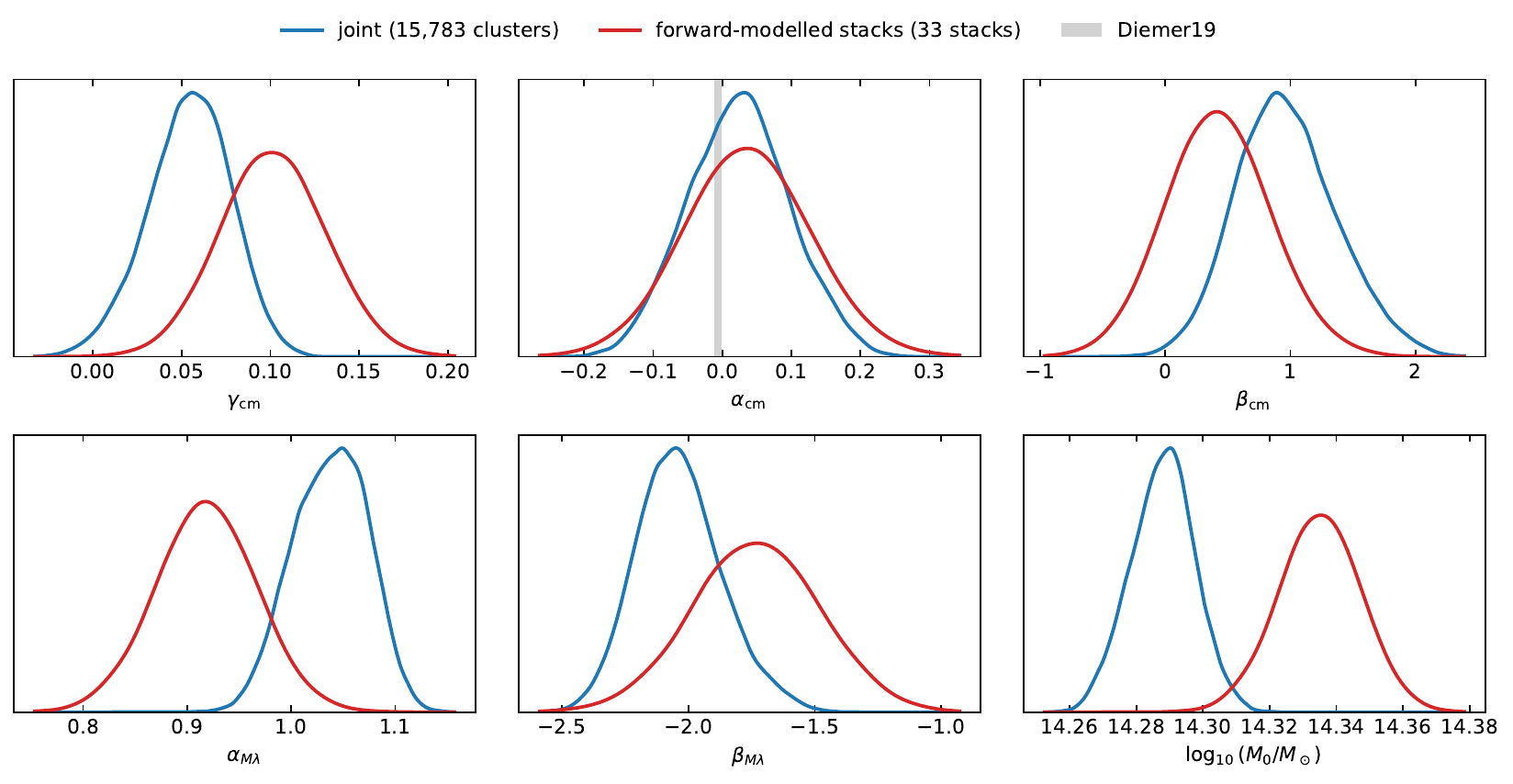}
    \caption{Real WaZP+DES-Y3 data: distribution of the joint fit per-bootstrap posterior medians (blue; smoothed, Section~\ref{sec:likeboot}) and marginal posteriors of the forward-modelled stacked fit (red), for the magnitude-gap slope, the mass-concentration and mass-richness slopes, and the mass normalisation. The grey band in the $\alpha_{\rm cm}$ panel spans the unweighted and Fisher-weighted slopes of the \citet{diemer2019} relation over the mass distribution of the sample (Section~\ref{sec:alphacm}).}
    \label{fig:real_stacked_full_param}
\end{figure}

Figure~\ref{fig:joint_fit_posterior} shows the 9-parameter posterior of the joint fit (bootstrap covariance with floor and the likelihood of Eq.~\ref{eq:shlike}), and Table~\ref{tab:real_results} gives its marginalised 68\% intervals, the Kullback-Leibler divergence of each marginal posterior from its prior, a measure of how much the data inform each parameter beyond the prior, and the comparison with the forward-modelled stacked fit of Section~\ref{sec:stackedresults} (also Figure~\ref{fig:real_stacked_full_param}). The scatter of the $50$ per-bootstrap medians is comparable to the width of a single posterior (for $\gamma_{\rm cm}$, $0.024$ against $0.027$~mag$^{-1}$).

\begin{table}
\footnotesize
\setlength{\tabcolsep}{3pt}
\centering
\caption{Real WaZP+DES-Y3 data: joint fit (median and 16th--84th percentiles of the 50 per-bootstrap posterior medians, Section~\ref{sec:likeboot}; 6-bin radial grid, bootstrap covariance with shape-noise floor, likelihood of Eq.~\ref{eq:shlike}, two-halo term; 50/50 bootstraps converged), the Kullback-Leibler divergence of its marginal posteriors from their priors (medians over bootstraps, in nats), the forward-modelled stacked fit of Section~\ref{sec:stackedresults}, and the difference between the two in units of their combined uncertainty.}
\label{tab:real_results}
\resizebox{\columnwidth}{!}{%
\begin{tabular}{lrcrr}
\toprule
Parameter & Joint & KL [nats] & Stacked & Difference \\
\midrule
$\gamma_{\rm cm}$ [mag$^{-1}$] & $0.057^{+0.020}_{-0.026}$ & 3.12 & $0.101^{+0.030}_{-0.029}$ & $-1.2\sigma$ \\
$\alpha_{\rm cm}$ & $0.033^{+0.052}_{-0.091}$ & 1.59 & $0.036^{+0.090}_{-0.089}$ & $-0.0\sigma$ \\
$\beta_{\rm cm}$ & $0.90^{+0.42}_{-0.30}$ & 1.39 & $0.416^{+0.416}_{-0.408}$ & $+0.9\sigma$ \\
$c_0$ & $8.35^{+1.39}_{-1.42}$ & 0.83 & $6.93^{+1.46}_{-1.21}$ & $+0.7\sigma$ \\
$\alpha_{M\lambda}$ & $1.042^{+0.029}_{-0.045}$ & 2.47 & $0.918^{+0.048}_{-0.047}$ & $+2.1\sigma$ \\
$\beta_{M\lambda}$ & $-2.05^{+0.18}_{-0.15}$ & 2.43 & $-1.732^{+0.246}_{-0.247}$ & $-1.1\sigma$ \\
$\log_{10}(M_0/M_\odot)$ & $14.289^{+0.008}_{-0.011}$ & 3.47 & $14.335^{+0.012}_{-0.012}$ & $-3.0\sigma$ \\
$f_{\rm mis}$ & $0.357^{+0.053}_{-0.040}$ & 1.14 & $0.302^{+0.079}_{-0.096}$ & $+0.6\sigma$ \\
$\sigma_{\rm off}$ [Mpc] & $0.390^{+0.025}_{-0.037}$ & 1.79 & $0.417^{+0.051}_{-0.044}$ & $-0.5\sigma$ \\
\midrule
All 9 parameters & -- & 21.36 & -- & -- \\
Residual (mutual information) & -- & 3.12 & -- & -- \\
\bottomrule
\end{tabular}}
\end{table}

The joint fit agrees with the stacked fit on the concentration and miscentering parameters, and finds a weaker magnitude-gap coupling (Table~\ref{tab:real_results}), with flat miscentering priors in both fits and neither fit using the other's result. The concentration normalisation is degenerate with the miscentering parameters, and the degeneracy depends on where $c_0$ is defined: it is the concentration at $M_{\rm piv}$ and $\Delta m=0$, below the sample medians ($M\approx10^{14.3}\,M_\odot$, $\Delta m\approx0.63$~mag), and re-expressing it at the medians, which changes only $c_0$ and its prior, changes its posterior correlation with $\alpha_{\rm cm}$ from $-0.47$ to $-0.24$, and with $\gamma_{\rm cm}$ from $-0.07$ to $+0.14$. The mass-richness parameters differ more between the methods (Table~\ref{tab:real_results}; Section~\ref{sec:mrmis}).

\section{Discussion}
\label{sec:discussion}

In the following, we discuss what the comparison of the methods and the real-data fits show, and address some of the results. Section~\ref{sec:precision} compares the precision of the joint and stacked analyses, in the simulations and on the real data, and Section~\ref{sec:simoffsets} interprets the offsets of the simulated fits. Section~\ref{sec:mrmis} discusses the mass-richness relation and miscentering that the two methods measure for WaZP clusters, and Section~\ref{sec:alphacm} shows that the nearly flat mass-concentration slope is explained by the non-power-law shape of the \citet{diemer2019} relation. Sections~\ref{sec:dmr}--\ref{sec:dmrinterp} turn to the magnitude-gap coupling: whether a magnitude-gap dependence of the mass at fixed richness can account for the stacked coupling, where the stacked profiles depart from the joint fit, and how such a dependence can be interpreted. Section~\ref{sec:robust} tests the coupling against the choices of the analysis and shows that the joint and stacked values agree when both use the same data products.

\subsection{Precision of the methods}
\label{sec:precision}

The joint fit is the more precise method in the simulations, and reports the smaller uncertainties on the real data (Table~\ref{tab:width_ratios}). In the simulations its estimates scatter less between independent realisations than the forward-modelled stacked ones for every scaling-relation slope, by $1.1$--$1.5\times$, and the reported uncertainties of both methods match that scatter. For these two fits, data density matters more than the method: both constrain $\gamma_{\rm cm}$ five to six times more precisely at LSST-like than at DES-like density (Table~\ref{tab:gamma_recovery}). The measurement of $\gamma_{\rm cm}$ is therefore a matter of statistical power and of controlling the miscentering parameters. In terms of concentration, the posterior width of $\gamma_{\rm cm}$ times the standard deviation of $\Delta m$, the uncertainty on the gap-driven spread of $\log_{10}c$, is $0.025$ (joint) and $0.030$~dex (forward-modelled) at DES-like density and $0.004$ and $0.005$~dex at LSST-like density. The gain over the standard two-step analysis is larger, $1.5$--$6.5\times$, and grows with data density, because the two-step $\gamma_{\rm cm}$ gains only a factor of $1.8$ in precision between the two densities: the per-stack fits of the two-step analysis are limited by the degeneracy between concentration and miscentering within each stack rather than by shape noise. On the real data, where the precision cannot be measured, the ratio of the reported uncertainties is $1.1$--$1.5$ for the slopes relative to the forward-modelled fit and $1.5$--$3.3$ relative to the two-step analysis, and the stacked fit reports a tighter interval only on $c_0$ (Table~\ref{tab:real_results}), which is degenerate with the miscentering parameters. Beyond precision, the joint fit works cluster by cluster: it needs no binning choices or minimum stack size, uses every cluster, and can be extended to per-cluster quantities that stacking averages away, such as intrinsic scatter about the scaling relations or individual observables beyond the binned ones.

\begin{table*}
\centering
\caption{Ratio of the uncertainty of the forward-modelled stacked fit (F) and of the two-step stacked analysis (2S, Appendix~\ref{app:twostep}) to that of the joint fit, per parameter. Simulations: ratio of the standard deviations of the estimates between independent realisations, the measured precision. Real data, where no truth is known: ratio of the reported uncertainties (68\% half-widths of the stacked posterior and of the distribution of the joint fit's per-bootstrap medians; weighted-least-squares errors of the two-step analysis). Values below 1 mean the stacked analysis is the more precise.}
\label{tab:width_ratios}
\begin{tabular}{lcccccc}
\toprule
 & \multicolumn{2}{c}{LSST-like} & \multicolumn{2}{c}{DES-like} & \multicolumn{2}{c}{Real data} \\
Parameter & F & 2S & F & 2S & F & 2S \\
\midrule
$\gamma_{\rm cm}$   & $1.28$ & $6.50$ & $1.23$ & $2.03$ & $1.29$ & $3.26$ \\
$\alpha_{\rm cm}$   & $1.37$ & $5.14$ & $1.27$ & $1.62$ & $1.25$ & $2.63$ \\
$\beta_{\rm cm}$    & $1.29$ & $5.27$ & $1.08$ & $1.49$ & $1.14$ & $2.56$ \\
$\alpha_{M\lambda}$ & $1.35$ & $1.76$ & $1.53$ & $1.80$ & $1.28$ & $1.46$ \\
$\beta_{M\lambda}$  & $1.27$ & $1.71$ & $1.27$ & $1.59$ & $1.47$ & $1.88$ \\
\bottomrule
\end{tabular}
\end{table*}

\subsection{Offsets in the simulated fits}
\label{sec:simoffsets}

Part of the low mass normalisation in the simulations (Section~\ref{sec:jointvsstacked}) is the bin-centre evaluation of the model (Section~\ref{sec:fullmodel}), which depends on the radial grid: it lowers the mass normalisation by about $0.019$~dex on the 8-bin grid, $0.027$~dex on the 6-bin grid and only $0.004$~dex on the 16-bin grid of the stacks (Appendix~\ref{app:binning}). This accounts for the difference between the two methods, $0.015$ and $0.020$~dex, and after it both are low by about $0.025$~dex. The rest is the expected response of a spherical model to triaxial haloes, whose orientation-averaged lensing signal is lower than that of a spherical halo of the same mass and concentration by a nearly radius-independent amount (Appendix~\ref{ap:triaxial}). Most of this part is definitional: the injected mass is that within an ellipsoid, which exceeds the spherical-overdensity mass of the same halo by about $0.017$~dex (Appendix~\ref{ap:triaxial}).

A plausible explanation of the low $c_0$ and $f_{\rm mis}$ of the LSST-like joint fits is that the spherical, miscentered model absorbs part of the triaxial projection (which changes both the amplitude and the inner slope of each cluster's profile) through the concentration-miscentering degeneracy, while stacking averages orientations before the fit. We have not tested this further.

\subsection{Mass-richness relation and miscentering}
\label{sec:mrmis}

Both methods measure the mass-richness relation of WaZP clusters (Table~\ref{tab:real_results}). The two methods differ more on the mass-richness parameters than on the concentration ones. About $0.02$~dex of the $0.046$~dex difference in normalisation is the different bin-centre offset of the two radial grids (Appendix~\ref{app:binning}). The runs share the same data, so these differences are a guide and not independent significances. In the simulations both methods recover these parameters (Section~\ref{sec:jointvsstacked}), so the remaining differences arise from properties of the real data that the simulations do not include (member contamination and the boost correction, richness scatter and selection, and the two-halo term) or from the weighting of individual clusters by their covariance (Appendix~\ref{app:analytic}).

The WaZP mass-richness values are not directly comparable with the DES-Y1 calibration of \citet{mcclintock2019}, which we use only to centre the priors: richness is defined by each cluster finder, and redMaPPer's $\lambda$ and WaZP's $N_{\rm gals}$ differ in definition, in their scaling with mass and in their projection effects, so the slopes need not agree. A calibration of WaZP richness against redMaPPer's is left to a forthcoming comparison.

The miscentered fractions are similar to the one found for redMaPPer clusters. The stacked fit's miscentered fraction, $f_{\rm mis}=0.30^{+0.08}_{-0.10}$, is of the same order as the roughly one quarter that \citet{zhang2019} find for redMaPPer clusters, although such fractions are specific to each cluster finder's centring and are not a benchmark for WaZP (Section~\ref{sec:misc}); the joint fit's $0.36^{+0.05}_{-0.04}$, obtained with a flat prior and without using the stacked result, is similar. These fractions are constrained together with the concentration normalisation: a higher normalisation raises the inner profile of every cluster, more miscentering dilutes it again, and $\gamma_{\rm cm}$ redistributes the remaining inner signal between magnitude-gap terciles (Section~\ref{sec:jointresults}, Appendix~\ref{app:analytic}).

\subsection{The mass-concentration slope}
\label{sec:alphacm}

\begin{figure*}
    \centering
    \begin{minipage}[c]{0.47\linewidth}\centering\includegraphics[width=\linewidth]{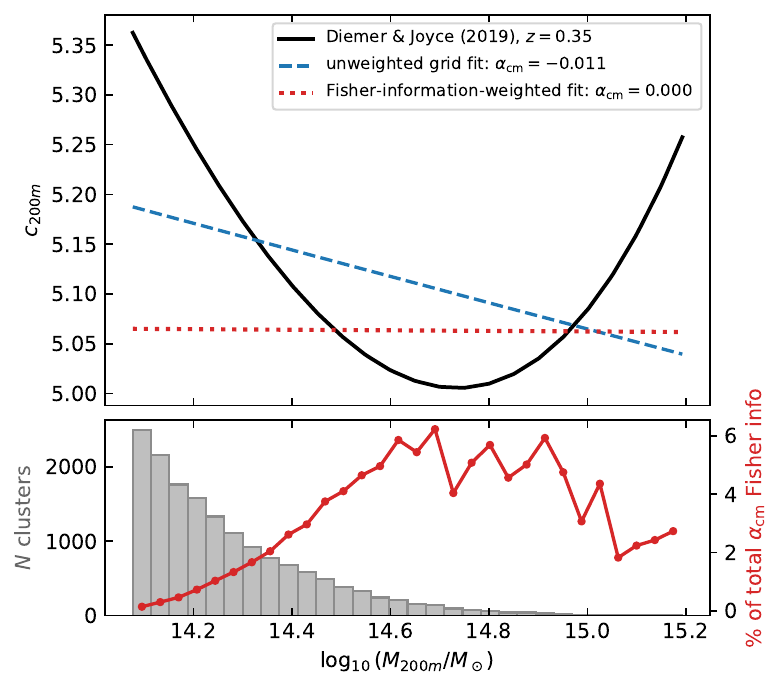}\\(a)\end{minipage}\hfill
    \begin{minipage}[c]{0.5\linewidth}\centering\includegraphics[width=\linewidth]{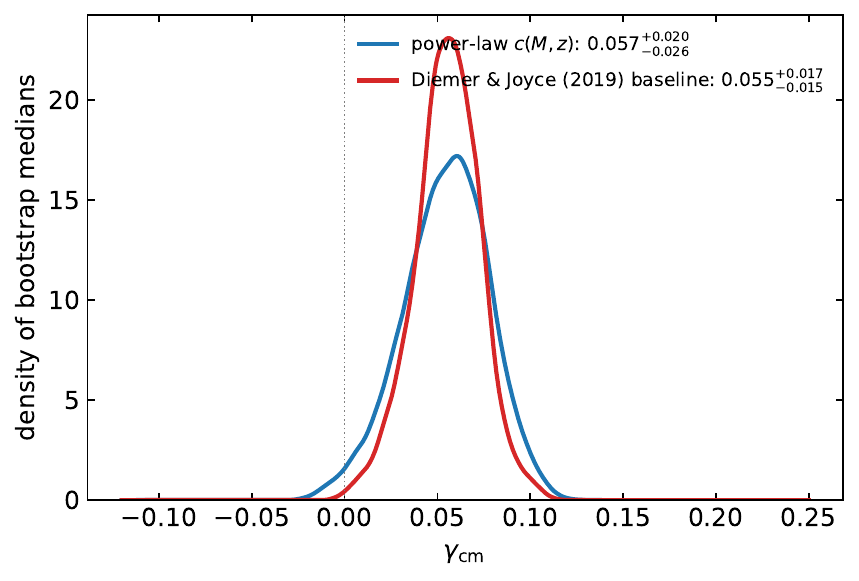}\\(b)\end{minipage}
    \caption{(a) Top: the \citet{diemer2019} concentration-mass relation at the pivot redshift ($z=0.35$, solid black), which is non-monotonic over the WaZP+DES-Y3 mass range, together with power-law fits to that same prediction restricted to the mass distribution of the sample -- unweighted (blue dashed) and weighted by the share of each mass bin in the Fisher information on $\alpha_{\rm cm}$ (red dotted). (a) Bottom: the cluster-count distribution of the sample in mass (grey) against the fraction of total $\alpha_{\rm cm}$ Fisher information contributed per mass bin (red), showing that the sample's statistical weight is concentrated near the curve's minimum and beyond, while its raw counts are not. (b) The magnitude-gap slope from the joint fit (distribution of the 50 per-bootstrap posterior medians) under two different mass-concentration baselines, both fit to the same real, boost-corrected WaZP+DES-Y3 sample: the free power-law relation of Eq.~\ref{eq:c-m} (the joint fit of Section~\ref{sec:jointresults}) and the \citet{diemer2019} prediction with only $\gamma_{\rm cm}$ (Eq.~\ref{eq:diemer-gamma}) added on top, every other parameter refit freely in both cases. The two posteriors agree within $1\sigma$.}
    \label{fig:diemer19_alpha_cm}
\end{figure*}

The mass-concentration slope is consistent with zero in both methods: $\alpha_{\rm cm}=0.033^{+0.052}_{-0.091}$ in the joint fit and $0.036\pm0.090$ in the stacked fit. This is what the concentration-mass relation of \citet{diemer2019} predicts for this sample, because that relation is not a power law over this mass range: it declines with mass at the low-mass end, reaches a minimum near $\log_{10}(M/M_\odot)\approx14.7$ at the pivot redshift, and turns upward beyond it (Fig.~\ref{fig:diemer19_alpha_cm}a, top panel). A single power-law slope fit to such a curve depends on where the fit's effective weight sits across the mass range. Evaluated over the mass distribution of the WaZP+DES-Y3 sample, the Diemer19 curve gives a slope close to zero however it is weighted -- $\alpha_{\rm cm}=-0.011$ treating each mass bin equally, and $0.000$ weighting each bin by its share of the Fisher information on $\alpha_{\rm cm}$, which peaks near the minimum of the curve and beyond (Fig.~\ref{fig:diemer19_alpha_cm}a, bottom panel). Both measured slopes lie within $1\sigma$ of this prediction. The per-stack concentrations of Figure~\ref{fig:stacked_cm_dm} lie above the \citet{diemer2019} prediction; their normalisation is degenerate with the miscentering parameters (Section~\ref{sec:jointresults}), so we do not interpret this offset.

The non-power-law shape raises a direct question about the magnitude-gap result itself. If the true mass dependence of concentration turns over rather than following a single power law, does $\gamma_{\rm cm}$ survive when the power law is replaced by a relation with that turnover built in? To test this, we refit the joint model on the same real, boost-corrected sample, with the same covariance and priors as the joint fit of Section~\ref{sec:jointresults}. The free power-law mass-concentration relation (Eq.~\ref{eq:c-m}) is replaced by the \citet{diemer2019} prediction as the baseline, with only the same magnitude-gap term added on top of it,
\begin{equation}
\log_{10} c_{200} = \log_{10} c_{\rm Diemer19}(M_{200}, z) + \gamma_{\rm cm}\,\Delta m\,.
\label{eq:diemer-gamma}
\end{equation}
Every other parameter (the mass-richness normalisation and slopes, the miscentered fraction, and the miscentering scale) is refit freely alongside $\gamma_{\rm cm}$, and the two-halo term is included, exactly as in that fit. This is not a fit of the Diemer19 relation's own shape to the data (its amplitude, pivot mass, and higher-order behaviour are all fixed to the literature prediction). It only asks the narrower question above: do $\gamma_{\rm cm}$'s sign and approximate size depend on assuming a power law for the mass dependence in the first place? A full treatment of a curved, non-power-law mass-concentration relation, with its own parameters left free, is deferred to following work.

Figure~\ref{fig:diemer19_alpha_cm}b shows the result: under the Diemer19 baseline $\gamma_{\rm cm}=0.055^{+0.017}_{-0.015}$~mag$^{-1}$, in close agreement with the power-law joint fit ($0.057^{+0.020}_{-0.026}$) and $1.4\sigma$ below the stacked value. The joint coupling is the same under either baseline, so it does not depend on assuming a power law for the mass dependence of concentration.

\subsection{Magnitude gap and mass at fixed richness}
\label{sec:dmr}

The joint and stacked couplings differ by $1.2\sigma$ ($0.057^{+0.020}_{-0.026}$ against $0.101\pm0.030$~mag$^{-1}$). The stacked coupling is positive, in the direction expected if large gaps trace early-assembled systems. The joint value stays within its uncertainty when the covariance estimator or the priors are changed (Appendix~\ref{app:analytic}). On the joint fit's source selection and radial range, the stacked fit gives $0.072^{+0.042}_{-0.040}$, in agreement with it (Section~\ref{sec:robust}). We look for an assumption of the model that could account for the stacked coupling. The joint and stacked fits share the model of Section~\ref{sec:Model}, in which the magnitude gap enters only the concentration (Eq.~\ref{eq:c-m}) and the mass depends only on richness and redshift (Eq.~\ref{eq:mass-richness}). If clusters with different gaps also differ in mass at fixed richness, the model can represent that difference only through $\gamma_{\rm cm}$, and the two methods need not absorb it in the same way.

A gap-dependent mass at fixed richness is plausible on two grounds. First, in the fossil-group picture \citep{jones2003}, which motivated the use of the magnitude gap in \citet{vitorelli2018} and here, a large gap forms when the brightest satellites merge into the central galaxy, which removes members, so that at fixed mass a large-gap cluster is poorer, and at fixed richness it is more massive. \citet{hearin2013} found this dependence in mock catalogues and SDSS groups. Second, in a photometric cluster finder, projected interlopers raise the richness at fixed mass and can also provide a spurious brightest satellite, lowering $\Delta m$, so projection alone gives small-gap clusters a lower mass at fixed richness \citep{sunayama2020}. The per-stack masses of Figure~\ref{fig:stacked_cm_dm} already hint at a positive dependence (Section~\ref{sec:stackedresults}).

\begin{figure*}
    \centering
    \includegraphics[width=0.95\linewidth]{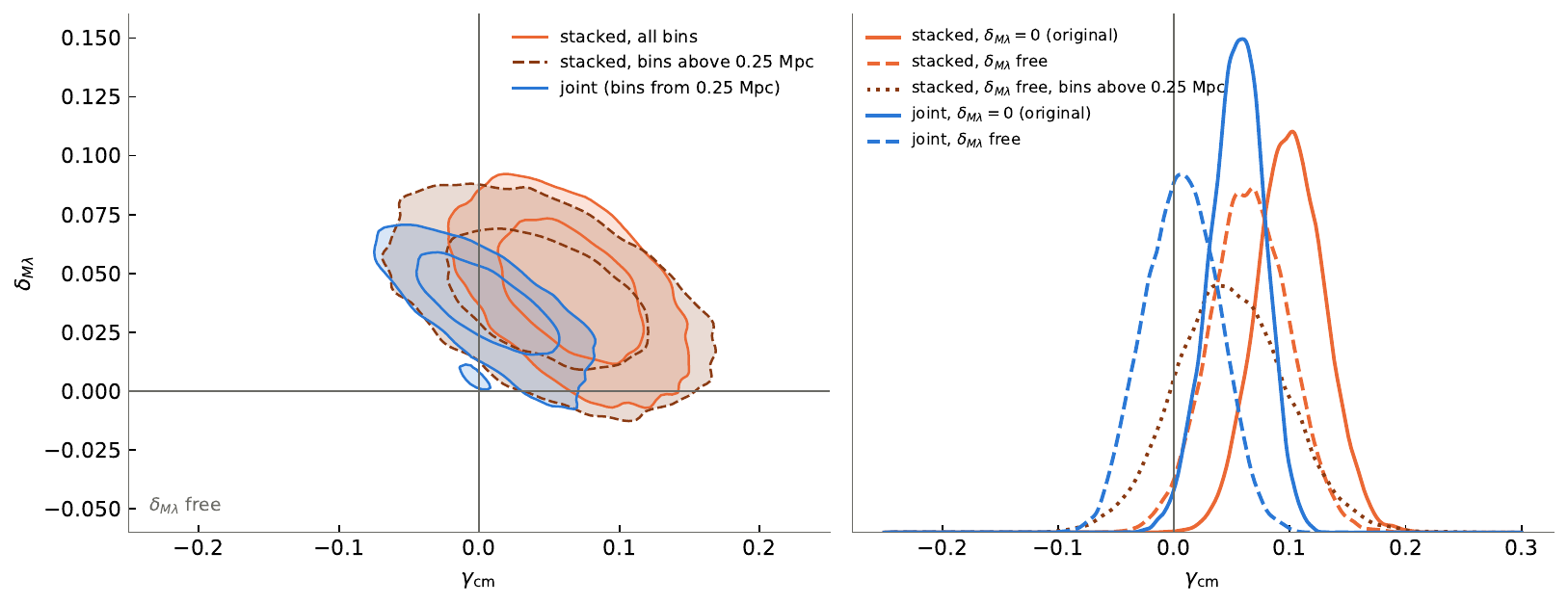}
    \caption{Left: 68\% and 95\% posterior regions of the magnitude-gap slopes of concentration, $\gamma_{\rm cm}$, and of mass at fixed richness, $\delta_{M\lambda}$, with $\delta_{M\lambda}$ free, for the forward-modelled stacked fit (all radial bins, and only bins above $0.25$~Mpc as in the joint fit) and the joint fit of the real WaZP+DES-Y3 data. Right: marginal posteriors of $\gamma_{\rm cm}$ in the original model, $\delta_{M\lambda}=0$ (solid, as in the rest of the paper), and with $\delta_{M\lambda}$ free (dashed; dotted: stacked fit with bins above $0.25$~Mpc only).}
    \label{fig:dmr}
\end{figure*}

We test it by adding a magnitude-gap term to the mass-richness relation,
\begin{equation}
\begin{split}
\log_{10} M_{200} ={}& \log_{10} M_0 + \alpha_{M\lambda}\log_{10}\frac{\lambda}{\lambda_{\rm piv}}\\
&+ \beta_{M\lambda}\log_{10}\frac{1+z}{1+z_{\rm piv}} + \delta_{M\lambda}\,\Delta m\,,
\end{split}
\label{eq:mass-richness-dm}
\end{equation}
with a prior $\delta_{M\lambda}\sim\mathcal N(0,1)$, in both fits, everything else unchanged. Each cluster's magnitude gap then enters its mass through $\delta_{M\lambda}$ and its concentration through $\gamma_{\rm cm}$ (and through $c(M)$). A larger mass raises a cluster's whole profile, including the two-halo term, while a larger concentration raises mainly the inner profile, so the two are separated by the radial shape of the profiles. The joint fit uses the bootstrap covariance and priors of Section~\ref{sec:jointresults}, with 50 cluster bootstraps summarised as in Section~\ref{sec:likeboot}; the stacked fit is that of Section~\ref{sec:stackedresults}. Table~\ref{tab:dmr} and Figure~\ref{fig:dmr} give the results.

\begin{table*}
\centering
\caption{Magnitude-gap slopes of concentration ($\gamma_{\rm cm}$, in mag$^{-1}$) and of mass at fixed richness ($\delta_{M\lambda}$, in dex~mag$^{-1}$, Eq.~\ref{eq:mass-richness-dm}), with $\delta_{M\lambda}$ fixed at zero (the original model, used in the rest of the paper) and free. The joint fit uses bins from $0.25$~Mpc; the stacked fit uses bins from $0.16$~Mpc, or only those above $0.25$~Mpc. Medians and 68\% intervals.}
\label{tab:dmr}
\begin{tabular}{lcc}
\toprule
Fit & $\gamma_{\rm cm}$ & $\delta_{M\lambda}$ \\
\midrule
Stacked, $\delta_{M\lambda}=0$ (original) & $0.101^{+0.030}_{-0.029}$ & -- \\
Stacked, $\delta_{M\lambda}$ free & $0.065^{+0.033}_{-0.033}$ & $0.043^{+0.020}_{-0.020}$ \\
Stacked, $\delta_{M\lambda}$ free, bins above $0.25$~Mpc & $0.048^{+0.048}_{-0.046}$ & $0.039^{+0.020}_{-0.019}$ \\
Joint, $\delta_{M\lambda}=0$ (original) & $0.057^{+0.020}_{-0.026}$ & -- \\
Joint, $\delta_{M\lambda}$ free & $0.007^{+0.032}_{-0.033}$ & $0.035^{+0.016}_{-0.012}$ \\
\bottomrule
\end{tabular}
\end{table*}

With $\delta_{M\lambda}$ free, both couplings fall by similar amounts: the stacked one from $0.101$ to $0.065\pm0.033$, and the joint one from $0.057$ to $0.007^{+0.032}_{-0.033}$. Both fits prefer a positive $\delta_{M\lambda}$, $0.043\pm0.020$ (stacked) and $0.035^{+0.016}_{-0.012}$~dex~mag$^{-1}$ (joint), and in both the two slopes are anti-correlated (Figure~\ref{fig:dmr}). The two methods then differ by $1.2\sigma$ in $\gamma_{\rm cm}$ and $0.3\sigma$ in $\delta_{M\lambda}$. On the joint fit's radial range, using only stacked bins above $0.25$~Mpc, the stacked coupling is $0.048^{+0.048}_{-0.046}$, within $0.7\sigma$ of the joint value. The Bayesian evidence of the stacked fit does not require the additional parameter (it slightly prefers $\delta_{M\lambda}=0$ under the wide prior used), so the data do not demand a gap-dependent mass; but once one is allowed, neither coupling is robust.

\subsection{Posterior predictive check}
\label{sec:ppc}

A posterior predictive check of the stacks against the joint fit shows where the two methods depart. With the joint fit's parameters (medians of Table~\ref{tab:real_results}, $\delta_{M\lambda}=0$), the stacked model of Eq.~\ref{eq:stackmodel} gives $\chi^2=550.3$ for the $528$ stacked data points, against $509.9$ at the stacked fit's own median. Of the difference, $32$ comes from the high-gap terciles, whose profiles lie above the joint prediction at every radius: inverse-variance weighted over the stacks, the data exceed the prediction by $14\pm3\%$ beyond $1$~Mpc, where the concentration has little effect, against $7\pm3\%$ and $10\pm3\%$ for the low- and middle-gap terciles (the common part reflects the different mass normalisations of the two fits, Section~\ref{sec:mrmis}). An excess of the large-gap stacks at large radius is the signature of a magnitude-gap dependence of the mass at fixed richness, which the original model cannot represent and partly absorbs through $\gamma_{\rm cm}$.

\subsection{Interpreting a gap-dependent mass-richness relation}
\label{sec:dmrinterp}

A non-zero $\delta_{M\lambda}$ is necessary but not sufficient evidence of assembly bias: scatter in richness at fixed mass alone makes any mass-correlated observable carry mass information beyond the richness, and the positive $\delta_{M\lambda}$ preferred by the real data is expected from projection effects as well as from the fossil-group picture. These data do not separate these possibilities, which we will address in a following work.

At the signal-to-noise of these data, a $\Delta \Sigma$ amplitude difference between clusters of different $\Delta m$ can be absorbed by a change of concentration. A larger mass raises the entire profile and a larger concentration its inner part, but neither the stacked profiles nor the individual ones separate the two well, so when the mass is not allowed to depend on the gap, both methods fit an amplitude difference by a change of concentration. Allowing a magnitude-gap-dependent mass is the model counterpart of the comparison in \citet{vitorelli2018}. There, each magnitude-gap stack was fit with its own mass and concentration, and the concentrations were moved along the \citet{duffy2008} mass-concentration relation to a common mass before being compared. Leaving the mass of each stack free is what $\delta_{M\lambda}$ restores in the forward model, while the slide along the mass-concentration relation is built into Eq.~\ref{eq:c-m}. A model in which the magnitude gap enters only the concentration can therefore return a coupling that does not survive a modest extension of the model, in a stacked as in a joint analysis.

We have not tested this mechanism in simulations yet: those of Section~\ref{sec:SimResults} contain no magnitude-gap dependence of mass, which is why the two methods agree there. Injecting a $\delta_{M\lambda}$ is left to future work, together with the assembly-bias question it raises for richness.

\subsection{Robustness of the real-data coupling}
\label{sec:robust}

Table~\ref{tab:robust} collects refits of both methods in which one choice of the data or the model is changed at a time. Each joint refit is a single posterior on the full sample, compared with a full-sample fit with the reference choices (the first row), so the shifts are paired; their widths include shape noise only (Section~\ref{sec:likeboot}). Each stacked refit is compared with a rerun of the stacked fit of Section~\ref{sec:stackedresults} with the same sampler settings (the first row), so these shifts are paired as well.

\begin{table*}
\centering
\caption{Magnitude-gap slope $\gamma_{\rm cm}$ [mag$^{-1}$] of the real data under changes of the analysis (medians and 68\% intervals). Joint: full-sample posteriors, to be compared with the first row; stacked: to be compared with the first row, a rerun of the fit of Section~\ref{sec:stackedresults} that differs from it only by sampling noise.}
\label{tab:robust}
\begin{tabular}{lcc}
\toprule
Change & Joint & Stacked \\
\midrule
None (reference) & $0.061^{+0.026}_{-0.026}$ & $0.100^{+0.030}_{-0.029}$ \\
\citet{sellentin2016} likelihood & -- & $0.101^{+0.030}_{-0.029}$ \\
Reduced shear in the model & $0.059^{+0.026}_{-0.026}$ & $0.097^{+0.029}_{-0.028}$ \\
Boost factors redrawn (8 draws, range of medians) & $0.057$--$0.064$ & -- \\
$\Delta m\leq0$ included ($15{,}925$ clusters) & $0.058^{+0.026}_{-0.026}$ & -- \\
$\Delta m<2\sigma_{\Delta m}$ removed ($13{,}181$ clusters) & $0.070^{+0.031}_{-0.030}$ & -- \\
Joint fit's source selection ($f=1.1$) & -- & $0.083^{+0.028}_{-0.028}$ \\
Bins above $0.25$~Mpc & -- & $0.094^{+0.044}_{-0.041}$ \\
Joint fit's source selection and bins above $0.25$~Mpc & -- & $0.072^{+0.042}_{-0.040}$ \\
$f_{\rm mis}$ free per magnitude-gap tercile & $0.059^{+0.051}_{-0.036}$ & $0.052^{+0.088}_{-0.095}$ \\
$f_{\rm mis}$ and $\sigma_{\rm off}$ free per tercile & $0.063^{+0.051}_{-0.044}$ & $0.055^{+0.090}_{-0.092}$ \\
Redshift $0.1$--$0.3$ only ($\beta$'s fixed at 0) & $0.112^{+0.039}_{-0.037}$ & -- \\
Redshift $0.3$--$0.5$ only & $0.004^{+0.037}_{-0.037}$ & -- \\
Redshift $0.5$--$0.7$ only & $0.070^{+0.101}_{-0.098}$ & -- \\
\bottomrule
\end{tabular}
\end{table*}

Changes of the likelihood, of the boost factors and of the treatment of small gaps each move $\gamma_{\rm cm}$ by at most $0.009$ (Table~\ref{tab:robust}). The \citet{sellentin2016} likelihood for the stacks, the uncertainty of the joint fit's boost factors and the inclusion of the clusters with $\Delta m\leq0$, which the linear term allows without a floor, each move it by $0.004$ or less, and removing the clusters whose gap is below twice its uncertainty raises the joint value by $0.009$, well within its uncertainty.

The reduced shear, which the model neglects (Section~\ref{sec:wl}), acts mainly on the concentration. With the parameters of the joint fit and the source redshift distribution, it raises $\Delta\Sigma$ of a well-centred cluster by $5\%$ at $0.25$~Mpc, $2\%$ at $0.44$~Mpc and $1\%$ or less beyond, and differs by less than $0.3\%$ between magnitude-gap terciles. Including it lowers the stacked $c_0$ from $6.9$ to $6.6$ and the joint $\alpha_{\rm cm}$ from $0.03$ to $0.00$, and changes $\gamma_{\rm cm}$ by $0.003$ or less.

With the same data products, the two methods agree: with the joint fit's source selection and radial range, the stacked coupling is $0.072^{+0.042}_{-0.040}$, consistent with the joint $0.061\pm0.026$. The stricter source selection and the innermost stacked bins add about $0.02$ and $0.01$ to the stacked value, shifts within the uncertainty of the stacked fit.

A magnitude-gap dependence of the miscentering separates the two methods. With $f_{\rm mis}$ free per tercile, the stacked fit prefers the smallest miscentered fraction for the largest gaps ($0.23\pm0.12$ against $0.30$ and $0.33$), as expected if large-gap clusters are more often centred on the true potential minimum. Its $\gamma_{\rm cm}$ then falls to $0.052$, with an uncertainty three times larger ($\pm0.09$): the stacks cannot tell a gap-dependent centring from a gap-dependent concentration. The joint fit, which models every cluster's profile with its own gap, finds no such preference ($0.35^{+0.17}_{-0.14}$ against $0.34$ and $0.37$) and keeps $\gamma_{\rm cm}=0.059^{+0.051}_{-0.036}$, unchanged from its reference value. The same holds with $\sigma_{\rm off}$ also free per tercile. A magnitude-gap dependence of the centring therefore does not explain the joint coupling, and only the joint fit can bound it.

Split by redshift, the coupling differs between $0.1<z<0.3$ ($0.112^{+0.039}_{-0.037}$) and $0.3<z<0.5$ ($0.004\pm0.037$) by $2.0\sigma$, which may point to a redshift-dependent systematic or population difference that the model does not capture. The mass normalisations of the three bins, fit without redshift evolution, imply $\beta_{M\lambda}\approx-2.6$, consistent with the full fit: the strong evolution of mass at fixed richness is shared by all bins rather than driven by one.

Our analytic covariance (Eq.~\ref{eq:covariance}) is a simple shape-noise model, with a single dispersion $\sigma_e$ and independent galaxies, and it does not describe the noise of the real measurement. In the DES-like simulations, where shape noise is the only source of noise, the bootstrap and analytic variances of a cluster agree, but in the real data the bootstrap variance is $1.8$ times the analytic one at $0.25$~Mpc and $1.3$ times at $0.44$~Mpc, falling to the analytic value beyond $0.7$~Mpc, even with $\sigma_e$ estimated separately for classes of source weight (Appendix~\ref{app:analytic}). The real inner bins therefore carry noise that this model misses. We have not identified its source; it may be member galaxies and blending in the crowded cluster cores, the variation of the shear response between galaxies, or the scatter of the photometric redshifts that enter $\Sigma_{\rm crit}$, none of which the simulations contain. An analytic covariance that misses this noise overstates the information in the inner bins, which constrain the concentration and the miscentering, and a fit with it moves along their degeneracy to $f_{\rm mis}\approx0.5$ and $\sigma_{\rm off}\approx0.5$~Mpc, but leaves $\gamma_{\rm cm}$ within its uncertainty (Appendix~\ref{app:analytic}). The bootstrap follows the real noise, at the cost of a weighting bias in the sparsest bins (Appendix~\ref{app:recommend}) that the likelihood of Eq.~\ref{eq:shlike} keeps from reaching $\gamma_{\rm cm}$ in the simulations.

The choices of the analysis therefore matter little: apart from a gap-dependent miscentering, which removes the stacked constraint, no change of the analysis choices we tested moves either coupling beyond its uncertainty, although the coupling differs between redshift bins.

\section{Conclusion}
\label{sec:conclusion}

We have developed a joint fit of individual-cluster weak-lensing profiles under mass-richness and mass-concentration scaling relations, with the magnitude gap as an additional observable of the concentration, and compared it with a forward-modelled stacked analysis that fits the same relations, with the same model and priors, to stacked profiles, and with the standard two-step analysis of the same stacks. In simulations at DES-like and LSST-like density, the joint and forward-modelled fits recover the injected magnitude-gap coupling with uncertainties that match the scatter between realisations.

Of the three methods, the joint fit is the most precise in the simulations. Its estimates scatter less between realisations than the forward-modelled stacked ones, by $1.2$--$1.3\times$ in $\gamma_{\rm cm}$ and $1.1$--$1.5\times$ on the other scaling-relation slopes, with neither method biased beyond a few per cent. The standard two-step analysis scatters $2$--$6.5\times$ more than the joint fit in $\gamma_{\rm cm}$ and $1.5$--$5.3\times$ more on the other slopes, and is biased by 3 to 6 standard errors, except for $c_0$, in several mass-concentration and mass-richness parameters (Tables~\ref{tab:width_ratios} and~\ref{tab:twostep_sims}). On the real data, the joint fit reports $1.1$--$1.5\times$ smaller uncertainties on every slope. Data density matters more, both methods being five to six times more precise at LSST-like density (Table~\ref{tab:gamma_recovery}). The precision of the joint fit rests on the model, however: a dependence the model omits, such as a gap-dependent mass, enters $\gamma_{\rm cm}$ in both methods but with the joint fit's smaller uncertainty, while only the joint fit keeps its constraint under a gap-dependent miscentering.

Applied to WaZP clusters with DES~Y3 shears, our main results are:
\begin{itemize}
  \item The joint and forward-modelled fits agree on the magnitude-gap coupling of the concentration once the mass at fixed richness may also depend on the gap, with $\gamma_{\rm cm}=0.065\pm0.033$ (forward-modelled) and $0.007^{+0.032}_{-0.033}$~mag$^{-1}$ (joint); the joint value is consistent with no coupling and the forward-modelled one is $2\sigma$ from zero, so neither is a detection.
  \item Both fits prefer a positive magnitude-gap dependence of the mass at fixed richness, $\delta_{M\lambda}=0.035^{+0.016}_{-0.012}$ (joint) and $0.043\pm0.020$~dex~mag$^{-1}$ (forward-modelled), although the evidence of the stacked fit does not require the extra parameter. Since freeing $\delta_{M\lambda}$ removes most of both couplings, a coupling measured in a model where the magnitude gap enters only the concentration need not survive a modest extension of the model, in stacked and joint analyses alike.
  \item The mass-concentration slope is consistent with zero in both fits, $\alpha_{\rm cm}=0.033^{+0.052}_{-0.091}$ (joint) and $0.036\pm0.090$ (forward-modelled), as expected from the upturn of the \citet{diemer2019} relation over the sample's mass range; with that relation as the baseline of the concentration, the joint magnitude-gap coupling is unchanged.
  \item Both fits measure the mass-richness relation of WaZP clusters, with $\alpha_{M\lambda}=1.042^{+0.029}_{-0.045}$ and $\beta_{M\lambda}=-2.05^{+0.18}_{-0.15}$ (joint) and $0.918\pm0.048$ and $-1.73\pm0.25$ (forward-modelled), specific to the WaZP richness definition.
\end{itemize}

The analysis rests on simplifications that future work should remove:
\begin{itemize}
  \item Neither model includes intrinsic scatter about the mass-richness relation or the selection on richness \citep{mantz2010}, so the WaZP mass-richness parameters are not corrected for scatter or selection.
  \item The simulations contain no realistic cluster-finder selection or orientation bias \citep{dietrich2014}, no substructure and no line-of-sight projection of unrelated structures \citep{sunayama2020}; their redshifts are drawn uniformly rather than from the comoving volume and the halo abundance, and the fits receive the noiseless richness, so the assumption of no scatter at fixed richness holds exactly in them.
  \item Source and cluster redshifts and magnitude gaps are used without their uncertainties \citep{kelly2007}.
  \item The per-cluster covariances contain shape noise only \citep{hoekstra2003,gruen2015}, and the boost factors are estimated from source counts rather than from the redshift distribution of the sources \citep{varga2019}.
  \item Neither analysis accounts for correlated noise between clusters whose apertures overlap and share background sources: $92\%$ of the real clusters share part of their $5.3$~Mpc aperture with at least one other cluster, mostly in the outer bins, which affects mainly the mass normalisation and the two-halo term.
\end{itemize}
These will be addressed in two forthcoming papers: one with real data only, comparing WaZP and redMaPPer clusters and adding further concentration-related observables, and one with simulations only, testing a fully Bayesian hierarchical model of the scaling relations and of the selection processes.

Beyond these extensions, this work is intended as the first step of a larger effort to model galaxy cluster scaling relations in a full Bayesian, per-cluster framework, rather than through stacked or otherwise pre-averaged summaries. A natural extension is to make the causal structure between richness, magnitude gap, mass, and concentration explicit, rather than implicit in the choice of which relations to include, so that questions about formation history can be posed as causal questions and answered with the tools built for them, instead of being inferred indirectly from correlations that a purely predictive model cannot by itself distinguish from confounding.

\section*{Data availability}

The lensing measurement and modelling code, \textsc{xlensing}, is publicly available at \url{https://github.com/andrevitorelli/xlensing}. The joint and forward-modelled stacked fitting code, \textsc{destacking}, will be made publicly available at \url{https://github.com/andrevitorelli/destacking}. The DES~Y3 \textsc{metacalibration} shape catalogue and the DES~Y6 WaZP cluster catalogue used here are public and can be obtained from the DES data release website, \url{https://des.ncsa.illinois.edu/releases}.

\section*{Acknowledgments}

The authors would like to thank Giulia Cerini, Eric Huff, Katarina Markovič, Mariana Penna-Lima, and Sandro Vitenti for insightful discussions. This work was made possible by the support of several institutions and countries. The authors are grateful for funding from JPL's R\&TD 7x `Future of Dark Sector Cosmology: Systematic Effects and Joint Analysis' (01STRS/R.23.312.005), and FAPESP for the LSC infrastructure at the PMR EP-USP. ESC acknowledges support from CNPq grant 302592/2025-3 and FAPESP grant 2023/05085-6. This work made heavy use of the computing facilities of the Centro de Processamento de Dados do IAG/USP (CPD-IAG), whose purchase was made possible by the Brazilian agency FAPESP (grants 2019/25950-8,  2017/24954-4 and 2009/54006-4).

Due to the large scope of coding involved in this work, most of the software development was performed with the help of a large language model, specifically Claude Code (Anthropic). The xlensing package was developed before \citet{vitorelli2018} but highly improved in performance with LLM assistance. The code was then read, verified and tested by humans.

\bibliographystyle{apsrev4-1}

\bibliography{oja_template}

\begin{appendix}

\section{Simulating triaxial haloes}
\label{ap:triaxial}

This appendix details the method summarised in Section \ref{sec:triaxial}: drawing a triaxial halo shape and orientation for each simulated cluster, projecting it onto the sky, and computing the shear it induces on the background galaxies.

\subsection{Shape and orientation sampling}

Each halo is an NFW profile in the ellipsoidal radius
\begin{equation}
    m^2 = \frac{x^2}{\alpha^2} + \frac{y^2}{\beta^2} + \frac{z^2}{\gamma^2}\,,
    \qquad \alpha\beta\gamma = 1\,,
    \label{eq:ellradius}
\end{equation}
with $(x,y,z)$ in the halo's principal frame and $\alpha\geq\beta\geq\gamma$ proportional to its semi-axes $a\geq b\geq c$. The unit-determinant normalisation keeps the volume inside every isodensity surface equal to that of a sphere of the same radius, so the mass enclosed within $m=r_{200}$ is the halo's $M_{200}$, and its concentration keeps its spherical meaning \citep{oguri2003,corless2007}. This $M_{200}$, the truth against which the fits are compared, is the mass within the ellipsoid, not a spherical-overdensity mass: for the axis ratios drawn here, the mass within the sphere $r<r_{200}$ is lower by $2.8\%$ on average ($0.012$~dex), and the mass within the radius where the enclosed mean density equals $200\bar\rho_m$ is lower by about $0.017$~dex for $c_{200}=4$. The axis ratios are drawn as
\begin{equation}
    b/a \sim \mathrm{Uniform}[0.5,\,1.0]\,, \qquad c/a \sim \mathrm{Uniform}[0.3,\,b/a]\,,
\end{equation}
with the second draw conditioned on the first so that $c \leq b$ by construction. These ranges are chosen to mimic the axis-ratio distributions found for CDM haloes in $N$-body simulations \citep[e.g.][]{jingsuto2002}. The line-of-sight direction $\mathbf{n}$ is drawn independently and isotropically in the halo's principal frame, $\cos\theta_{\rm los} \sim \mathrm{Uniform}[0,\,1]$ and $\phi_{\rm los} \sim \mathrm{Uniform}[0,\,2\pi]$, so no orientation is preferred over any other; Section \ref{sec:SimResults} lists the caveat that this does not include the orientation selection a cluster finder may introduce.

\subsection{Projection onto the sky}

With $Q = \mathrm{diag}(\alpha^{-2}, \beta^{-2}, \gamma^{-2})$, a point $\mathbf{x}$ has $m^2=\mathbf{x}^\top Q\,\mathbf{x}$. Writing a point along the line of sight as $\mathbf{x}_\perp + s\,\mathbf{n}$ and completing the square in $s$ gives $m^2 = (\mathbf{n}^\top Q\,\mathbf{n})\,s'^2 + \xi^2$, with
\begin{equation}
\begin{split}
    \xi^2 &= \mathbf{x}_\perp^\top M\,\mathbf{x}_\perp\,,\\
    M_{ij} &= \mathbf{e}_i^\top Q\, \mathbf{e}_j - \frac{(\mathbf{e}_i^\top Q\, \mathbf{n})(\mathbf{e}_j^\top Q\, \mathbf{n})}{\mathbf{n}^\top Q\, \mathbf{n}}\,, \quad i,j \in \{1,2\}\,,
\end{split}
\end{equation}
in an orthonormal sky-plane basis $(\mathbf{e}_1, \mathbf{e}_2)$: $M$ is the Schur complement of $Q$ along $\mathbf{n}$. Integrating the density along the line of sight then gives an exactly elliptical surface density,
\begin{equation}
    \Sigma(\mathbf{x}_\perp) = \frac{\Sigma_{\rm NFW}(\xi)}{\sqrt{\mathbf{n}^\top Q\,\mathbf{n}}}\,,
    \label{eq:triaxsigma}
\end{equation}
where $\Sigma_{\rm NFW}$ is the surface density of the spherical halo with the same $M_{200}$ and $c_{200}$. The factor $(\mathbf{n}^\top Q\,\mathbf{n})^{-1/2}$ is the line-of-sight projection amplitude: it exceeds unity when the halo is elongated along the line of sight and falls below it when the halo is viewed along its short axis. Since $\det M\;\mathbf{n}^\top Q\,\mathbf{n} = \det Q = 1$, the total projected mass is the same for every orientation. Diagonalising $M$, with eigenvalues $\lambda_1 \leq \lambda_2$, gives the projected axis ratio $q = (\lambda_1/\lambda_2)^{1/2}$ and the position angle of the projected major axis. This projection recovers the expected limits: a spherical halo projects to a circle of unit amplitude regardless of orientation, and a halo viewed along one of its principal axes projects to the ratio of the other two. It is the projection used to build the $q$ distribution of Fig.~\ref{fig:triaxiality_demo}.

\begin{figure}
    \centering
    \includegraphics[width=1.0\linewidth]{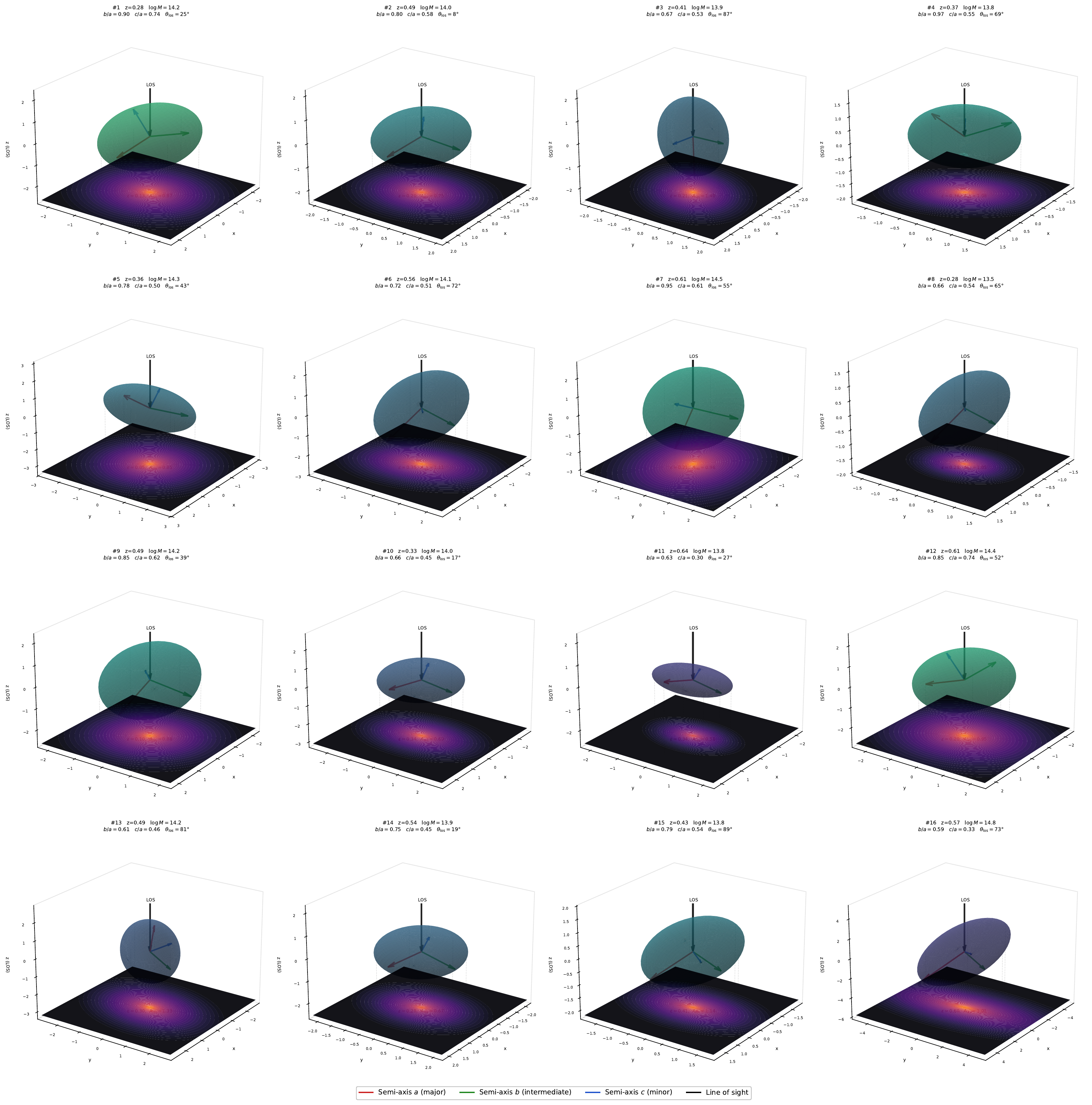}
    \caption{Sixteen example haloes drawn from the shape and orientation sampling described in this appendix, each showing the $m=r_{200}$ isodensity ellipsoid with its principal axes and line-of-sight direction (top) and the projected lensing convergence $\kappa$ of Eq.~\ref{eq:triaxsigma} (bottom), including the line-of-sight amplitude. Labels give the sky-plane ellipticity $\varepsilon=1-q$ and the amplitude $A_{\rm los}=(\mathbf{n}^\top Q\,\mathbf{n})^{-1/2}$.}
    \label{fig:triaxiality_examples}
\end{figure}

Figure \ref{fig:triaxiality_examples} shows sixteen individual haloes drawn from this procedure, with their 3D ellipsoid and projected convergence map.

\subsection{Shearing the background catalogue}

In the frame of the projected major ($u$) and minor ($v$) axes, the convergence of Eq.~\ref{eq:triaxsigma} depends on position only through $\xi_K^2 = u^2 + v^2/q^2$. For any such elliptical convergence $\kappa(\xi_K)$, the second derivatives of the lensing potential $\psi$ reduce to one-dimensional integrals \citep{schramm1990,keeton2001},
\begin{equation}
\begin{split}
    \psi_{uu} &= 2q\,u^2 K_0 + q\,J_0\,, \qquad
    \psi_{vv} = 2q\,v^2 K_2 + q\,J_1\,,\\
    \psi_{uv} &= 2q\,u\,v\,K_1\,,\\
    J_n &= \int_0^1 \frac{\kappa\!\left(\xi^2(w)\right)\,\mathrm{d}w}{\left[1-(1-q^2)w\right]^{n+1/2}}\,,\\
    K_n &= \int_0^1 \frac{w\,\kappa'\!\left(\xi^2(w)\right)\,\mathrm{d}w}{\left[1-(1-q^2)w\right]^{n+1/2}}\,,
\end{split}
\end{equation}
with $\xi^2(w) = w\left[u^2 + v^2/\left(1-(1-q^2)w\right)\right]$ and $\kappa' = \mathrm{d}\kappa/\mathrm{d}(\xi_K^2)$. The shear is $\gamma_1 = (\psi_{uu}-\psi_{vv})/2$, $\gamma_2 = \psi_{uv}$, rotated from the $(u,v)$ frame to the sky. Because the NFW profile has the same shape at every mass once lengths are measured in units of its scale radius, the dimensionless shear is computed once on a grid of scaled radius, position angle, and axis ratio, and interpolated for each galaxy; the interpolation reproduces the direct integrals to a median of $4\times10^{-4}$ in relative terms. The shear is then multiplied by $\Sigma_{\rm NFW}$'s normalisation, the line-of-sight amplitude, and $1/\Sigma_{\rm crit}$ for the galaxy's redshift, and added to its intrinsic ellipticity. Where a per-galaxy shear response is available (Section \ref{sec:Sims}), the injected shear is scaled by that response, following the metacalibration convention $\delta e = R\,\delta\gamma$, so that a response-corrected estimator recovers the injected signal unbiased.

We validated the implementation in four ways: for $b=c=a$ it reproduces the spherical NFW shear; the azimuthally averaged tangential shear on any circle equals the mean convergence inside it minus the convergence on it, as it must for any mass distribution; the shear field agrees point by point with a Fourier-space inversion of the same convergence map; and the cross shear vanishes on the projected principal axes.

\begin{figure*}
    \centering
    \includegraphics[width=1.0\linewidth]{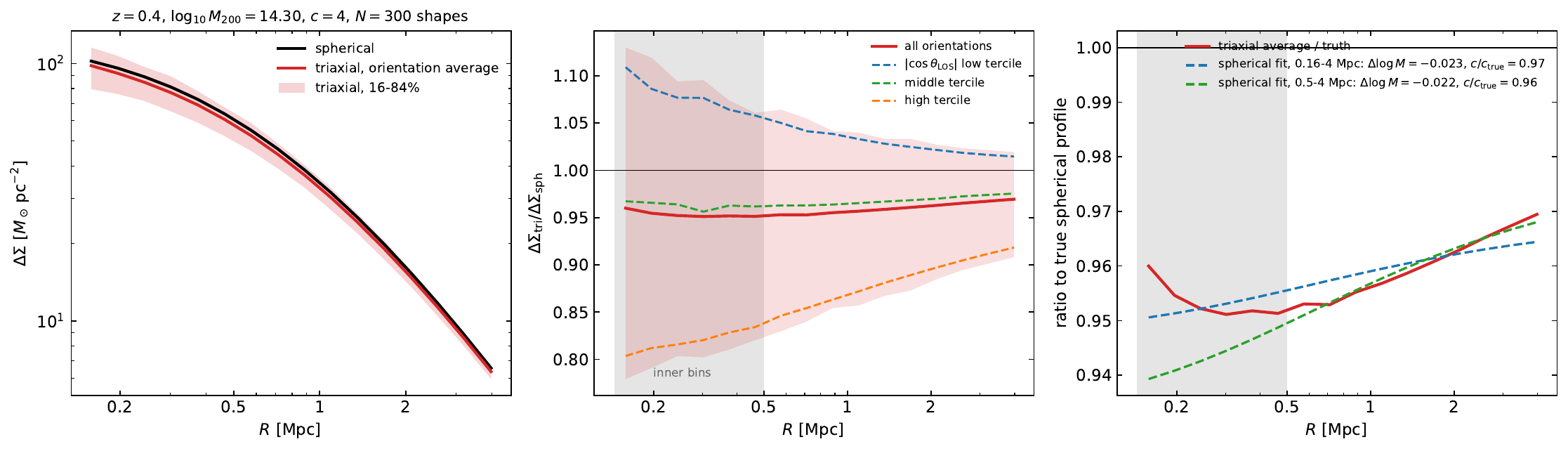}
    \caption{Effect of triaxiality on the measured $\Delta\Sigma$ of one representative halo ($z=0.4$, $M_{200}=2\times10^{14}\,M_\odot$, $c_{200}=4$), from 300 random shapes and isotropic orientations sheared through the full measurement pipeline at zero shape noise. Left: spherical profile and the orientation-averaged triaxial profile with its 16--84\% range. Middle: ratio of the triaxial to the spherical profile, for all orientations and by tercile of $|\cos\theta_{\rm los}|$ (large values: line of sight close to the short axis); the shaded region marks the bins inside $0.5$~Mpc. Right: the orientation-averaged ratio, and spherical NFW fits to the averaged profile over $0.16$--$4$ and $0.5$--$4$~Mpc.}
    \label{fig:triaxiality_dsigma}
\end{figure*}

Figure~\ref{fig:triaxiality_dsigma} shows the effect on the measured $\Delta\Sigma$, for one representative halo ($z=0.4$, $M_{200}=2\times10^{14}\,M_\odot$, $c_{200}=4$) sheared through the full measurement pipeline in 300 random shapes and orientations. Individual orientations differ from the spherical profile by up to $\sim\!10$--$20\%$, most strongly in the innermost bins: a halo elongated along the line of sight has a higher signal, and one viewed along its short axis a lower one. Averaged over isotropic orientations the signal is $\approx5\%$ below the spherical one, nearly independent of radius, so a spherical NFW fit to the average profile recovers a mass $0.023$~dex low and a concentration $3\%$ low, whether or not the bins inside $0.5$~Mpc are included.

\section{Radial binning choices}
\label{app:binning}

The radial grid enters the analysis in three ways: how many clusters contribute to each bin, whether the model is evaluated at the bin centre or averaged over the annulus, and how much information is lost if the innermost bins are cut. We summarise each.

\begin{figure}
    \centering
    \includegraphics[width=0.9\linewidth]{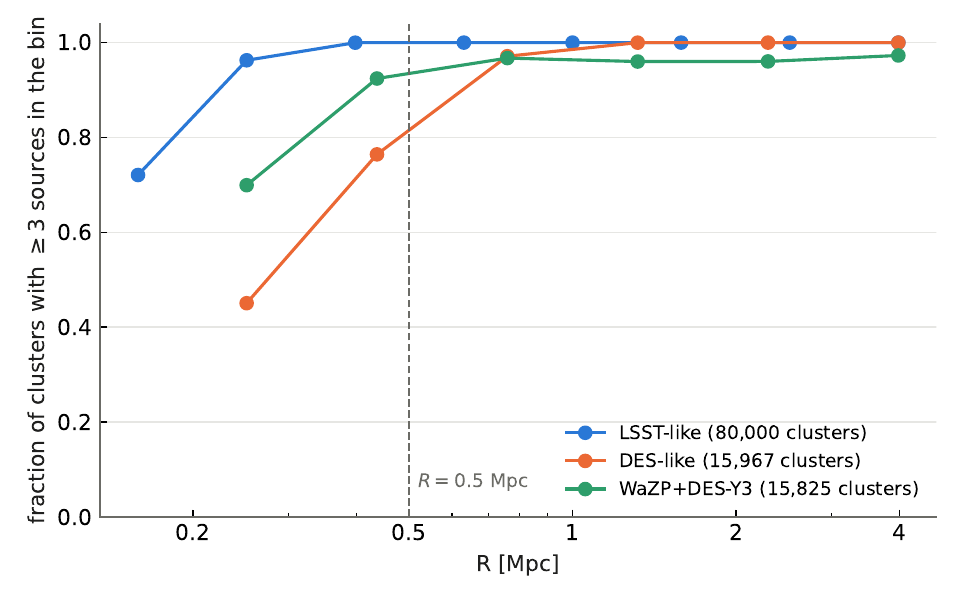}
    \caption{Fraction of clusters with at least three background sources in each radial bin, for the LSST-like and DES-like simulated samples and the real WaZP+DES-Y3 sample (15{,}825 clusters with a positive magnitude gap). The dashed line marks $0.5$~Mpc.}
    \label{fig:bin_occupancy}
\end{figure}

\paragraph{Bin occupancy.} In every bin, a cluster's measurement with the analytic covariance is used only if the bin holds at least three background sources; otherwise the bin is masked for that cluster (the bootstrap covariance applies the slightly looser criterion of Section~\ref{sec:covariance}). Figure~\ref{fig:bin_occupancy} shows the fraction of clusters that pass this criterion. At LSST-like density the innermost bin ($0.16$~Mpc) is valid for $72\%$ of the clusters and the next for $96\%$. At DES-like density the two innermost bins of the 6-bin grid ($0.25$ and $0.44$~Mpc) are valid for only $45\%$ and $76\%$ of the clusters, and the real WaZP+DES-Y3 sample is better, with $70\%$ and $92\%$ at those radii and $96$--$97\%$ beyond. Because the inner bins are those that constrain the concentration, the grid is matched to what the source density supports: $8$ bins from $0.16$ to $3.98$~Mpc for LSST-like data and $6$ bins from $0.25$ to $3.98$~Mpc for DES-like and real data.

\paragraph{Cost of cutting the inner bins.} Excluding all bins below $0.5$~Mpc would discard most of the inner, concentration-sensitive information. Table~\ref{tab:fisher_cut} gives the resulting increase of the Fisher uncertainty on $\gamma_{\rm cm}$ and $c_0$, with the other parameters held fixed. We therefore keep every bin and rely on the per-cluster masking and, for the bootstrap covariance, on the variance floor of Section~\ref{sec:covariance}.

\begin{table*}
\centering
\caption{Factor by which the Fisher uncertainty grows when all bins below $0.5$~Mpc are removed, other parameters held fixed, for the miscentered model at the injected (simulations, one realisation each) or fitted (real data) parameters, with the per-cluster variances of each data set.}
\label{tab:fisher_cut}
\begin{tabular}{lccc}
\toprule
 & LSST-like (3 bins removed) & DES-like (2 bins) & Real data (2 bins) \\
\midrule
$\sigma(\gamma_{\rm cm})$ & $1.71$ & $1.66$ & $1.89$ \\
$\sigma(c_0)$ & $1.54$ & $1.51$ & $1.83$ \\
\bottomrule
\end{tabular}
\end{table*}

\begin{figure*}
    \centering
    \includegraphics[width=0.9\textwidth]{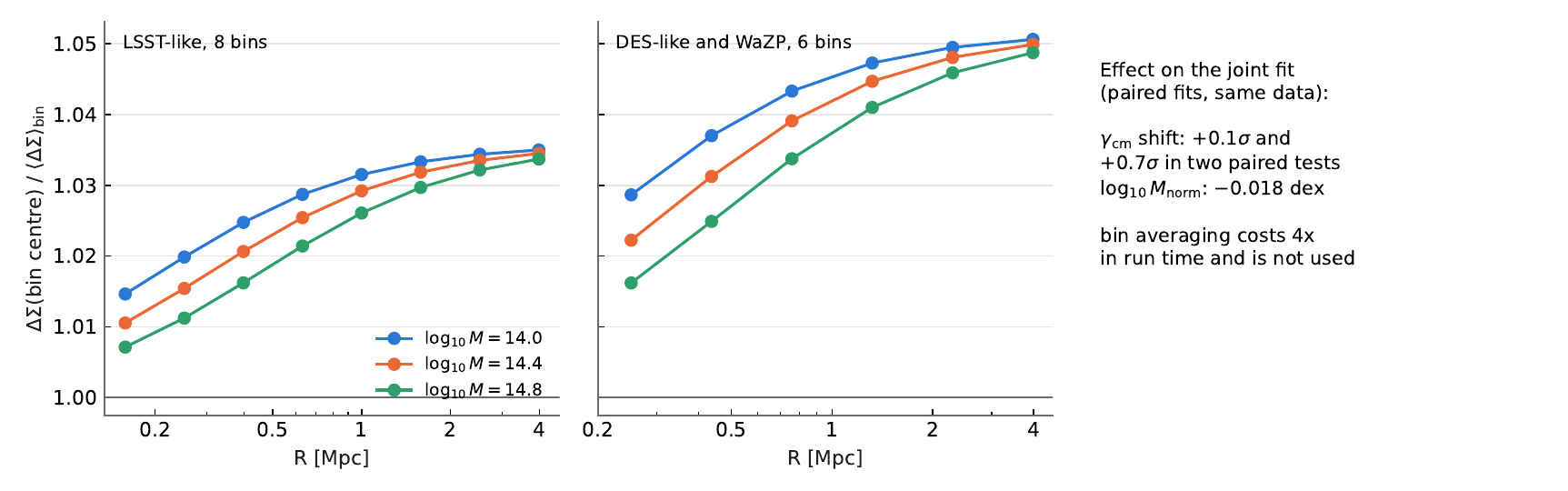}
    \caption{Ratio of the bin-centre model to the bin-averaged model $\Delta\Sigma$ for the LSST-like (left) and DES-like and WaZP (middle) radial grids, for three halo masses at $z=0.4$ with a miscentering offset of $0.47$~Mpc, and a summary of the effect on the joint fit (right).}
    \label{fig:bin_centre_vs_average}
\end{figure*}

\paragraph{Bin centre versus bin average.} Figure~\ref{fig:bin_centre_vs_average} shows the ratio of the model $\Delta\Sigma$ evaluated at the logarithmic bin centre to its area-weighted average over the annulus, for a miscentered NFW halo at $z=0.4$. The bin-centre value is high by $0.7$--$3.5\%$ on the 8-bin grid and $1.6$--$5\%$ on the 6-bin grid, increasing with radius and decreasing slightly with mass. Fitting identical simulated data with the exact bin average and with the bin-centre model changes $\gamma_{\rm cm}$ by only $0.1\sigma$ and $0.7\sigma$ in two paired tests; the difference is absorbed by the mass normalisation, which the bin-centre model places $\approx0.018$~dex lower. Bin averaging costs four times more per likelihood evaluation, so the analysis uses the bin-centre model, and the logarithmic mass normalisations quoted in this paper carry that offset. Its size depends on the grid: weighting all bins equally, the bin-centre model is high on average by $4.0\%$ on the 6-bin grid, $2.6\%$ on the 8-bin grid and $0.6\%$ on the 16-bin grid of the stacks, which lowers the fitted mass normalisation by about $0.027$, $0.019$ and $0.004$~dex, consistent with the $0.018$~dex measured in the paired fits. The joint and stacked mass normalisations therefore carry different offsets, of which the difference is about $0.02$~dex.

\section{Magnitude-gap dependence of the boost factor}
\label{app:boostdm}

\begin{figure}
    \centering
    \includegraphics[width=1.0\linewidth]{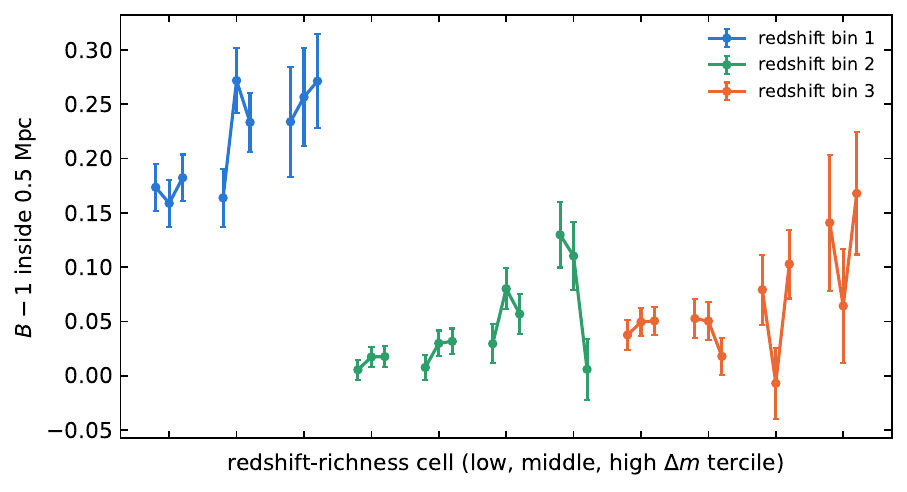}
    \caption{Boost excess $B-1$ inside $0.5$~Mpc (inverse-variance mean over the radial bins) of the low, middle and high magnitude-gap tercile of each redshift-richness cell of the stacked analysis ($f=1.2$ background selection), for the 11 cells in which all three terciles have at least 50 members; colours mark the redshift bin.}
    \label{fig:boost_vs_dm}
\end{figure}

The boost correction removes the dilution of the signal by cluster members and foreground galaxies that pass the background selection, and a magnitude-gap dependence of that contamination at fixed richness and redshift would enter $\gamma_{\rm cm}$ directly. Figure~\ref{fig:boost_vs_dm} shows the boost excess $B-1$ inside $0.5$~Mpc for the three magnitude-gap terciles of each redshift-richness cell of the stacked analysis. Within each cell, the high-gap tercile exceeds the low-gap one by $0.010\pm0.007$ on average (inverse-variance weighted over the 11 cells with three usable terciles), against a typical $B-1\approx0.06$, so the contamination depends at most weakly on the magnitude gap, and the two magnitude-gap bins of the boost fits used by the joint fit (Section~\ref{sec:boost}) are sufficient.

\section{A paired test of per-cluster covariances}
\label{app:recommend}

\begin{figure*}
    \centering
    \includegraphics[width=0.9\textwidth]{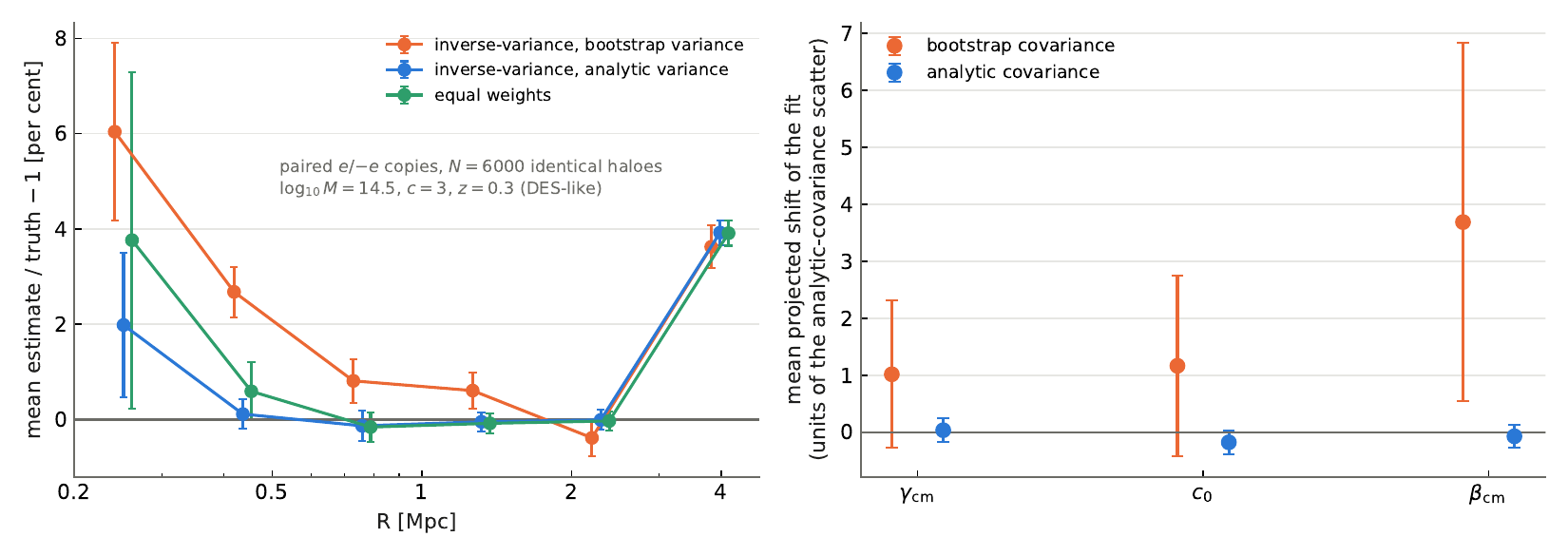}
    \caption{Left: bias of the mean $\Delta\Sigma$ estimate, relative to the true profile, from paired $e$/$-e$ simulations of identical haloes, for three weightings (see text); the outermost bin is truncated by the $5.0$~Mpc source aperture of the test. Right: mean projected shift of $\gamma_{\rm cm}$, $c_0$ and $\beta_{\rm cm}$ over $24$ DES-like realisations with centred, spherical haloes, for bootstrap and analytic per-cluster covariances, in units of the realisation-to-realisation scatter of the analytic case; error bars are standard errors of the mean.}
    \label{fig:paired_cov}
\end{figure*}

The analytic covariance (Eq.~\ref{eq:covariance}) cannot respond to the fluctuation of a cluster's own data, but the bootstrap covariance can: for a ratio estimator such as Eq.~\ref{eq:deltasigma}, responses and critical densities vary between galaxies and sheared shapes have a skewed distribution, so a bin that fluctuates high tends to receive a smaller bootstrap variance and therefore a larger weight. To test an estimator and its covariance on identical haloes, simulate each halo's sources twice, once with intrinsic ellipticities $e$ and once with all of them reversed, $-e$, through the same pipeline. The shear signal is the same in both copies, while the shape noise has opposite sign, so averaging the two cancels it to first order, and a residual deviation of the average from the true profile is a bias of the estimator or of the weighting. Figure~\ref{fig:paired_cov} (left) applies this to $6000$ identical DES-like haloes of $\log_{10}M=14.5$, $c=3$ at $z=0.3$. Inverse-variance means with the per-cluster bootstrap variances are biased high by $6.0\pm1.9\%$ at $0.25$~Mpc, $2.7\pm0.5\%$ at $0.44$~Mpc and $0.8\pm0.5\%$ at $0.76$~Mpc, while the same means with the analytic variance are consistent with zero beyond $0.25$~Mpc ($+2.0\pm1.5\%$, $+0.1\pm0.3\%$, $-0.1\pm0.3\%$ in the first three bins). The outermost bin is high by $\approx3.9\%$ for every weighting, because the test collects sources only within $5.0$~Mpc and so truncates that bin; it is unrelated to the covariance. Errors are from a 20-block jackknife.

The same difference propagates to the fit. For $24$ DES-like realisations with centred, spherical haloes, with the first-order shift of a fit of $c_0$, $\alpha_{\rm cm}$, $\beta_{\rm cm}$, $\gamma_{\rm cm}$ and the mass normalisation (the other parameters held at their true values) computed without refitting, bootstrap weights give a mean shift of $+0.13\pm0.16$~mag$^{-1}$ in $\gamma_{\rm cm}$ (realisation-to-realisation scatter $0.80$), $+0.7\pm1.0$ in $c_0$ and $+1.8\pm1.5$ in $\beta_{\rm cm}$, against $+0.004\pm0.026$ ($0.13$), $-0.11\pm0.12$ and $-0.03\pm0.10$ with the analytic variance (Figure~\ref{fig:paired_cov}, right). These test realisations have narrower magnitude gaps than the simulations of Section~\ref{sec:Sims} (standard deviation $0.16$~mag), so $\gamma_{\rm cm}$ is poorly determined in them; what the test shows is that the bootstrap weights multiply the scatter of every parameter several times over, while the analytic weights leave it at the shape-noise level. The test needs no fit and is applicable to any lensing pipeline that can regenerate its sources with reversed ellipticities. Full refits of the simulations of Section~\ref{sec:SimResults} with their bootstrap covariances, which test the configuration used for the real data, are described in Appendix~\ref{app:analytic}.

\section{Choice of covariance for the real data}
\label{app:analytic}

\begin{figure}
    \centering
    \includegraphics[width=1.0\linewidth]{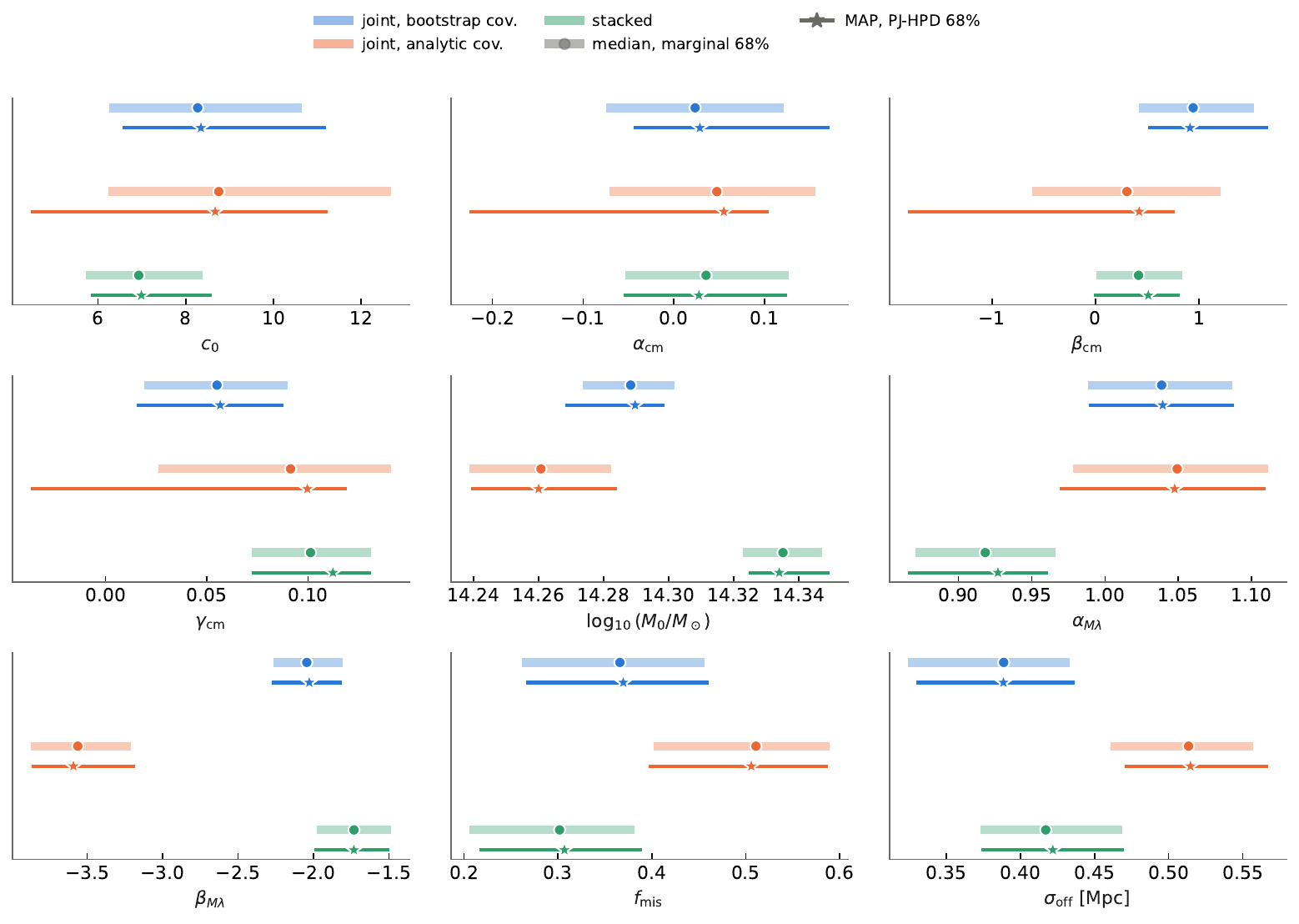}
    \caption{Marginal posterior median with its $68\%$ interval (thick bar and circle) and MAP with its PJ-HPD $68\%$ interval (thin bar and star) for each parameter, for the joint fit with bootstrap covariance, the joint fit with analytic covariance and the forward-modelled stacked fit of the real WaZP+DES-Y3 data.}
    \label{fig:map_vs_median}
\end{figure}

Repeating the real-data joint fit with the analytic covariance moves the degenerate parameters but not $\gamma_{\rm cm}$. For this fit we use the analytic covariance of Eq.~\ref{eq:covariance}, with the direct estimator of Eq.~\ref{eq:deltasigma} as the data vector and $\sigma_e$ estimated separately for $50$ quantile classes of the source weight ($0.155$--$0.224$, against $0.195$ for the whole selected DES~Y3 sample; in the response-uncorrected units of Eq.~\ref{eq:deltasigma}). The bootstrap variance is larger than the analytic one, most of all in sparse inner bins and at high redshift, so the two estimates weight the clusters differently. Repeating the fit with the analytic covariance leaves $\gamma_{\rm cm}$ and $\alpha_{M\lambda}$ unchanged within their uncertainties, but moves the posterior along the degeneracy between the concentration normalisation and the miscentering parameters, towards higher $c_0$, $f_{\rm mis}$ and $\sigma_{\rm off}$, and steepens the redshift slope of the mass-richness relation (Appendix~\ref{app:map}, Figure~\ref{fig:map_vs_median}).

The shift is caused by the weighting of the individual clusters, not by the priors: it is the same with flat miscentering priors, with Gaussian miscentering priors centred on the stacked fit (used for the analytic-covariance fit shown in Appendix~\ref{app:map}) and with either $c_0$ prior, while changing the priors with the bootstrap covariance does not produce it. The analytic covariance gives relatively more weight to high-redshift, high-richness clusters, and the shift is driven by their profiles at intermediate radii ($\sim1$~Mpc), not by the innermost bins. The likelihood along $c_0$ is shallow for both covariance estimators, with $f_{\rm mis}$ rising and $\alpha_{\rm cm}$ falling along the valley, so the two estimators select different points of a weakly constrained direction. We conclude that $c_0$, $f_{\rm mis}$ and $\alpha_{\rm cm}$ are degenerate and weakly constrained by the individual profiles, so their values should not be given a physical interpretation, and that $\beta_{M\lambda}$ depends on how high-redshift, high-richness clusters are weighted; $\gamma_{\rm cm}$ and $\alpha_{M\lambda}$ do not. The maximum a posteriori points and projected joint highest-posterior-density intervals \citep{joachimi2021} of both fits are given in Appendix~\ref{app:map}; they do not change this picture.

The paired test of Appendix~\ref{app:recommend} shows that the bootstrap covariance correlates with the measured signal, so that inverse-variance weights bias the mean profile high in the sparse inner bins, and that the variance floor inside $0.5$~Mpc does not remove this. We nevertheless use the bootstrap covariance for the real data, for two reasons.

First, the analytic covariance does not describe the noise of the real measurement. In the DES-like simulations, where shape noise is the only source of noise, the median ratio of the bootstrap to the analytic variance of a cluster is $0.95$--$1.00$ in every bin; in the real data it is $1.80$ at $0.25$~Mpc and $1.32$ at $0.44$~Mpc, falling to $0.93$--$1.02$ beyond $0.7$~Mpc, even with $\sigma_e$ estimated separately for each class of source weight. The real inner bins therefore scatter $30$--$80\%$ more than shape noise predicts, so the analytic covariance overstates the information in the bins that constrain $c_0$, $f_{\rm mis}$ and $\sigma_{\rm off}$, and its posteriors for them are too narrow. The resulting fit places half of the clusters miscentered by $\approx0.5$~Mpc ($f_{\rm mis}=0.51$ and $\sigma_{\rm off}=0.51$~Mpc, against $0.30$ and $0.42$~Mpc from the stacks), which we regard as a symptom of the misspecified covariance rather than a physical result.

Second, the bootstrap covariance, combined with the likelihood of Eq.~\ref{eq:shlike}, recovers $\gamma_{\rm cm}$ without bias in the simulations. Refitting all $84$ DES-like and $60$ LSST-like realisations (the $48$ of Section~\ref{sec:SimResults} and twelve more at $\gamma_{\rm cm}^{\rm true}=5/36$ and $6/36$) with their galaxy-bootstrap covariances (floored inside $0.5$~Mpc, as for the real data) gives recovered-against-injected slopes of $0.96\pm0.10$ and $0.94\pm0.02$, against $0.98\pm0.09$ and $0.95\pm0.02$ with the analytic covariance, and shifts $\gamma_{\rm cm}$ relative to the analytic-covariance fit of the same realisation by $+0.04$ and $+0.09$ of its uncertainty on average; the scatter between realisations matches the posterior widths. As on the real data, the covariance moves the degenerate parameters more: at LSST-like density $\beta_{\rm cm}$, $\sigma_{\rm off}$ and $c_0$ shift by $+1.0$, $+0.6$ and $+0.5$ of their uncertainties, and by $0.3$ or less at DES-like density. With a Gaussian likelihood instead, the same DES-like fits are dominated by a few clusters whose bootstrap variance collapses in the first unfloored bin, and their scatter exceeds the posterior widths several times over. The likelihood of Eq.~\ref{eq:shlike}, which marginalises over the uncertainty of each estimated variance, is therefore required whenever the bootstrap covariance is used.

\section{Two-step versus forward-modelled stacking}
\label{app:twostep}

The forward-modelled stacked fit of Section~\ref{sec:stackmodel} replaces the common two-step analysis of stacked profiles. To compare them, we also perform the two-step analysis on the same real WaZP+DES-Y3 stacks and on the stacks of every simulated realisation of Section~\ref{sec:SimResults} (in the simulations without the two-halo term). In the first step each stack is fit on its own with an NFW profile plus miscentering and the two-halo term, with its mass, concentration, $f_{\rm mis}$ and $\sigma_{\rm off}$ free under flat priors, at the stack's mean redshift (these are the per-stack values of Figure~\ref{fig:stacked_cm_dm}). In the second step the per-stack values are regressed, by weighted least squares, on the stacks' mean observables,
\begin{align}
\log_{10}M_s ={}& \log_{10}M_0 + \alpha_{M\lambda}\log_{10}\frac{\bar\lambda_s}{\lambda_{\rm piv}}\nonumber\\
&+ \beta_{M\lambda}\log_{10}\frac{1+\bar z_s}{1+z_{\rm piv}}\,,\\
\log_{10}c_s ={}& \log_{10}c_0 + \alpha_{\rm cm}\log_{10}\frac{M_s}{M_{\rm piv}}\nonumber\\
&+ \beta_{\rm cm}\log_{10}\frac{1+\bar z_s}{1+z_{\rm piv}} + \gamma_{\rm cm}\,\overline{\Delta m}_s\,,
\end{align}
with the fitted $\log_{10}M_s$ as the regressor of the second relation.

Table~\ref{tab:twostep} compares the two methods.

\begin{table}[H]
\centering
\caption{Two-step and forward-modelled stacked fits of the same 33 real stacks. Two-step: weighted-least-squares coefficients and their errors; forward-modelled: posterior medians and 68\% half-widths (Section~\ref{sec:stackedresults}).}
\label{tab:twostep}
\begin{tabular}{lcc}
\toprule
Parameter & Two-step & Forward-modelled \\
\midrule
$\gamma_{\rm cm}$ [mag$^{-1}$] & $0.074\pm0.075$ & $0.101\pm0.030$ \\
$\alpha_{\rm cm}$ & $-0.013\pm0.188$ & $0.036\pm0.090$ \\
$\beta_{\rm cm}$ & $0.89\pm0.92$ & $0.42\pm0.41$ \\
$c_0$ & $6.8\pm1.5$ & $6.9\pm1.3$ \\
$\alpha_{M\lambda}$ & $0.958\pm0.054$ & $0.918\pm0.048$ \\
$\beta_{M\lambda}$ & $-1.76\pm0.31$ & $-1.73\pm0.25$ \\
$\log_{10}(M_0/M_\odot)$ & $14.337\pm0.012$ & $14.335\pm0.012$ \\
\bottomrule
\end{tabular}
\end{table}

On the real data the two methods agree on every central value, and the forward-modelled fit constrains the concentration slopes two to three times more tightly; the mass-richness parameters are similar. The forward model does not extract more signal from the same profiles; the two-step analysis discards some of it, while the forward model makes stronger assumptions that let it use the rest:
\begin{enumerate}
\item \emph{Fewer free parameters.} The two-step analysis gives every stack its own mass, concentration and miscentering ($4\times33$ parameters), and within one stack the concentration and the miscentering are strongly degenerate, so each stack's concentration is poorly determined (Figure~\ref{fig:stacked_cm_dm}). The forward model has nine parameters in total; $f_{\rm mis}$ and $\sigma_{\rm off}$ are shared by all stacks and constrained by all of them, which breaks the per-stack degeneracy. This is the largest gain.
\item \emph{Mass tied across stacks.} In the forward model each member's mass follows from the mass-richness relation constrained by all stacks together, which removes the per-stack correlation between mass and concentration.
\item \emph{A lossy summary.} The regression keeps only a median and a symmetric error per stack, ignores the correlation between the fitted mass and concentration of a stack, and treats the fitted mass as an exact regressor, while the per-stack posteriors are skewed. The forward model also uses every member's own redshift, richness and magnitude gap rather than one mean value per stack.
\end{enumerate}
In the simulations the two-step analysis is unbiased in $\gamma_{\rm cm}$ but far less precise, and biased in other parameters (Table~\ref{tab:twostep_sims}). Its recovered-against-injected slope of $\gamma_{\rm cm}$ is $1.08\pm0.14$ (LSST-like) and $1.01\pm0.19$ (DES-like), and its scatter between realisations is $6.5$ and $2.0$ times that of the joint fit. Its redshift slope of the mass-concentration relation, however, is off by $-0.35\pm0.07$ (LSST-like) and $+0.44\pm0.10$ (DES-like). At LSST-like density its concentration normalisation is off by $-0.87\pm0.08$, and at DES-like density its mass-concentration slope and the redshift slope of the mass-richness relation are off by $-0.075\pm0.022$ and $+0.35\pm0.06$. Apart from $c_0$, these are 3 to 6 standard errors, while the joint and forward-modelled fits show no comparable bias (Table~\ref{tab:sim_bias}). The bias arises in the first step. Replacing each stack's fitted mass and concentration by the true mean $\log_{10}M$ and $\log_{10}c$ of its members recovers every parameter, while using the members' mean observables instead of the stacks' means in the regression changes nothing. The single-halo fits return masses $0.030$ (LSST-like) and $0.036$~dex (DES-like) and concentrations $0.07$ and $0.03$~dex below the members' means, with a mass offset that varies with redshift at DES-like density and biases $\beta_{M\lambda}$. This is the difference between the mass of a stacked profile and the mean mass of its members that calibrations of two-step analyses correct with simulations \citep{mcclintock2019}; the forward-modelled and joint fits avoid it by modelling every member. The reported weighted-least-squares errors match the scatter at LSST-like density and are $1.4$--$1.5$ times too large on the mass-concentration parameters at DES-like density, consistent with the larger two-step to joint ratio of reported uncertainties on the real data (Table~\ref{tab:width_ratios}).

\begin{table*}
\centering
\caption{Two-step analysis of the simulated realisations: mean bias (fit minus truth) $\pm$ its standard error, and the ratio of its scatter between realisations to that of the joint fit of the same realisations. LSST-like: 48 realisations; DES-like: 84.}
\label{tab:twostep_sims}
\begin{tabular}{lrcrc}
\toprule
 & \multicolumn{2}{c}{LSST-like} & \multicolumn{2}{c}{DES-like} \\
Parameter & Bias & Scatter ratio & Bias & Scatter ratio \\
\midrule
$\gamma_{\rm cm}$ [mag$^{-1}$] & $+0.005\pm0.008$ & $6.50$ & $-0.019\pm0.010$ & $2.03$ \\
$\alpha_{\rm cm}$ & $+0.017\pm0.014$ & $5.14$ & $-0.075\pm0.022$ & $1.62$ \\
$\beta_{\rm cm}$ & $-0.35\pm0.07$ & $5.27$ & $+0.44\pm0.10$ & $1.49$ \\
$c_0$ & $-0.87\pm0.08$ & $2.52$ & $+0.28\pm0.14$ & $0.85$ \\
$\log_{10}M_0$ & $-0.034\pm0.001$ & $1.22$ & $-0.011\pm0.003$ & $0.96$ \\
$\alpha_{M\lambda}$ & $+0.002\pm0.005$ & $1.76$ & $+0.026\pm0.017$ & $1.80$ \\
$\beta_{M\lambda}$ & $-0.015\pm0.011$ & $1.71$ & $+0.35\pm0.06$ & $1.59$ \\
\bottomrule
\end{tabular}
\end{table*}

The extra precision therefore rests on the assumptions that the miscentering is common to all stacks and that the mass is set by richness and redshift alone. When these hold, the forward model is the full likelihood of the stacked data and the two-step analysis a degraded version of it; when they do not, the extra precision can turn into bias. Section~\ref{sec:dmr} shows an example: tying the mass to richness alone forces a magnitude-gap dependence of the mass into $\gamma_{\rm cm}$, whereas the two-step analysis, with every stack's mass free, is protected from it, as was the comparison at equal mass in \citet{vitorelli2018}. The two-step $\gamma_{\rm cm}$ agrees with the forward-modelled one on the real data only because its uncertainty is large. The same trade-off between precision and robustness to a misspecified model separates the joint fit from stacked analyses.

\section{Point estimates: maximum a posteriori and projected joint credible intervals}
\label{app:map}

Marginal posterior medians, which we quote throughout, are unambiguous summaries of one parameter after marginalising over all others, and transform trivially under a change of variables. When parameters are degenerate and the priors are wide, however, the marginal posterior need not peak at the best fit to the data. As a complement, Table~\ref{tab:map} gives the multivariate maximum a posteriori (MAP) point and the projected joint highest-posterior-density (PJ-HPD) interval of \citet{joachimi2021}. The PJ-HPD interval of a parameter is the range spanned, along that parameter, by the highest-posterior-density region of the full multivariate posterior that contains $68\%$ of that parameter's marginal mass; it contains the MAP by construction, reduces to the usual highest-posterior-density interval in one dimension and to the $1\sigma$ interval for a Gaussian posterior, and is generally somewhat wider than the marginal interval for degenerate posteriors.

We evaluate them directly from the posterior samples, using for each sample the log-likelihood returned by the sampler plus the log prior density. The MAP is the highest-posterior sample, which is approximate. For the joint fits, which have $50$ bootstrap posteriors, we compare here with the marginal of their pooled mixture, so that all intervals in this appendix are posterior intervals (wider than the bootstrap-median intervals of Section~\ref{sec:jointresults}, see Section~\ref{sec:likeboot}). The MAP is then the median over the bootstrap MAPs. The PJ-HPD interval is computed on the pooled samples, ordered by their log posterior density relative to each bootstrap's own maximum, with each bootstrap carrying equal mass. This is an approximation, since the density of the pooled mixture is not available. For the stacked fit, a single posterior, the MAP and PJ-HPD interval are those of \citet{joachimi2021} up to the sampling of the posterior peak.

\begin{table}
\footnotesize
\setlength{\tabcolsep}{3pt}
\centering
\caption{MAP and PJ-HPD $68\%$ interval, in brackets, of the joint fit (bootstrap covariance), the analytic-covariance joint fit (with Gaussian miscentering priors centred on the stacked fit) and the forward-modelled stacked fit of the real WaZP+DES-Y3 data.}
\label{tab:map}
\resizebox{\columnwidth}{!}{%
\begin{tabular}{llll}
\toprule
Parameter & Joint, bootstrap cov. & Joint, analytic cov. & Stacked \\
\midrule
$c_0$ & $8.346\;[6.560,\,11.192]$ & $8.670\;[4.460,\,11.236]$ & $6.988\;[5.833,\,8.600]$ \\
$\alpha_{\rm cm}$ & $0.029\;[-0.045,\,0.172]$ & $0.055\;[-0.226,\,0.105]$ & $0.028\;[-0.055,\,0.125]$ \\
$\beta_{\rm cm}$ & $0.913\;[0.509,\,1.669]$ & $0.422\;[-1.816,\,0.765]$ & $0.510\;[-0.016,\,0.816]$ \\
$\gamma_{\rm cm}$ [mag$^{-1}$] & $0.057\;[0.015,\,0.088]$ & $0.100\;[-0.037,\,0.119]$ & $0.112\;[0.072,\,0.131]$ \\
$\log_{10}(M_0/M_\odot)$ & $14.290\;[14.268,\,14.299]$ & $14.260\;[14.239,\,14.284]$ & $14.334\;[14.325,\,14.349]$ \\
$\alpha_{M\lambda}$ & $1.039\;[0.989,\,1.088]$ & $1.047\;[0.969,\,1.110]$ & $0.927\;[0.865,\,0.961]$ \\
$\beta_{M\lambda}$ & $-2.028\;[-2.279,\,-1.812]$ & $-3.592\;[-3.868,\,-3.184]$ & $-1.732\;[-1.996,\,-1.499]$ \\
$f_{\rm mis}$ & $0.370\;[0.266,\,0.461]$ & $0.506\;[0.397,\,0.587]$ & $0.307\;[0.217,\,0.390]$ \\
$\sigma_{\rm off}$ [Mpc] & $0.389\;[0.330,\,0.437]$ & $0.515\;[0.471,\,0.567]$ & $0.422\;[0.374,\,0.470]$ \\
\bottomrule
\end{tabular}}
\end{table}

The MAP agrees with the marginal median within $0.4$ of the posterior width for every parameter of the three fits, and within $0.15$ for the joint fits, so the choice of point estimate does not change any conclusion drawn from the medians. The PJ-HPD intervals have widths similar to the marginal ones, but for the degenerate parameters ($c_0$, $f_{\rm mis}$, $\alpha_{\rm cm}$, $\beta_{\rm cm}$) they are shifted along the degeneracy discussed in Section~\ref{sec:jointresults}; $\gamma_{\rm cm}$ and $\alpha_{M\lambda}$ agree within their intervals between the two covariance estimators, the mass normalisation shifts by $0.030$~dex between the two, and $\beta_{M\lambda}$ depends strongly on it.

\end{appendix}

\end{document}